\documentclass[%
groupedaddress,
unsortedaddress,
nofootinbib,
 amsmath,amssymb,
 aps,
twocolumn,
]{revtex4-2}

\usepackage[utf8]{inputenc}
\usepackage[english]{babel}
\usepackage{titlesec}
\usepackage{etoolbox}

\usepackage{amsmath}
\usepackage{amssymb} 
\usepackage{booktabs}

\usepackage{graphicx} 
\usepackage{xcolor}
\usepackage{natbib}
\usepackage{braket}
\usepackage{float}
\usepackage[hidelinks]{hyperref}

\usepackage[normalem]{ulem}
\newcommand{\stkout}[1]{\ifmmode\text{\sout{\ensuremath{#1}}}\else\sout{#1}\fi}

\usepackage[all]{nowidow}
\usepackage{siunitx}
\usepackage{multirow}

\newcommand{\Mg}{\ensuremath{\mathrm{Mg}^{2+}}}
\newcommand{\Cl}{\ensuremath{\mathrm{Cl}^{-}}}

\begin{document}

\preprint{APS/123-QED}

\title{
Can DFT-trained neural network potentials reproduce structure, solvation, and water-exchange properties in aqueous magnesium solutions?
}

\date{\today}

\author{Sebastian Falkner 
}
\affiliation{Institute of Physics, University of Augsburg, Universit\"atsstraße 1, 86159 Augsburg, Germany.}
\affiliation{Faculty of Physics, University of Vienna, 1090 Vienna, Austria}

\author{Pablo Montero de Hijes
}
\affiliation{Faculty of Physics, University of Vienna, 1090 Vienna, Austria}

\author{Christoph Dellago 
}
\affiliation{Faculty of Physics, University of Vienna, 1090 Vienna, Austria}
\affiliation{Research Platform on Accelerating Photoreaction Discovery (ViRAPID), University of Vienna, 1090 Vienna, Austria}

\author{Nadine Schwierz
}
\email{nadine.schwierz@uni-a.de}
\affiliation{Institute of Physics, University of Augsburg, Universit\"atsstraße 1, 86159 Augsburg, Germany.}

\begin{abstract}
Magnesium ions play an essential role in many biological processes but remain challenging to model in biomolecular simulations. Despite considerable scientific effort, classical force fields fail to simultaneously reproduce  key structural, thermodynamic and kinetic solution properties, likely due to their inability to explicitly account for quantum many-body effects. 
Here, we develop and systematically benchmark MACE neural network potentials (NNPs) for aqueous MgCl$_2$ solutions trained on revPBE-D3/zd and revPBE0-D3/zd density functional theory reference data and assess their ability to reproduce a broad range of experimental solution properties including the structure of the first hydration shell, diffusion coefficient, activity derivative, water-exchange rate and mechanism as well as  solvation free energy. 
Both NNPs accurately reproduce the octahedral structure of the first hydration shell, ion pairing properties and diffusion coefficients. Combining the NNPs with transition path sampling and other enhanced sampling techniques allows us to capture the rare event of water exchange in the first hydration shell of Mg$^{2+}$ revealing a dissociative exchange mechanism. Transition interface sampling yields exchange rates within one order of magnitude of experiment, representing a substantial improvement over classical dissociative force fields.
In contrast, the NNP-derived solvation free energy significantly underestimates the experimental value, revealing a limitation of the present local NNP architectures for describing ion solvation thermodynamics. Our results demonstrate that DFT-trained NNPs can accurately describe Mg$^{2+}$ hydration structure, diffusion, ion pairing, and exchange kinetics, while highlighting the need for explicit long-range electrostatic treatments to achieve quantitative agreement with experimental ion solvation free energies.
\end{abstract}
\maketitle

\section{Introduction}
Magnesium ions play a fundamental role in numerous biological processes, serving as stabilizing agents for proteins and RNA and as essential cofactors in a wide range of enzymatic reactions~\cite{Misra1998,Williams2000,Cowan2002,Pyle2002,Born2009,Stachura2017}. Classical molecular dynamics (MD) simulations have become a standard tool for investigating Mg$^{2+}$ in complex aqueous and biological environments, and considerable effort has been devoted to the development and optimization of ion force fields to reproduce experimental data ~\cite{Allnr2012,Li2014,Dubou-Dijon2018,Mamatkulov2018,Grotz2021,Soniat2015,Jing2018}. 
Despite these efforts, classical force field simulations often fail to simultaneously reproduce key experimental structural and thermodynamic properties such as the hydration free energy and the structure of the first hydration shell  \cite{Mamatkulov2013,Li2013,Mamatkulov2018,Grotz2021}. Additional limitations include unrealistically slow exchange kinetics in the first hydration shell of Mg$^{2+}$ and excessive binding of divalent ions to nucleic acids \cite{Schwierz2020,Cruz-Len2020}. These limitations arise from the functional form of standard, fixed-charge classical force fields, which cannot explicitly capture quantum many-body effects such as polarization and charge transfer.
Although approximate corrections, including modified combination rules~\cite{Fyta2012}, charge scaling approaches~\cite{Leontyev2011,Kohagen2014,Zeron2019}, and additional empirical terms in the pair potential~\cite{Li2015}, have been proposed, these methods do not provide a rigorous treatment of the underlying many-body effects.

Machine-learned interatomic potentials have emerged as a powerful alternative to conventional classical force fields~\cite{Schran2021,Kocer2022,Omranpour2024}. In particular, neural network potentials (NNPs) provide a promising framework for extending near electronic-structure accuracy to the time and length scales required for the simulation of complex condensed-phase and biomolecular systems~\cite{Kocer2022,Daru2022,Liu2022,Batatia2022}. A major breakthrough demonstrating the capabilities of NNPs has been achieved in simulations of liquid water, where NNP-based models trained on \textit{ab initio} reference data were shown to reproduce structural, thermodynamic, and dynamical properties with near first-principles accuracy~\cite{Morawietz2016,Schran2021,Omranpour2024,MonterodeHijes2024b}. In this way, NNPs can bridge the gap between electronic-structure calculations and experimentally relevant simulation scales, which remain inaccessible to conventional \textit{ab initio} molecular dynamics (AIMD) due to its prohibitively high computational cost~\cite{Lightstone2001,Bhattacharjee2012,Wang2020}. Recently, NNPs have also been developed for aqueous ionic systems and electrolyte solutions, including Mg$^{2+}$, MgCl$_2$, NaCl, Na$_2$SO$_4$, and ZnCl$_2$ solutions~\cite{Juraskova2025,Ferretti2025,ONeill2024,Soyemi2025,Cao2025}.

However, NNPs are not free of limitations, as they necessarily reflect the approximations and deficiencies of the underlying electronic-structure reference method used for training. For instance, the structural, thermodynamic, and dynamical properties of water and aqueous electrolytes are known to depend sensitively on the choice of exchange--correlation functional, the treatment of dispersion interactions, and subtle error cancellation effects~\cite{Gillan2016,Palos2022,Dasgupta2021,ONeill2024}. 
Hence, careful validation against experimental reference data is essential. For classical force fields, parameterization and validation are typically based on experimental ion properties such as the solvation free energy, Mg$^{2+}$--oxygen distances in the first hydration shell, hydration numbers, activity derivatives, and water exchange kinetics in the first solvation shell. 
However, a comprehensive and systematic validation of NNPs for aqueous electrolyte solutions against a broad range of experimental observables has not yet been performed.

In this work, we present a robust MACE-based NNP for aqueous MgCl$_{2}$ solutions and systematically investigate a broad range of structural, thermodynamic, and kinetic properties in comparison with established classical force fields and experimental reference data. To assess the sensitivity of the resulting properties to the underlying electronic-structure description, we train NNPs based on both the revPBE and hybrid revPBE0 exchange--correlation functionals~\cite{Perdew1996,Zhang1998,Adamo1999} with D3 zero-damping dispersion corrections~\cite{Grimme2010, Grimme2011,Lausch2025}, motivated by the well-known functional dependence of water and electrolyte properties~\cite{Gillan2016,Palos2022,ONeill2024}. 
As a stringent test of whether NNPs can simultaneously reproduce the structure, thermodynamics, and dynamics of aqueous Mg$^{2+}$ systems, we calculate solvation free energy, hydration structure, ion pairing, water-exchange rate and determine the mechanism of water exchange.

We find that both NNPs accurately reproduce the octahedral hydration shell structure of Mg$^{2+}$ and predict a dissociative water-exchange mechanism, with computed exchange rate constants within one order of magnitude of experimental values. At the same time, the NNP-derived solvation free energies are significantly underestimated, indicating that despite the significant improvements offered by NNPs, important challenges remain in accurately describing long-range electrostatic interactions and charge-transfer effects in aqueous electrolyte systems.

\section{Methods}

\subsection{Molecular Dynamics Simulations}
\textbf{Force Field Simulations:}
Initial classical force field molecular dynamics (MD) simulations were conducted using GROMACS~2023.05~\cite{Abraham2015} with the Mamatkulov-Schwierz \Mg parameters~\cite{Mamatkulov2018} and the TIP3P flexible water model~\cite{Jorgensen1983}. The simulation protocol comprised energy minimization, followed by \SI{1}{\nano\second} of NVT equilibration and \SI{1}{\nano\second} of NPT equilibration at \SI{298.15}{\kelvin} and \SI{1}{\bar}. Final classical force field production runs were executed in the NPT ensemble for \SI{5}{\nano\second}. Detailed numerical parameters, including cutoff schemes, electrostatic treatments, and thermostat/barostat settings, are provided in the Supplementary Information.
\medskip

\textbf{Ab-Initio MD:}
All ab initio molecular dynamics simulations were performed using CP2K~2025~\cite{Hutter2014} within the Gaussian and Augmented Plane Wave (GAPW) framework~\cite{Lippert1999}. Electronic structure calculations employed the revPBE exchange-correlation functional~\cite{Perdew1996, Zhang1998} supplemented by Grimme's DFT-D3 dispersion correction~\cite{Grimme2010, Grimme2011} with zero damping, using TZV2P-GTH basis sets and GTH-PBE pseudopotentials~\cite{Goedecker1996, Hartwigsen1998}. Self-consistent field convergence was achieved via the orbital transformation method with a DIIS minimizer. 

The AIMD simulations were initiated from force field equilibrated structures following brief geometry optimizations. Molecular dynamics were conducted in the NVT ensemble at \SI{298.15}{\kelvin} using the CSVR thermostat~\cite{Bussi2007}. Detailed electronic and numerical parameters are provided in the Supplementary Information. 

The initial NNP training dataset was produced by running five independent replicas for a pure water system ($\mathrm{w}256$) and a water system with solvated $\mathrm{MgCl}_2$ ($\mathrm{w}256\mathrm{MgCl}_2$). For later testing of NNP transferability, simulations of a larger $512$-water system ($\mathrm{w}512\mathrm{MgCl}_2$) and each ion separately in the larger system ($\mathrm{w}512\mathrm{Mg}$ and $\mathrm{w}512\mathrm{Cl}$) were performed. In order to obtain revPBE0 training data, we follow previous works by assuming that revPBE and revPBE0 phase spaces have significant overlap~\cite{MonterodeHijes2024}. Under this assumption, we computed single-point revPBE0-D3/zd energies and forces for configurations extracted from revPBE-D3/zd AIMD trajectories at intervals of \SI{5}{\femto\second}, which is the stride we later use to subsample our training data.
\medskip

\textbf{NNP Simulations:}
All NNP simulations were conducted using OpenMM version~$8.2.0$~\cite{Eastman2017}. The neural network potential was implemented via the Message Passing Atomic Cluster Expansion (MACE) model~\cite{Batatia2022}; parameter details are given in the next section. The Langevin equations of motion were integrated using the OVRVO integrator with time step rescaling~\cite{Sivak2014}. The integration parameters were set to a time step of \SI{1}{\femto\second} and a friction constant of \SI{1}{\per\pico\second}. Simulations were typically performed in the NPT ensemble at a temperature of \SI{298.15}{\kelvin} and a pressure of \SI{1.0}{\bar}, controlled by a Monte Carlo barostat. Enhanced sampling NNP-MD simulations, including metadynamics~\cite{Laio2002, Barducci2008} and umbrella sampling~\cite{Torrie1977}, were implemented via the PLUMED plugin~\cite{Bonomi2019, Tribello2014} to apply biasing potentials. Specific deviations from these standard simulation parameters, such as timestep adjustments or changes in ensemble conditions, are specified in the respective sections.

\subsection{Neural Network Potential}

\textbf{Model Architecture and Training:}
We used the MACE architecture~\cite{Batatia2022} to construct the NNP. MACE is an equivariant message passing neural network that uses higher-order body messages to describe atomic interactions. The model here consists of two message passing layers. The features are updated based on the local environment within a radial cutoff of \SI{5.0}{\angstrom}. The maximum symmetry order of the messages was set to $L=0$, corresponding to an invariant model. We found no significant improvement setting $L>0$ for our dataset. The number of feature channels was set to $32$. These parameters were selected based on initial benchmarks showing a good balance between simulation performance and accuracy. The AIMD-derived $\mathrm{w}256$ and $\mathrm{w}256\mathrm{MgCl}_2$ datasets were split into training, validation, and test sets for both revPBE-D3/zd and revPBE0-D3/zd potentials, with isolated atomic energies ($E_0$) determined from DFT calculations. Models were trained separately on a single NVIDIA A100 GPU using the MACE reference implementation; detailed hyperparameters, batch sizes, and dataset splits are provided in the Supplementary Information.
\medskip

\textbf{Iterative Training:}
To generate additional training data, we performed NNP molecular dynamics simulations using the initially trained NNPs to explore phase space regions beyond the initial dataset, in particular transition states of the water exchange process. These included standard NPT simulations, high-temperature NVT runs, a \SI{1}{M} $\mathrm{MgCl}_2$ solution simulation, and two well-tempered metadynamics~\cite{Laio2002, Barducci2008} simulations biased along Mg$^{2+}$-oxygen and Mg$^{2+}$-Cl$^-$ distances. We ran single-point DFT calculations on configurations extracted from these NNP-MD trajectories, added them to the initial dataset, and performed a second round of model training. Detailed simulation parameters are provided in the Supplementary Information.

\subsection{Ion Properties}

\textbf{Structure of First Hydration Shell:}
To determine the first-shell Mg$^{2+}$-oxygen distance ($R_1$) and Mg$^{2+}$ coordination number ($n_1$), we conducted ten independent equilibrium NNP-MD simulations for each system. These production runs were performed in the NPT ensemble at \SI{298.15}{\kelvin} and \SI{1.0}{\bar} for a duration of \SI{1}{\nano\second} each. Other simulation parameters remained as stated above. The final configurations of these runs were then used as initial configurations for most of the following calculations.
\medskip

\textbf{Diffusion Coefficient:}
Diffusion coefficients ($D_0$) of the Mg$^{2+}$ ion were computed from five additional NNP-MD equilibrium simulations performed in the NVT ensemble. These simulations utilized the same temperature (\SI{298.15}{\kelvin}) and time step (\SI{1}{\femto\second}) as the NPT runs, lasting \SI{1}{\nano\second} per replica. Trajectories were analyzed using the Einstein relation to extract the self-diffusion coefficients. Specifically, we follow the workflow as explained in Grotz~et~al.~\cite{Grotz2021}.
\medskip

\textbf{Activity Derivative:}
Activity derivatives ($a_{cc}$) were determined from five NNP-MD equilibrium simulations of a \SI{1.08}{m} ($\approx \SI{1}{M}$) $\text{MgCl}_2$ solution in the NPT ensemble at \SI{298.15}{\kelvin} and \SI{1}{\bar}. These production runs were performed for \SI{1}{\nano\second} using a simulation box with an edge length of \SI{2.5}{\nano\meter}.  While the integration timestep and temperature coupling were in line with the common NNP parameters stated above, the Monte Carlo barostat coupling was reduced to $25$ steps to ensure proper density equilibration at the high ion concentration. Activity derivatives were determined as described in previous works~\cite{Kirkwood1951}.
\medskip

\textbf{Free Energy Calculations:}
Absolute solvation free energies ($\Delta G_{\mathrm{solv}}$) were computed using two distinct alchemical pathways to validate the NNP predictions. The first method followed an indirect thermodynamic cycle approach~\cite{Karwounopoulos2024}, where the free energy difference between the classical force field and the NNP level of theory was calculated. The second method employed a direct decoupling protocol based on neighborlist manipulation~\cite{Picha2025}, allowing the solute interactions to be gradually switched off directly without transforming to a force field representation. The solvation free energy of neutral MgCl$_2$ ion pairs $\Delta G_{\mathrm{solv}}$ was obtained by first computing the single-ion solvation energies for Mg$^{2+}$ and Cl$^-$, applying corrections, and summing the results \cite{Grotz2021}. Detailed $\lambda$-window configurations, sampling durations, and correction schemes are provided in the Supplementary Information.
\medskip

\textbf{Exchange Mechanism from Transition Path Sampling:}
Flexible-length transition path sampling (TPS) simulations~\cite{Bolhuis2002} were performed to characterize the mechanism of water exchange in the first hydration shell. The underlying molecular dynamics parameters followed the standard NNP-MD protocols outlined above. Initial reactive pathways were generated using steered molecular dynamics with a moving harmonic restraint on the Mg$^{2+}$-oxygen distance. The reaction progress was monitored by using an order parameter $\xi$ to define two stable states: State A, where a water molecule resides in the first hydration shell, and State B, where an external water molecule has replaced it. Detailed parameters for the steering protocol, the exact mathematical definition of $\xi$, and sampling settings are provided in the Supplementary Information.
\medskip

\textbf{Exchange Rate from Transition Interface Sampling:}
Transition interface sampling (TIS) simulations~\cite{vanErp2003} were performed using the DFT-trained NNPs to calculate the rate constants for water exchange. Following the TPS section, we employed the same underlying NNP molecular dynamics parameters, order parameter $\xi$, and stable state definitions. The flux through the first interface was estimated from unbiased MD trajectories. Initial reactive pathways for the interface ensembles were generated using steered molecular dynamics with harmonic restraints on two separate Mg$^{2+}$-oxygen distances and coordination numbers. Path sampling was conducted across multiple interfaces with parallel path swapping between ensembles to enhance sampling efficiency~\cite{vanErp2007}. Detailed interface spacing, the functional form of the shooting weight function, and all numerical parameters are provided in the Supplementary Information.
\medskip

\textbf{Potential of Mean Force:}
The potential of mean force (PMF) along the Mg$^{2+}$–oxygen distance was computed using umbrella sampling~\cite{Torrie1977} in  PLUMED~\cite{Bonomi2019, Tribello2014} combined with the readily available NNP-MD equilibrium simulations. To reduce the computational load, umbrella windows were only run for regions missing in the PMF reconstructed from equilibrium data. The free energy profiles were reconstructed using the weighted histogram analysis method (WHAM)~\cite{Kumar1992}. Detailed parameters regarding the window placement, spacing, and biasing force constants are provided in the Supplementary Information.
\medskip

\textbf{Metadynamics:}
Well-tempered metadynamics~\cite{Laio2002, Barducci2008} was used to explore the free energy landscape along the Mg$^{2+}$–oxygen distance and the ion coordination number. The simulation parameters followed the standard NNP protocol outlined above and were performed using PLUMED~\cite{Bonomi2019, Tribello2014}. Detailed simulation and biasing parameters are provided in the Supplementary Information.
\medskip

\textbf{Comparison to Classical Force Fields:}
For comparison of the NNP results, we employed two established non-polarizable Mg$^{2+}$ force fields: the Mamatkulov-Schwierz Mg$^{2+}$ parameters~\cite{Mamatkulov2018} and the microMg parameters~\cite{Grotz2021} combined with the TIP3P water model. Both force fields were optimized to reproduce experimental solution properties including 
 Mg$^{2+}$–oxygen distances, coordination number, activity derivative and solvation free energy.
 In addition these models provide a particularly useful comparison since microMg closely reproduces the experimental water exchange rate but follows an associative water exchange mechanism. The Mamatkulov-Schwierz model follows a dissociative exchange mechanism but significantly underestimates the exchange rate \cite{Schwierz2020}.

\section{Results and Discussion}
\begin{figure*}[tb!]
    \centering
    \includegraphics[width=\linewidth]{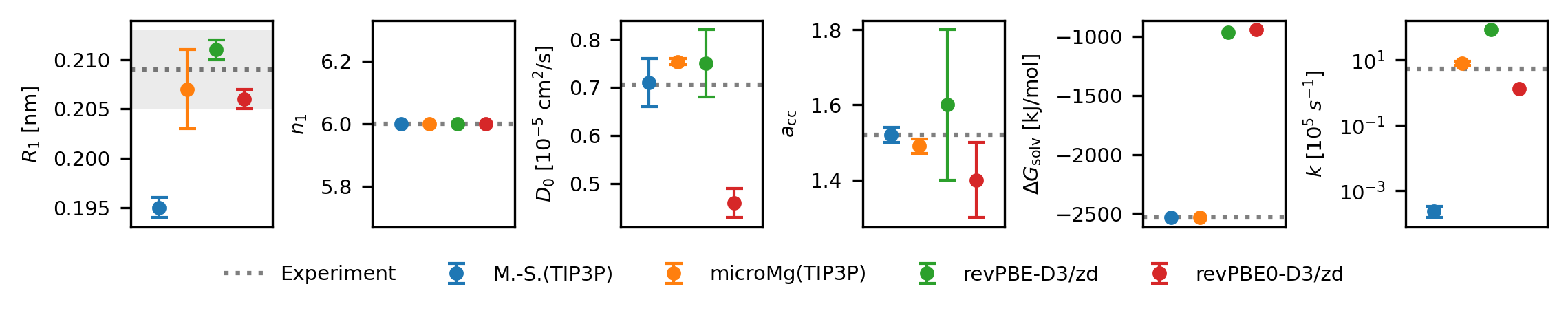}
    \caption{Comparison of structural, dynamical, kinetic, and thermodynamic properties of aqueous Mg$^{2+}$ obtained from classical force fields, DFT-trained neural network potentials to the experimental reference. Reported properties are the first-shell Mg$^{2+}$-oxygen distance $R_1$, first-shell coordination number $n_1$, Mg$^{2+}$ self-diffusion coefficient $D_0$, activity derivative $a_{cc}$ for \SI{1.08}{m} MgCl$_2$ concentration, 
 solvation free energy $\Delta G_{\mathrm{solv}}$ of neutral MgCl$_2$ ion pairs and the water-exchange rate constant $k$. The experimental rate is indicated as the dashed line and shaded areas indicate experimental uncertainties if available.}
    \label{fig:Mg2_properties}
\end{figure*}

We first evaluate the accuracy and transferability of the DFT-trained NNPs. Subsequently, we present  a systematic benchmark against experimental data and established classical force fields. The analysis covers Mg$^{2+}$ hydration structure, self-diffusion, first-shell water exchange kinetics and mechanism, Mg$^{2+}$--Cl$^{-}$ ion pairing, activity derivative, and solvation free energy (Fig.~\ref{fig:Mg2_properties}, Tab.~\ref{tab:Mg2_properties}),  
 thereby testing whether the NNPs can simultaneously describe structure, solvation and water-exchange properties in aqueous Mg$^{2+}$ solutions.

\subsection{Neural Network Potential Training}

To obtain an accurate and transferable description of aqueous Mg$^{2+}$, we develop a neural network potential (NNP) using the MACE framework. The training is performed in an iterative manner to progressively improve the robustness of the model and extend its coverage of relevant configurational space. An initial model is first trained only on configurations from unbiased AIMD simulations and is subsequently used to efficiently sample additional regions of phase space that are not represented in the original training data. The initial model showed good accuracy for equilibrium validation configurations. However, enhanced sampling using the preliminary NNP revealed significant errors in predicted energies and forces for configurations outside the configuration space regions represented in the initial training set (see Supplementary Information Fig.~S1 for a comparison before and after iterative refinement).
This behavior is consistent with previous NNP studies, which showed that models can appear accurate while failing in rare-event or high-free-energy regions if such configurations are not represented in the training data~\cite{Mondal2023,Park2025,Juraskova2025,Ferretti2025}. 
In the present case, the relevant rare event is water exchange in the first hydration shell of Mg$^{2+}$, which occurs on the microsecond timescale and transiently gives rise to configurations with increased or decreased hydration numbers of the Mg$^{2+}$ ion. Accurate representation of these configurations is essential for correctly describing both the exchange kinetics and the underlying exchange mechanism, as discussed in detail below.

After retraining on configurations generated through enhanced sampling with the preliminary NNP, the model showed a substantially improved description of transient water-exchange configurations with altered first-shell hydration numbers.
Similar active-learning strategies were used by Juraskova et al.~\cite{Juraskova2025} and Ferretti et al.~\cite{Ferretti2025}, where ligand-exchange or coordination-number sampling was explicitly included to improve the description of metal-solvent exchange regions.
Importantly, the refined model also successfully extrapolates to larger and charged simulation boxes, particularly with respect to force predictions, as validated against DFT single-point calculations (see Supplementary Information Fig.~S1).

\begin{table*}[t!]
\caption{
Comparison of structural, dynamical, thermodynamic and kinetic  properties of aqueous Mg$^{2+}$ obtained from classical force fields, DFT-trained neural network potentials, and experiments. Reported properties include the first-shell Mg$^{2+}$-oxygen distance $R_1$, first-shell coordination number $n_1$, Mg$^{2+}$ self-diffusion coefficient $D_0$, activity derivative $a_{cc}$ for 1.08m MgCl$_2$ concentration, 
solvation free energy $\Delta G_{\mathrm{solv}}$ of neutral MgCl$_2$ ion pairs, 
water-exchange rate constant $k$, and the corresponding water-exchange mechanism. Results obtained in the present work are highlighted in \textbf{bold}.}

\begin{ruledtabular}
\footnotesize
\def\arraystretch{1.8}
\begin{tabular}{lccccccl}
\multirow{2}{*}{Description} & 
\multirow{2}{*}{$R_\mathrm{1}$ [nm] }& 
\multirow{2}{*}{$n_\mathrm{1}$} & 
\multirow{2}{*}{$D_\mathrm{0}$ [$10^{-5}$ cm$^2$/s]} & 
\multirow{2}{*}{$a_\mathrm{cc}$ } & 
\multirow{2}{*}{$\Delta G_\mathrm{solv}$ [kJ/mol]} &
\multirow{2}{*}{$k$ [$10^{5}$ s$^{-1}$]} & 
\multirow{2}{*}{Exch. Mech.}  
\\ \\
\hline
M.-S.(TIP3P)~\cite{Mamatkulov2018}& $0.195 \pm 0.001$ & $6$ & $0.71 \pm 0.05^a$ & $1.52 \pm 0.02$ & $-2531.1$ & $ (24.0 \pm 8.8) \times 10^{-5}$ $^b$ & Dissoc.\\
microMg(TIP3P)~\cite{Grotz2021}& $0.207 \pm 0.004$ & $6$ & $0.754 \pm 0.006^a$ & $1.49 \pm 0.02$ & $-2532.9 \pm 1$ & $ 8.04 \pm 1.20$ & Assoc.\\
\hline
revPBE-D$^{\mathrm{OPT}}$ ~\cite{Ferretti2025} & $0.207$ & $6$ & $-$ & $-$ & $-$ & $20.0$ $^b$ & Dissoc.\\
$\omega$B97X-D3BJ ~\cite{Juraskova2025} & $0.208$ & $6$ & $-$ & $-$ & $-$ & $-$ & Dissoc. \\
\textbf{revPBE-D3/zd} & $0.211 \pm 0.001$ & $6$ & $0.75 \pm 0.07$ & $1.6 \pm 0.2$ & $-961/-984$ $^c$ & $82.8$ & Dissoc.\\
\textbf{revPBE0-D3/zd} & $0.206 \pm 0.001$ & $6$ & $0.46 \pm 0.03$ & $1.4 \pm 0.1$ & $-943/-946$ $^c$ & $1.29$ & Dissoc.\\
\hline
Experiment & $0.209 \pm 0.004$ ~\cite{Marcus1988} & $6$ ~\cite{Marcus1988} & $0.706$ ~\cite{Marcus1997} & $1.52$ ~\cite{Robinson2002} & $-2532$ ~\cite{Marcus1997} & $5.3 / 6.7$ ~\cite{Bleuzen1997,Neely1970} & Dissoc. ~\cite{Helm1999,Bleuzen1997}\\
\end{tabular}
\end{ruledtabular}
\footnotetext[1]{Corrected for water viscosity }
\footnotetext[2]{From transition state theory}
\footnotetext[3]{Thermodynamic Cycle/Neighborlist Decoupling}
\label{tab:Mg2_properties}
\end{table*}

\subsection{Structure of the First Hydration Shell and Self-diffusion Coefficient}

The structure of the hydration shells is a key property of aqueous Mg$^{2+}$. In particular, the strongly bound six water molecules in the first hydration shell govern many of the structural and kinetic properties of the ion. The distance to the oxygen atoms in the first hydration shell, $R_1$, and the first-shell coordination number, $n_1$, therefore provide direct measures of whether the model accurately reproduces the characteristic octahedral hydration structure.

Both the revPBE-D3/zd and revPBE0-D3/zd NNPs accurately reproduce the DFT (Fig.~S2)
and experimental first-shell properties of aqueous Mg$^{2+}$ (Tab.~\ref{tab:Mg2_properties}, Fig. ~\ref{fig:Mg2_properties}).
The predicted average first-shell distances, $R_1$, of $0.211$ nm for revPBE-D3/zd and $0.206$ nm for revPBE0-D3/zd are both within the experimental uncertainty of $0.209 \pm 0.004$ nm. In addition, both models yield a first-shell coordination number $n_1 = 6$, consistent with previous simulations and experimental studies reporting an octahedral first hydration shell for the Mg$^{2+}$ ion~\cite{Marcus1988,Lightstone2001,Jiao2006,Juraskova2025,Ferretti2025}. Overall, both DFT-trained NNPs accurately capture the key structural features of the Mg$^{2+}$ first hydration shell.

The self-diffusion coefficient of Mg$^{2+}$ provides a complementary measure of model accuracy, as it probes the mobility of the ion in solution. It therefore serves as an important test of whether a model that reproduces the correct hydration structure also yields realistic transport properties.
The diffusion coefficient obtained with revPBE-D3/zd is in good agreement with experiment, yielding $D_0 = 0.75 \times 10^{-5}$ cm$^2$/s compared to the experimental value of $0.706 \times 10^{-5}$ cm$^2$/s \cite{Marcus1997}. In contrast, revPBE0-D3/zd predicts a lower diffusion coefficient of $0.46 \times 10^{-5}$ cm$^2$/s, indicating significantly reduced Mg$^{2+}$ mobility in solution. 
This trend is consistent with previous DFT studies of liquid water, which showed that revPBE-type GGAs tend to produce comparatively less structured and more mobile water, whereas revPBE0 does not necessarily improve the description of condensed-phase water dynamics despite the inclusion of exact exchange~\cite{Gillan2016,ONeill2024}.

Previous studies reported that revPBE-D3/zd reproduces bulk-water properties well, at least partly due to favorable error compensation, whereas revPBE0-D3/zd yields similar performance for bulk water and Cl$^{-}$ hydration without systematically improving the description of cation--water interactions~\cite{ONeill2024}. In addition, revPBE-D3/zd was shown to reproduce the experimental water density isobar more accurately than its hybrid counterpart~\cite{MonterodeHijes2024}. In the context of \textit{ab initio} Mg$^{2+}$ simulations, Ferretti et al.~\cite{Ferretti2025} similarly noted that revPBE-D4 provides a reasonable description of aqueous Mg$^{2+}$, although its performance depends sensitively on the treatment of dispersion interactions and associated error compensation. Overall, our diffusion results are consistent with the broader picture that revPBE-D3/zd yields favorable water-like dynamics, whereas revPBE0-D3/zd does not provide a systematic improvement for the structural and transport properties of Mg$^{2+}$.

Table~\ref{tab:Mg2_properties} and Fig.~\ref{fig:Mg2_properties} also include a comparison to classical force fields, many of which reproduce the experimental Mg$^{2+}$ diffusion coefficient with good accuracy. For such comparisons, however, it is important to note that diffusion coefficients obtained from classical force fields are commonly corrected for the viscosity of the underlying water model~\cite{Yeh2004}. 
This procedure compensates for deviations in the water self-diffusion and is intended to isolate the contribution of the Mg$^{2+}$ force field itself to the ion transport properties. In the present work, no viscosity correction was applied, such that the reported diffusion coefficients reflect the combined description of both Mg$^{2+}$ and the underlying DFT-based water dynamics.

\begin{figure}[tb!]
    \centering
    \includegraphics[width=\linewidth]{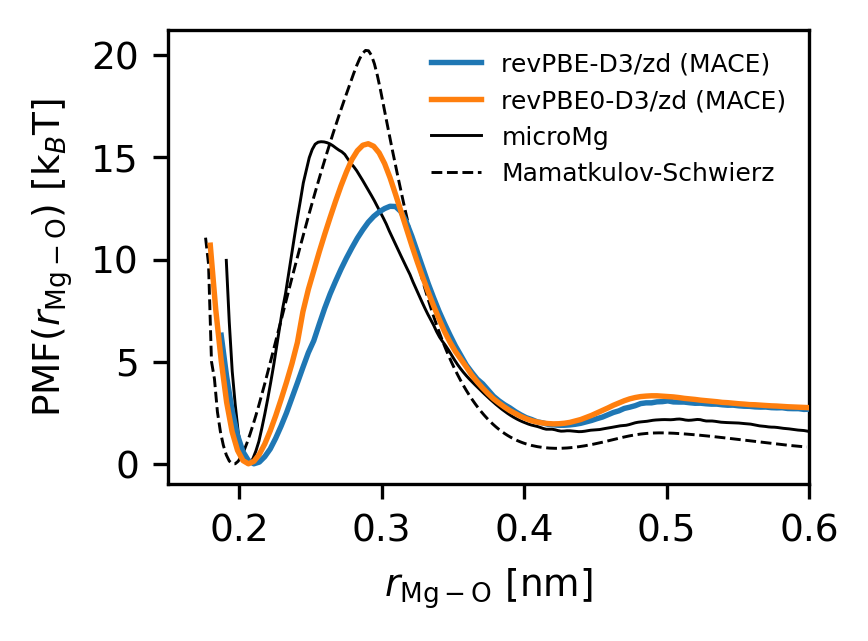}
    \caption{Comparison of PMF as function of the Mg$^{2+}$-water oxygen distance $r_\mathrm{Mg-O}$ derived from DFT-based NNPs (revPBE-D3/zd and revPBE0-D3/zd) against two selected classical force fields Mg$^{2+}$ (microMg \cite{Grotz2021} and Mamatkulov-Schwierz\cite{Mamatkulov2018}).}
    \label{fig:Mg2_PMF}
\end{figure}

\begin{figure}[tb!]
    \centering
    \includegraphics[width=\linewidth]{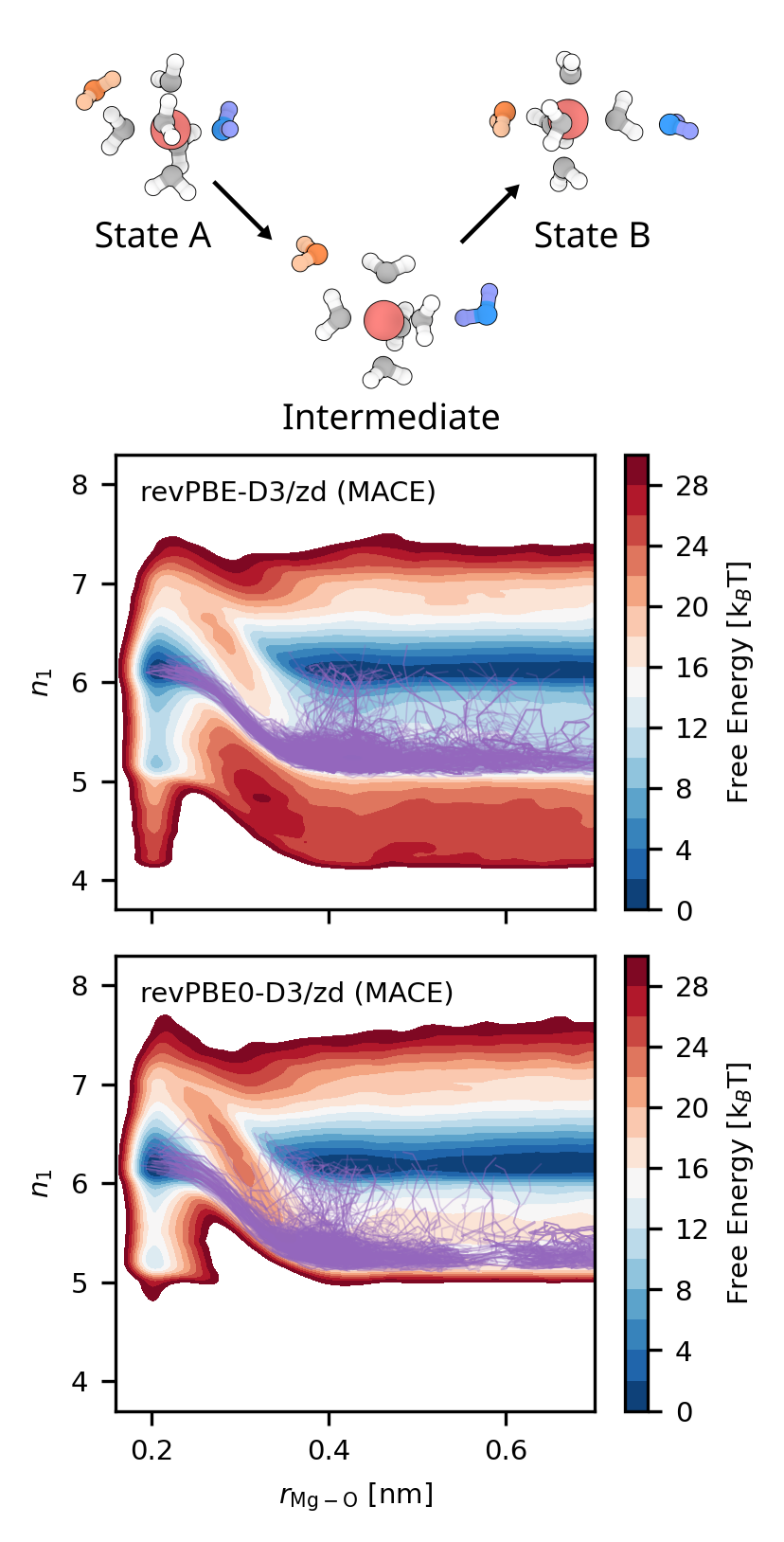}
    \caption{Top: Schematic representation of water exchange in the first hydration shell of Mg$^{2+}$: Initial state A, representative five-fold coordinated intermediate with the exchanging water molecules outside of the first hydration shell, and the final state B. The incoming water molecule is shown in orange and the leaving water molecule in blue. Bottom: Free energy landscapes as functions of the Mg$^{2+}$–oxygen distance of the exchanging water molecule, $r_{\mathrm{Mg-O}}$, and the first-shell coordination number, $n_1$, for the revPBE-D3/zd and revPBE0-D3/zd MACE potentials. Purple lines represent individual water-exchange trajectories sampled by Transition Path Sampling.}
    \label{fig:Mg2_TPS}
\end{figure}

\subsection{Water Exchange in the First Hydration Shell}
Water exchange between the first and second hydration shells of metal ions is a fundamental kinetic process governing reactions in aqueous solution and biological systems. 
Reproducing both the rate and mechanism of water exchange represents a particularly stringent test for the accuracy of aqueous Mg$^{2+}$ descriptions obtained with neural network potentials, since classical force fields have so far generally failed to reproduce both properties simultaneously~\cite{Grotz2021,Schwierz2020,Falkner2021}.

In general, water exchange mechanisms are classified as associative or dissociative depending on whether the exchange proceeds through an intermediate or transition state with increased or decreased coordination number~\cite{CHLangford1965,Falkner2021}. Experimentally, the mechanism is inferred indirectly from the activation volume obtained from the pressure dependence of the exchange rate measured by $^{17}$O NMR spectroscopy, which indicates an interchange-dissociative ($I_d$) mechanism for Mg$^{2+}$ ions ~\cite{Neely1970,Bleuzen1997}.

A common first step for characterizing Mg$^{2+}$ water exchange is the calculation of the potential of mean force (PMF) as a function of the Mg$^{2+}$--water oxygen distance $r_\mathrm{Mg-O}$, which provides an initial estimate of the water-exchange free-energy barrier (Fig.~\ref{fig:Mg2_PMF}).

The PMF obtained with revPBE-D3/zd exhibits a barrier height of $12.6\,k_\mathrm{B}T$ (Fig.~\ref{fig:Mg2_PMF}). In contrast, revPBE0-D3/zd predicts a substantially higher barrier of $15.7\,k_\mathrm{B}T$. The revPBE-D3/zd barrier is in good agreement with results reported by Juraskova et al., who obtained a barrier of $12.83\,k_\mathrm{B}T$ from umbrella sampling simulations ~\cite{Juraskova2025}. This agreement is notable because Juraskova et al. employed a cluster-trained MACE model based on a $\omega$B97X-D3BJ reference, whereas the present revPBE-D3/zd model was trained and applied for periodic MgCl$_2$ solutions.

However, a one-dimensional PMF along $r_\mathrm{Mg-O}$ alone is insufficient to resolve the underlying water-exchange mechanism~\cite{Schwierz2020}. To obtain a complete molecular picture of the water exchange, we therefore combine two-dimensional free energy landscapes along $r_\mathrm{Mg-O}$ and $n_1$ with transition path sampling and transition interface sampling simulations. Here, the choice of the oxygen atom in $r_\mathrm{Mg-O}$ is arbitrary but remains fixed during the simulation.

The $r_\mathrm{Mg-O}$-$n_1$ free energy landscapes (Fig.~\ref{fig:Mg2_TPS}) indicate a dissociative pathway for revPBE-D3/zd and revPBE0-D3/zd due to the lower barrier at a reduced coordination number of $n_1=5$. The TPS simulations confirm that the true reactive trajectories pass through a transient five-fold coordinated state (Fig.~\ref{fig:Mg2_TPS}, top). 
This observation is in agreement with the interchange dissociative mechanism proposed by experiments~\cite{Helm1999,Bleuzen1997}. Based on the free energy landscape, the five-fold coordination state is lower in free energy for revPBE-D3/zd than for revPBE0-D3/zd. TPS shows a slightly longer-lived five-fold transition region for revPBE0-D3/zd which results in longer path lengths. 
These findings agree qualitatively with previous works using different functionals, which observed dissociation from the octahedral first shell before another water molecule entered during pulling~\cite{Juraskova2025}, and that DFT-based potentials favor the five-fold state over heptacoordination~\cite{Ferretti2025}.

We use TIS to obtain water-exchange rates based on true dynamical trajectories, thereby eliminating the need for any transition state theory (TST) approximation. For Mg$^{2+}$ this is particularly important as TST along $r_\mathrm{Mg-O}$ significantly overestimates the true exchange rate \cite{Schwierz2020}. The TIS rate constants are $82.8 \times 10^5$ s$^{-1}$ for revPBE-D3/zd and $1.29 \times 10^5$ s$^{-1}$ for revPBE0-D3/zd, compared to the experimental values of ($5.3 - 6.7) \times 10^5$ s$^{-1}$~\cite{Bleuzen1997, Neely1970}.
Even though revPBE-D3/zd overestimates the experimental exchange rate and revPBE0-D3/zd slightly underestimates it, both results are within an order of magnitude, indicating good agreement.

In a previous study, Ferretti et al.~\cite{Ferretti2025} estimated rate constants from coordination-number free-energy barriers by exponential scaling relative to microMg, which was parametrized to reproduce the experimental exchange rate~\cite{Ferretti2025}. This approach results in approximately $8.4 \times 10^7$ s$^{-1}$ for revPBE-D4 and $2.0 \times 10^6$ s$^{-1}$ for revPBE-D$^\mathrm{OPT}$. However, this approach based on TST assumes that the pre-exponential constant in the Arrhenius equation does not differ between force field and DFT descriptions, which may explain the deviation from our rates, since microMg water exchange is associative while the revPBE-D4/D$^\mathrm{OPT}$ water exchange occurs through a dissociative mechanism.

In summary, the results show that water exchange in the first hydration shell of Mg$^{2+}$ using NNPs with revPBE-D3/zd and revPBE0-D3/zd follows a dissociative mechanism, consistent with the experimentally assigned interchange-dissociative mechanism. Moreover, the calculated water-exchange rates are in reasonable agreement with experiments, representing an important step forward, as classical and polarizable force fields have generally failed to reproduce both properties simultaneously.~\cite{Grotz2021,Schwierz2020,Falkner2021}

\begin{figure}[tb!]
    \centering
    \includegraphics[width=\linewidth]{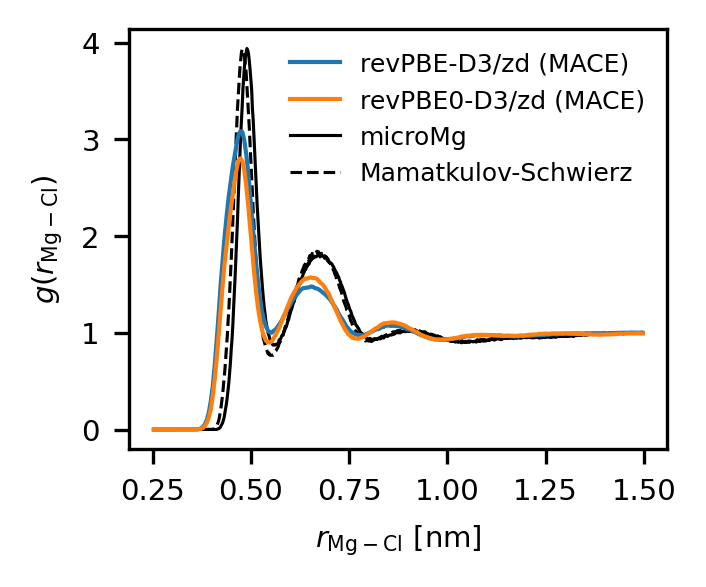}
    \caption{Mg$^{2+}$–Cl$^{-}$ radial distribution functions at a salt concentration of \SI{1.08}{m} for DFT-based NNPs (revPBE-D3/zd and revPBE0-D3/zd) and classical force fields 
    (microMg \cite{Grotz2021} and Mamatkulov-Schwierz\cite{Mamatkulov2018}).}
    \label{fig:Mg2_RDF}
\end{figure}

\subsection{Mg$^{2+}$–Cl$^{-}$ Ion Pairing and Activity Derivatives}

To assess ion pairing in MgCl$_2$ solutions, we analyze the Mg$^{2+}$–Cl$^{-}$ radial distribution function (RDF) and the activity derivative. The Mg$^{2+}$–Cl$^{-}$ radial distribution function provides direct insight into local structure and the propensity for contact and solvent-shared ion pairs. 

The Mg$^{2+}$–Cl$^{-}$ RDF shows no stable inner-shell contact ion pair and no ion clustering under the studied conditions (Fig.~\ref{fig:Mg2_RDF}). Instead, Mg$^{2+}$–Cl$^{-}$ association occurs predominantly through solvent-mediated ion-pairs, as expected~\cite{Schwaab2019}. This result highlights the strong binding of first-shell waters to Mg$^{2+}$ and indicates that Mg$^{2+}$–Cl$^{-}$ association is primarily mediated by the hydration shell rather than direct contact. 

Complementary to the Mg$^{2+}$–Cl$^{-}$ RDF, the activity derivative provides a thermodynamic measure of ion association that can be directly compared with experiments. 
Moreover, it depends on the balance of ion–ion and ion–water interactions in solution. It hence provides a particularly stringent test of whether the NNP captures ion association with the correct balance between Mg$^{2+}$–Cl$^{-}$ and hydration interactions.

The activity derivatives at salt concentration \SI{1.08}{m} 
align well with experimental data (Tab.~\ref{tab:Mg2_properties}, Fig. ~\ref{fig:Mg2_properties}). 
Interestingly, the classical microMg and MS force fields with modified combination rules also reproduce the experimental activity derivative even though the propensity of Mg$^{2+}$–Cl$^{-}$ solvent-shared ion pairs is significantly higher (Fig.~\ref{fig:Mg2_RDF}). This effect is compensated by stronger hydration interactions, resulting in a similar overall thermodynamic response.
The inclusion of many-body quantum effects therefore alters the balance between Mg$^{2+}$–Cl$^{-}$ association and ion hydration relative to the classical force fields, while yielding a comparable activity derivative.

Still, it should be noted that the relative stability of contact and solvent-shared ion pairs can be highly sensitive to the electronic-structure reference, as shown for NaCl, and that revPBE-D3/zd and revPBE0-D3/zd can give similar ion-pair barriers while differing in structural details~\cite{ONeill2024}.

\subsection{Solvation Free Energy}

Solvation free energies provide a stringent thermodynamic test for ion models, since they depend on ion-water interactions, long-range electrostatics, polarization, and charge-transfer effects. They therefore constitute a common target in ion force field optimization. In contrast, solvation free energies have been explored much less for neural network descriptions, partly because absolute ion solvation free energies are challenging to compute for NNPs using alchemical transformation methods.

We calculate the solvation free energy $\Delta G_{\mathrm{solv}}$ of neutral MgCl$_2$ ion pairs using two different methods: one based on neighborlist decoupling and the other based on a thermodynamic cycle involving a transformation to a classical force field description (see Methods). 

The NNP-derived $\Delta G_{\mathrm{solv}}$ for revPBE-D3/zd and revPBE0-D3/zd shows significant deviations from experiment (Tab.~\ref{tab:Mg2_properties}, Fig. ~\ref{fig:Mg2_properties}), whereas the classical force fields reproduce the experimental values by construction, as solvation free energies were included among their parameterization targets~\cite{Jiao2006,Soniat2015,Mamatkulov2018,Grotz2021}.
Specifically, the $\Delta G_{\mathrm{solv}}$ is significantly smaller in magnitude, around $-943$ to $-984$ kJ/mol for revPBE-D3/zd and revPBE0-D3/zd, than the experimental value of about $-2532$ kJ/mol~\cite{Marcus1997}.
The deviations from experiment may arise from several factors, including limitations of the underlying DFT reference, the absence of explicit long-range electrostatic interactions in the current local MACE description, and NNP artifacts associated with the net charge of the simulation box during alchemical transformations. 
In particular, the lack of explicit long-range electrostatics may limit the accuracy with which the long-range solvent response encoded in the DFT reference is reproduced and therefore contribute significantly to the observed deviations in ion hydration thermodynamics.
This interpretation is consistent with recent discussions of long-range NNPs, which emphasize that local cutoff models can struggle with dilute ionic solutions, dielectric response, charge redistribution, and varying total charge states~\cite{Anstine2023,Maruf2025}. 

The large deviation of the NNP-derived Mg$^{2+}$ solvation free energies from experiment indicates a limitation of the present models for describing absolute ion hydration thermodynamics. At the same time, the NNPs accurately reproduce structural and dynamical properties, including hydration structure, diffusion, ion pairing, and water-exchange kinetics. These results suggest that ion hydration free energies constitute a particularly stringent benchmark for current NNPs. An important next step will be to assess whether recently developed long-range NNP architectures, such as MACE with charge equilibration and global charge states, can resolve this discrepancy while preserving the accurate local structure and dynamics obtained with the present models \cite{Anstine2023,Maruf2025,ONeill2024,Ko2021,Vondrk2026}.

\section{Conclusion}

Machine-learned interatomic potentials provide a promising route toward accurate simulations of ions in aqueous solution. 
For classical force field simulations, Mg$^{2+}$ has remained a particularly challenging case, as force fields have failed in the past to simultaneously reproduce solvation free energy, size of the first hydration-shell, water-exchange rate, and exchange mechanisms, likely reflecting the absence of an explicit treatment of quantum many-body effects.

The aim of this work was therefore 
to train MACE NNPs  on revPBE-D3/zd and revPBE0-D3/zd reference data and to systematically evaluate the ability of DFT-trained NNPs to reproduce key structural, dynamical and thermodynamic experimental properties of aqueous MgCl$_2$ solutions. For this purpose, NNPs provide a unique opportunity to calculate experimentally accessible macroscopic solution properties which are inaccessible for computationally demanding DFT calculations. Moreover, the direct comparison with experimental data enables critical assessment of the underlying DFT reference and current limitations of NNP descriptions. 

Both NNPs perform well with respect to the structure of the first hydration shell, the self-diffusion coefficient, and the activity derivative. The models 
reproduce the octahedral hydration shell structure and experimental Mg$^{2+}$-water oxygen distance, and revPBE-D3/zd yields a diffusion coefficient close to experiment. The slower diffusion obtained with revPBE0-D3/zd demonstrates that properties predicted by NNPs remain sensitive to the choice of reference density functional.

Transition path sampling and transition interface sampling provided unbiased mechanistic insight and approximation-free rate calculations for  water exchange in the first hydration shell of Mg$^{2+}$. This represents a remarkable achievement, as the combination of path-sampling techniques with NNPs extends DFT-level simulations to rare events occurring on the microsecond timescale, which remain inaccessible to conventional AIMD.

Both NNPs predict a dissociative water-exchange mechanism via a five-coordinated intermediate, in agreement with experimental findings \cite{Helm1999}. Moreover, the calculated exchange rates are within one order of magnitude of experiment, representing a substantial improvement over classical force fields, for which the rate deviates by four orders in magnitude when reproducing the correct dissociative exchange mechanism \cite{Schwierz2020}.

Overall, the NNPs accurately reproduce a range of structural and dynamical properties of aqueous Mg$^{2+}$.
However, they significantly underestimate the experimental solvation free energy, revealing a clear limitation of the present local NNP for describing ion solvation thermodynamics, where long-range electrostatics are essential and net charged systems must be considered. The results hence support the use of DFT-trained NNPs for Mg$^{2+}$ hydration structure, transport, and exchange kinetics, while demonstrating that accurate local structure and dynamics alone are insufficient to guarantee accurate hydration thermodynamics. Future work should therefore incorporate explicit long-range electrostatic treatment and charge-equilibration schemes and further assess the dependence of the results on the underlying electronic-structure reference.

Our results demonstrate that rigorous validation against experimental thermodynamic, structural, and kinetic solution properties is essential for assessing the predictive power of ion NNPs and establish a benchmark for the development of next-generation machine-learned potentials for aqueous electrolytes.

\section*{Data and Software Availability}

MACE models, simulation scripts, and TPS/TIS code are publicly available at \url{https://git.rz.uni-augsburg.de/cbio-gitpub/Mg2-MACE}.

\section*{Acknowledgements}
The work was supported by the research support program (Forschungspotenziale besser nutzen!) of the University of Augsburg.
The authors gratefully acknowledge the scientific support and HPC resources provided by the Erlangen National High Performance Computing Center (NHR@FAU) of the Friedrich-Alexander-Universit\"at Erlangen-N\"urnberg (FAU) under the NHR project b119ee and b253ee and the HPC resources provided on the LiCCA HPC cluster of the University of Augsburg, co-funded by the Deutsche Forschungsgemeinschaft under Project-ID 499211671. Support of the Austrian Science Fund (FWF) [10.55776/COE5] (Cluster of Excellence MECS) is gratefully acknowledged.

\newpage

\onecolumngrid
\normalsize
\patchcmd{\large}{15}{15}{}{}
\begin{center}
  \textbf{\LARGE Supplementary Information: Can DFT-trained neural network potentials reproduce structure, solvation, and water-exchange properties in aqueous magnesium solutions?}\\[.2cm]
  Sebastian Falkner$^{1,2}$, Pablo Montero de Hijes$^{2}$, Christoph Dellago$^{2,3}$, and Nadine Schwierz$^{1,*}$\\[.1cm]
  {\itshape ${}^1$Institute of Physics, University of Augsburg, Universit\"atsstraße 1, 86159 Augsburg, Germany.\\
  \itshape ${}^2$Faculty of Physics, University of Vienna, 1090 Vienna, Austria.\\
  \itshape ${}^3$Research Platform on Accelerating Photoreaction Discovery (ViRAPID), University of Vienna, 1090 Vienna, Austria.\\
  }
  ${}^*$Electronic address: nadine.schwierz@uni-a.de\\
(Dated: \today)\\[2cm]
\end{center}

\setcounter{equation}{0}
\setcounter{figure}{0}
\setcounter{table}{0}
\setcounter{page}{1}
\setcounter{section}{0}
\renewcommand{\theequation}{S\arabic{equation}}
\renewcommand{\thefigure}{S\arabic{figure}}
\renewcommand{\thetable}{S\arabic{table}}
\renewcommand{\bibnumfmt}[1]{[S#1]}
\renewcommand{\citenumfont}[1]{S#1}
\renewcommand{\thesection}{S\Roman{section}}
\renewcommand{\thepage}{S\arabic{page}}

\titleformat*{\section}{\Large\bfseries}

\section{Methods}

\subsection{Molecular Dynamics Simulations}

\subsubsection{Force Field Simulations}
Classical molecular dynamics simulations for the production of initial ab-initio simulation configurations were performed using the GROMACS package~\cite{SAbraham2015} with the Mamatkulov-Schwierz magnesium force field~\cite{SMamatkulov2018} and the flexible TIP3P water model~\cite{SJorgensen1983}. A flexible water model was chosen to better reproduce the behavior of water in ab initio simulations. Energy minimization was achieved via the steepest descent algorithm for a maximum of $50000$ steps, with convergence criteria set at \SI{100}{\kilo\joule\per\nano\meter}. The simulations utilized the Verlet cutoff scheme with non-bonded interaction cutoffs of \SI{1.0}{\nano\meter}, although if the simulation box was smaller, the cutoff was set to half the box length. Long-range electrostatic interactions were treated using the particle-mesh Ewald (PME)~\cite{SDarden1993} method with a Fourier spacing of \SI{0.12}{\nano\meter}.

A \SI{1}{\nano\second} NVT equilibration was performed at \SI{298.15}{\kelvin}, followed by a \SI{1}{\nano\second} NPT equilibration at the same temperature and \SI{1}{\bar} pressure with a timestep of \SI{1}{\femto\second}. Temperature coupling during both stages utilized the stochastic velocity-rescaling thermostat~\cite{SBussi2007} with a coupling time constant of $\tau_\mathrm{T} = \SI{1.0}{\pico\second}$. For the NPT equilibration, the Berendsen barostat~\cite{SBerendsen1984} was used for pressure coupling ($\tau_\mathrm{P} = \SI{5.0}{\pico\second}$).

Production runs were executed in the NPT ensemble for \SI{5}{\nano\second}. During these production trajectories, temperature was maintained at \SI{298.15}{\kelvin} using velocity rescaling ($\tau_\mathrm{T} = \SI{1.0}{\pico\second}$), while pressure was controlled via the C-rescale barostat~\cite{SBernetti2020} with a coupling time of $\tau_\mathrm{P} = \SI{5.0}{\pico\second}$.

\subsubsection{Ab-Initio MD}
All ab-initio simulations were performed using CP2K version 2025~\cite{SHutter2014}. The GAPW method~\cite{SLippert1999} was employed with a plane-wave cutoff of \SI{700}{Ry}, five grid levels, and a relative cutoff of \SI{50}{Ry}. Exchange-correlation effects were treated using the revPBE functional~\cite{SPerdew1996,SZhang1998} supplemented by Grimme's DFT-D3 dispersion correction with long-range corrections and zero damping~\cite{SGrimme2010,SGrimme2011}. TZV2P-GTH basis sets combined with GTH-PBE pseudopotentials were used for all elements (q10 for \Mg, q1 for H, q7 for \Cl, and q6 for O)~\cite{SGoedecker1996,SHartwigsen1998}. SCF convergence was achieved via the orbital transformation method with a DIIS minimizer, targeting an energy threshold of $5.0 \times 10^{-7}$ Hartree.

Molecular dynamics were initiated from force-field equilibrated structures following a brief BFGS geometry optimization (five iterations). Simulations proceeded in the NVT ensemble at \SI{298.15}{\kelvin} with a timestep of \SI{1}{\femto\second}. After an initial \SI{100}{steps} minimization phase, equilibration utilized a strongly coupled CSVR thermostat~\cite{SBussi2007} ($\tau = \SI{5}{\femto\second}$), followed by production runs with weaker coupling ($\tau = \SI{100}{\femto\second}$) lasting \SI{1250}{steps}. System charge was adjusted according to the simulated ion species.

\subsection{Neural Network Potential Training}
For Neural Network Potential (NNP) training, the $\mathrm{w}256$ and $\mathrm{w}256\mathrm{MgCl}_2$ datasets were split into training, validation, and test sets for both revPBE and revPBE0 potentials. The training set contained $2500$ configurations, the validation set contained $125$ configurations ($5\%$ of the training data), and the test set contained $250$ configurations. Models were trained separately on a single NVIDIA A100 GPU for $250$ epochs with a batch size of $6$ and a learning rate of $0.01$. Energy and force weights were set to $\lambda_E = 1$ and $\lambda_F = 1000$, and Stochastic Weight Averaging along with Exponential Moving Average of the weights were enabled.

\subsubsection{Iterative Training}
\begin{figure}[H]
    \centering
    \includegraphics[width=0.8\linewidth]{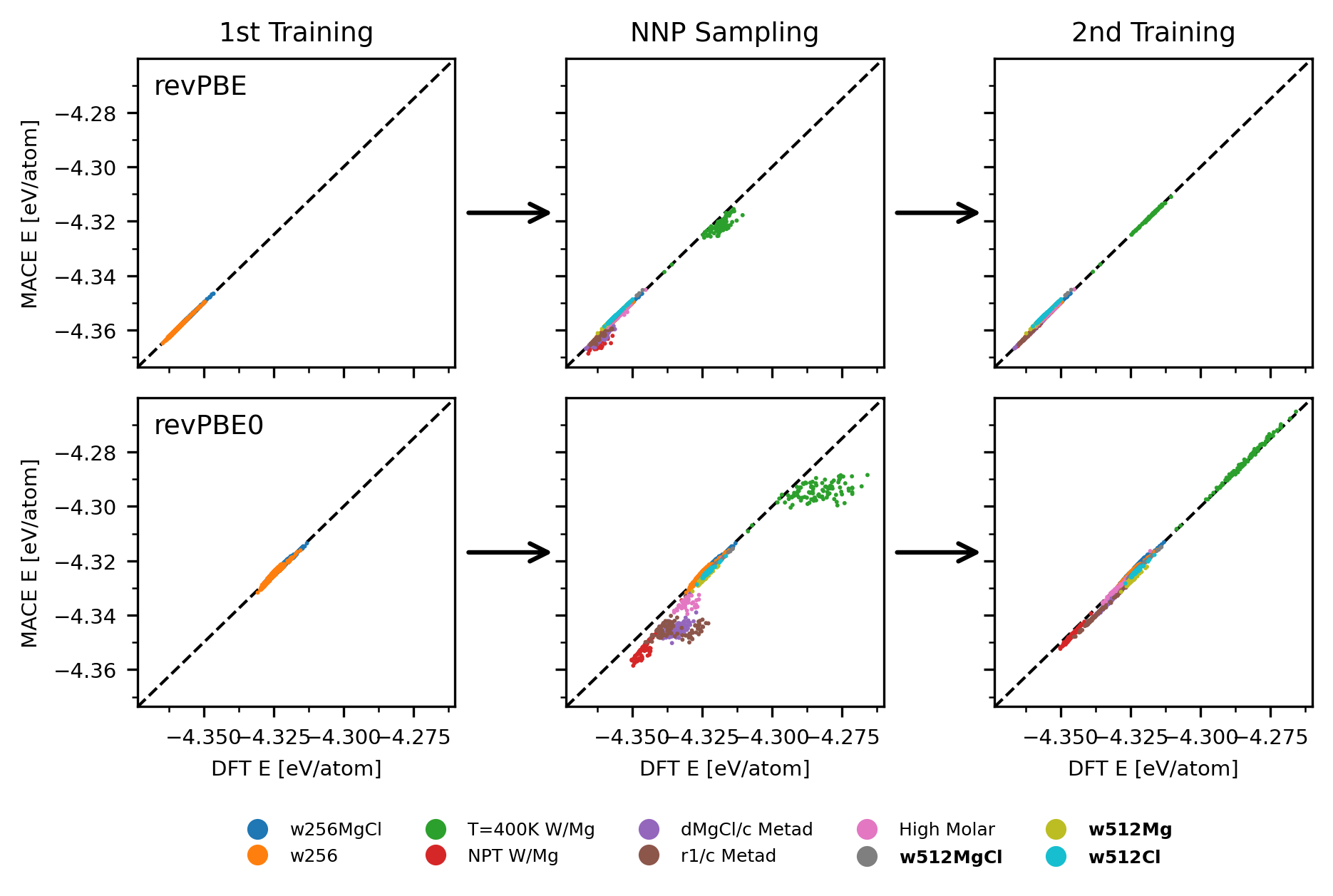}
    \includegraphics[width=0.8\linewidth]{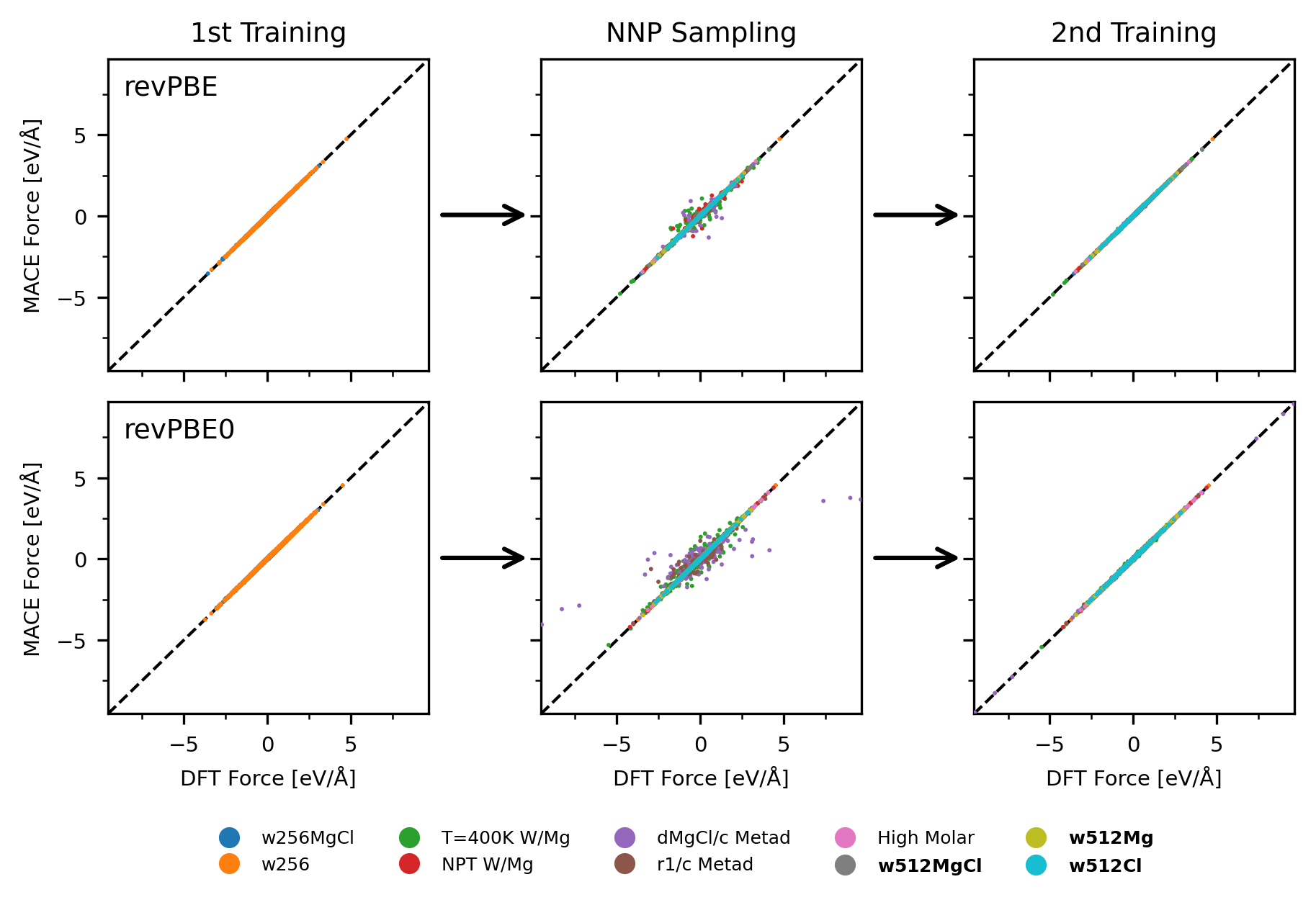}
    \caption{Comparison of DFT-predicted energies per atom and force components against MACE neural network potential predictions for the revPBE and revPBE0 functionals across successive training stages. Bold labels indicate separate test data not used during training.}
    \label{fig:Mg2_Energies}
\end{figure}

\begin{figure}[H]
    \centering
    \includegraphics[width=0.5\linewidth]{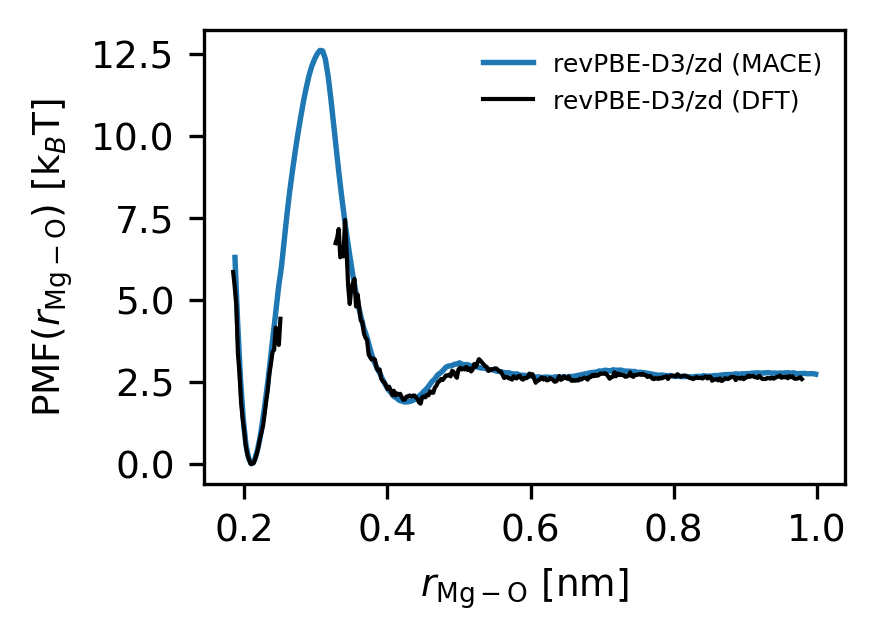}
    \caption{Comparison between the $r_\mathrm{Mg-O}$ PMF obtained from ab-initio MD and the NNP-derived PMF from umbrella sampling. Note that the ab-initio PMF was obtained from the radial distribution function of equilibrium runs via $-\ln g(r_\mathrm{Mg-O})$, hence there is a gap in unexplored regions.}
    \label{fig:SI_Mg2_PMF}
\end{figure}

The following simulations were performed using the initial revPBE and revPBE0 neural network potentials to generate new configurations:
\begin{itemize}
\item NPT: \SI{1}{\nano\second} at \SI{298.15}{\kelvin} and \SI{1.0}{\bar} for $\mathrm{w}256$ and $\mathrm{w}256\mathrm{MgCl}_2$ with stronger barostat coupling (every 25 steps).
\item High temperature: \SI{0.125}{\nano\second} NVT at \SI{500}{\kelvin} for $\mathrm{w}256$ and $\mathrm{w}256\mathrm{MgCl}_2$ with a \SI{0.5}{\femto\second} timestep.
\item High concentration: \SI{1}{\nano\second} NVT for a \SI{1}{M} $\text{MgCl}_2$ solution.
\item Metadynamics ($r_\mathrm{Mg-O}$ + $n_1$): \SI{2.5}{\nano\second} NVT biased along the Mg–O distance and coordination number (bias factor 10, hill height \SI{1.0}{\kilo\joule\per\mol}, stride 100 steps, adaptive width $\tau = \SI{500}{\femto\second}$, $\sigma_\mathrm{min}=\SI{0.02}{\nano\meter\squared}$ and 0.05).
\item Metadynamics (Mg-Cl Distance): \SI{2.5}{\nano\second} NVT biased along the Mg–Cl distance and coordination number, using identical metadynamics parameters.
\end{itemize}
From each run, 50 evenly spaced configurations (100 for metadynamics) were extracted, their energies and forces computed via single-point DFT calculations, and added to the initial dataset. This augmented dataset was then used to perform a second round of model training using the same protocol as in the first round.  Figure~\ref{fig:Mg2_Energies} shows the network's accuracy in predicting energies and forces after the first training cycle, before the second cycle, and after the second cycle. The middle column illustrates clearly the initial lack of performance in unknown regions of configuration space explored via enhanced sampling, while the last column shows that accurate predictions can be recovered in a second training round. As a further validation, Figure~\ref{fig:SI_Mg2_PMF} shows good agreement between the Potential of Mean Force (PMF) obtained via umbrella sampling with the NNP and the partial PMF calculated directly from distance histograms of the DFT simulation.

\subsection{Ion Properties}

\subsubsection{Free Energy Calculations}
Detailed numerical parameters for the solvation free energy calculations are provided here. For both alchemical approaches, equilibrium simulations were performed at discrete $\lambda$-windows using the NNP simulation parameters defined in the main text, such as temperature and pressure coupling. Each $\lambda$-window was simulated for \SI{50}{\pico\second} with a timestep of \SI{1}{\femto\second}. The following simulation protocols were repeated for w512Mg and w512Cl, based on which the solvation free energy of the Mg$^{2+}$-Cl$^{-}$ ion pair was estimated.

In the thermodynamic cycle method~\cite{SKarwounopoulos2024}, the correction term $\Delta G_{\mathrm{MM} \to \mathrm{NNP}}$ was obtained via thermodynamic integration. Sampling was conducted across $6$ equidistant windows with $\lambda \in \{0.0, 0.2, 0.4, 0.6, 0.8, 1.0\}$, where $\lambda=0$ represents the pure force field and $\lambda=1$ the pure NNP potential. The force field side of the thermodynamic cycle ($\Delta G_{\mathrm{MM}}$) was set up in the same way as the initial force field simulation setup described in section S1.1 employing Mamatkulov-Schwierz Mg$^{2+}$ with flexible TIP3P water. The term $\Delta G_{\mathrm{MM}}$ was recalculated using the protocol described in Mamatkulov et al.~\cite{SMamatkulov2018}, but did not show significant differences to the value reported for rigid TIP3P water. Correction terms as described in Grotz et al.~\cite{SGrotz2021} were applied to $\Delta G_{\mathrm{MM}}$. Therefore, no further corrections were applied to $\Delta G_{\mathrm{MM} \to \mathrm{NNP}}$.

For the neighborlist decoupling method~\cite{SPicha2025}, the solute-solvent interactions were scaled by modifying the interatomic distances provided to the NNP using a \texttt{linear\_to\_cutoff} shifting scheme. For this approach $21$ $\lambda$-windows were utilized, distributed as $\lambda \in \{0.0, 0.025, \dots, 0.25, 0.3, 0.35, 0.4, 0.5, \dots, 0.9, 1.0\}$, where $\lambda=1$ represents the decoupled state. Free energy differences were subsequently estimated using the BAR method~\cite{SBennett1976}. Corrections were applied as described in Grotz et al.~\cite{SGrotz2021}.

\subsubsection{Transition Path Sampling with NNP}

Initial reactive pathways were generated using steered molecular dynamics by applying a moving harmonic restraint on the distance between an inner-shell oxygen atom and the magnesium ion. The bias center was shifted linearly from \SI{0.23}{\nano\meter} to \SI{0.38}{\nano\meter} over \SI{100}{\pico\second} with a force constant of \SI{50000}{\kilo\joule\per\mol\nano\meter\squared}. We define an order parameter $\xi = \exp(-10 r_2) - \exp(-10 r_1)$, where $r_1$ denotes the distance between the magnesium ion and the pulled oxygen atom, and $r_2$ represents the distance to the sixth nearest oxygen neighbor excluding the oxygen assigned to $r_1$. Stable states A and B were defined by thresholds $\xi < -0.08$ and $\xi > 0.08$, respectively. Sampling was performed over \num{1250} trials with a maximum trajectory length of \SI{10}{\pico\second}, and shooting points were selected uniformly along the trajectories.

\begin{figure}[H]
    \centering
    \includegraphics[width=0.8\linewidth]{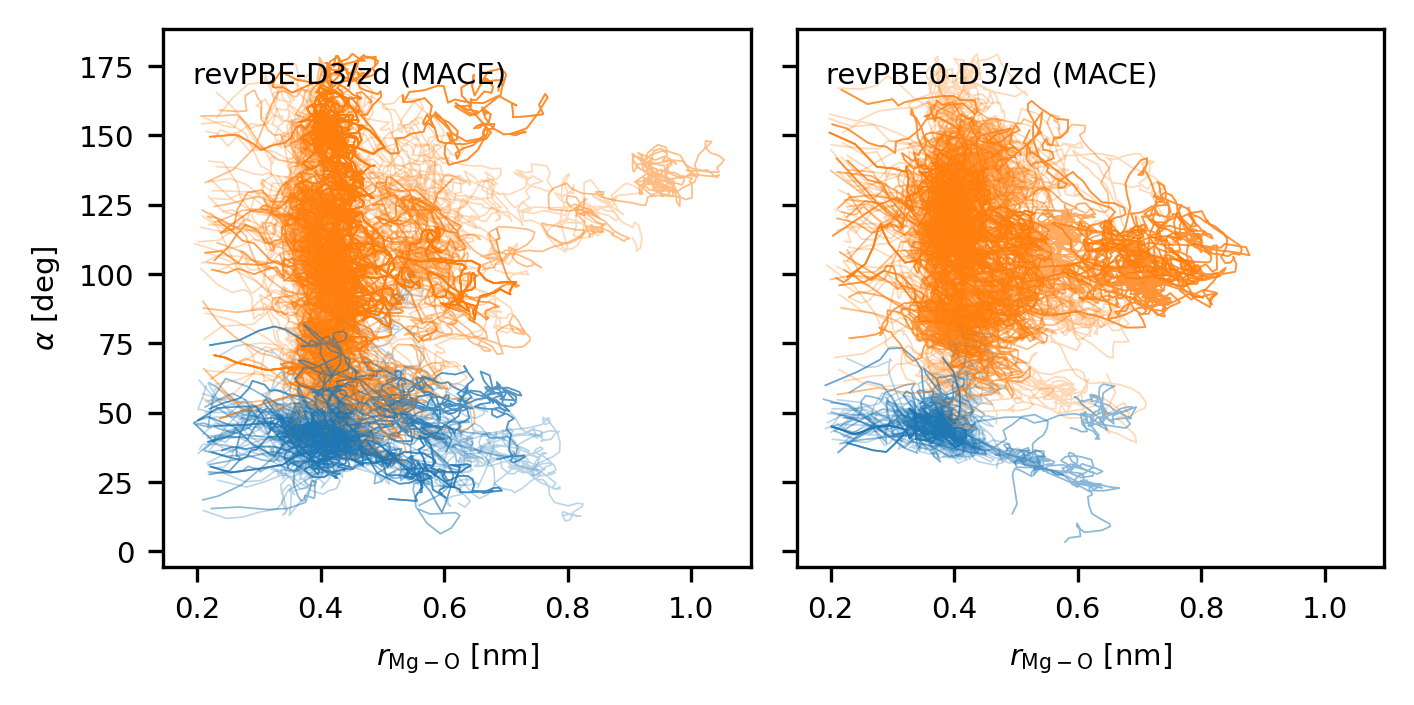}
    \caption{Distance between Mg$^{2+}$ and the oxygen atom of an exchanging water molecule, $r_{\mathrm{Mg-O}}$, plotted against the angle $\alpha$ between the incoming and leaving water molecules during exchange events. Two exchange pathways are observed. In the direct-exchange pathway (blue), the incoming and leaving water molecules are adjacent to each other, giving smaller $\alpha$ values. In the indirect-exchange pathway (orange), the two water molecules are located on opposite sides of Mg$^{2+}$, giving larger $\alpha$ values.}
    \label{fig:Mg2_TPS_directindirect}
\end{figure}

In addition to the associative/dissociative classification discussed in the main text, water exchange can also be categorized by the angle between the incoming and outgoing water molecules (Fig.~\ref{fig:Mg2_TPS_directindirect}). This results in two pathways, a direct mechanism (narrow angle) and an indirect mechanism (wide angle). Both pathways are observed across both DFT functionals, indicating that there is no clear preference for a specific pathway for the NNP-described magnesium ion.

\subsubsection{Transition Interface Sampling with NNP}

The flux through the first interface ($\lambda_0 = -0.08$) was determined from unbiased MD simulations with a duration of \SI{0.05}{\nano\second} per replica across $5$ replicas. Initial reactive pathways were generated using steered molecular dynamics over \SI{0.1}{\nano\second}, applying harmonic restraints to two magnesium–oxygen distances $r_1$ and $r_2$, and a coordination number restraint ($k = \SI{10000}{\kilo\joule\per\mol\nano\meter\squared}$, $n_\mathrm{ref}=5.1$). The bias centers for $r_2$ and $r_1$ were shifted from \SI{0.21}{\nano\meter} to \SI{0.5}{\nano\meter} and from \SI{0.5}{\nano\meter} to \SI{0.21}{\nano\meter}, respectively, over \SI{100}{\pico\second} with a force constant of \SI{15000}{\kilo\joule\per\mol\nano\meter\squared}. 

Path sampling was conducted across 11 interfaces ranging from $\lambda_0 = -0.08$ to $\lambda_{10} = 0.00$ with a spacing of $0.008$. Each interface ensemble was sampled for $1100$ trials with a maximum trajectory length of \SI{10}{\pico\second}. Shooting points were selected using a biased weight function~\cite{SJung2017} $w(\xi)$ defined by the interface position $\lambda_i$, a capping value $\xi_\mathrm{cap} = 0.02$, and a falloff parameter $\alpha = 0.015$:
\begin{align}
    w(\xi) = \begin{cases} 
    0 & \xi \geq \xi_\mathrm{cap} \\
    1 & \lambda_i \leq \xi < \xi_\mathrm{cap} \\
    \exp\left(-\frac{|\xi - \lambda_i|}{\alpha}\right) & \xi < \lambda_i
    \end{cases}
\end{align}
Parallel path swapping was employed between adjacent interface ensembles after every trial to improve efficiency and reduce correlation~\cite{SvanErp2007}.

\subsubsection{Activity Derivatives with NNP}

\begin{figure}[H]
    \centering
    \includegraphics[width=1\linewidth]{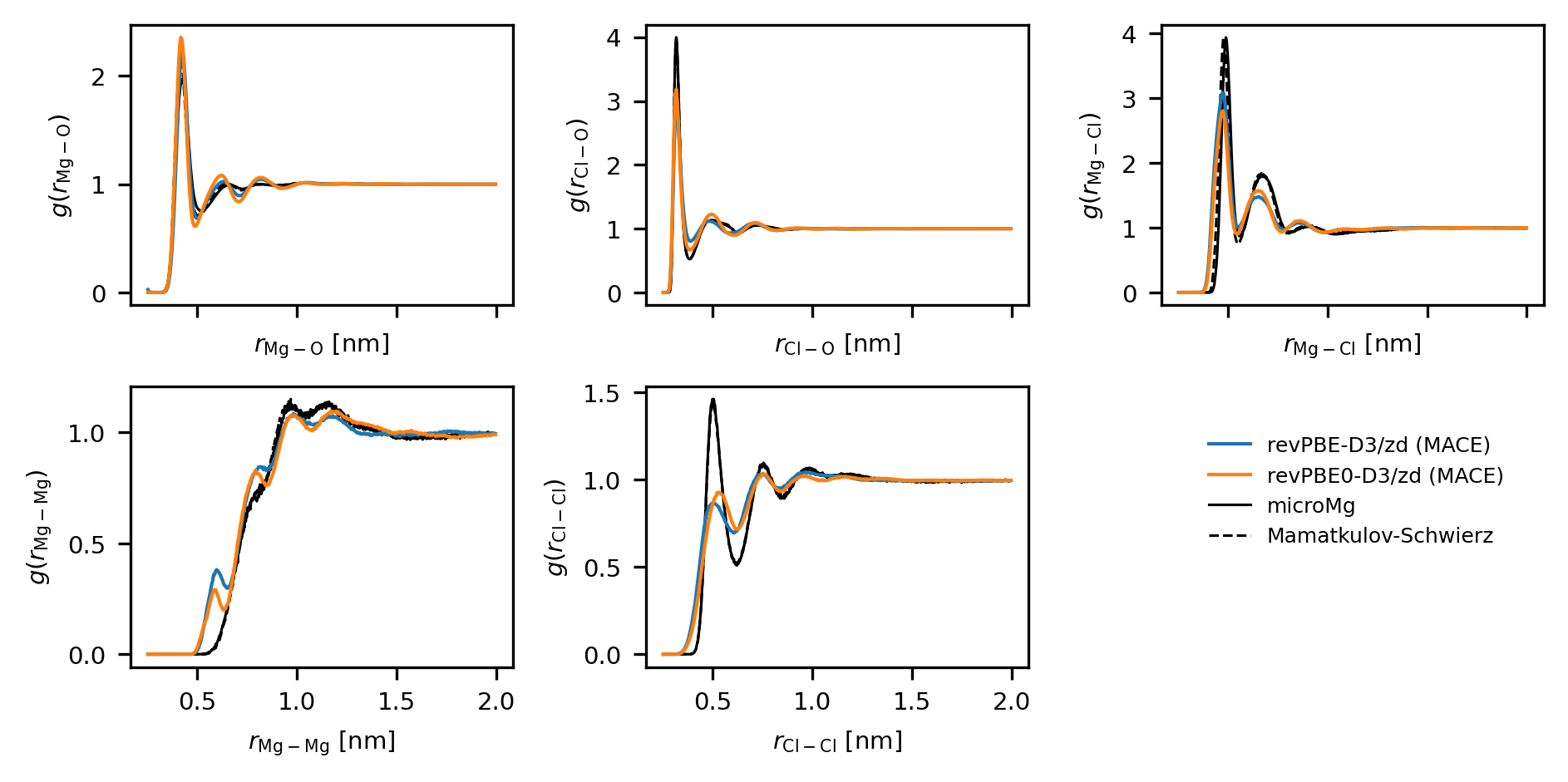}
    \caption{Radial distribution functions from NNP-MD simulations of a \SI{1.08}{m} MgCl$_2$ solution used for the activity derivative calculation.}
    \label{fig:Mg2_RDFS}
\end{figure}

Activity derivatives were calculated from NPT radial distribution functions using Kirkwood–Buff theory~\cite{SKirkwood1951}, as detailed for magnesium in a previous work by Grotz et al.~\cite{SGrotz2021}. All ion-ion and ion-water radial distribution functions involved in this calculation are shown in figure~\ref{fig:Mg2_RDFS}. We follow the protocol described in Grotz et al.~\cite{SGrotz2021}, only the normalization of the radial distribution functions prior to the numerical Kirkwood–Buff integral calculation is adjusted.

 In general, Radial Distribution Functions (RDF) from molecular dynamics simulations contain finite size effects and noise due to finite simulation boxes and time. This can lead to the RDFs not converging to unity at the cutoff distance. Without correct normalization, the Kirkwood–Buff integrals are numerically unstable and very prone to diverge at larger distances. Therefore, the common approach is to scale the simulated RDFs $g^\mathrm{sim}_{ij}$ using an estimated normalization factor $f$
\begin{align}
g_{ij}(r) = f g^\mathrm{sim}_{ij}(r)\, ,
\end{align}
so that $g_{ij}(r)$ is guaranteed to converge to unity at the end of the cutoff. The scaling factor $f$ is usually obtained numerically from an average over the tail of the RDF.

For NNP simulations, it is particularly challenging to obtain low-noise RDFs due to the high computational cost of simulations. Instead of averaging to obtain $f$ to improve convergence, we screened a grid of possible normalization values between $0.95$ and $1.05$ and picked the normalization value that minimizes the variance of the calculated Kirkwood–Buff integral after a cutoff of \SI{1.62}{\nano\meter}. While this has close to no effect on force field-derived RDFs from long MD simulations, it leads to a much more stable normalization and hence convergence of the numerical integrals for noisy RDFs.

\subsubsection{Potential of Mean Force with NNP}
Umbrella sampling simulations were performed along the magnesium–oxygen distance to compute the PMF~\cite{STorrie1977}. For the revPBE NNP, seven windows were employed with restraint centers positioned between \SI{0.285}{\nano\meter} and \SI{0.335}{\nano\meter}, spaced at intervals of \SI{0.005}{\nano\meter}. For revPBE0, windows were spaced by \SI{0.005}{\nano\meter} between \SI{0.275}{\nano\meter} and \SI{0.325}{\nano\meter}. Each window was biased using a harmonic restraint with a force constant of \SI{50000}{\kilo\joule\per\mol\nano\meter\squared}. The umbrella sampling simulations were run for \SI{2}{\nano\second} per window in the NPT ensemble. The weighted histogram analysis method (WHAM)~\cite{SKumar1992} was used to reconstruct the free energy profile from the combined equilibrium and umbrella sampling trajectory data. In most umbrella sampling analysis tools, this can be achieved by supplying the equilibrium data as a separate umbrella window with a zero force constant.

\subsubsection{Metadynamics with NNP}
Well-tempered metadynamics~\cite{SLaio2002,SBarducci2008} calculations were performed with a total duration of \SI{7.5}{\nano\second}. To constrain the sampling to the magnesium–oxygen distance of interest, a harmonic upper wall was applied to the $r_\mathrm{Mg-O}$ coordinate at \SI{0.7}{\nano\meter} with a force constant of \SI{25000}{\kilo\joule\per\mol\nano\meter\squared}. The metadynamics bias was added by depositing Gaussian hills of initial height \SI{0.8}{\kilo\joule\per\mol} every \num{100} steps using a bias factor of $10$. Fixed Gaussian widths of \SI{0.015}{\nano\meter} were used for the distance $r_\mathrm{Mg-O}$, and $0.05$ for the coordination number. The first-shell coordination number $n_1$ is evaluated using a rational switching function characterizing magnesium–oxygen contacts within a radial cutoff of \SI{0.3}{\nano\meter}, utilizing switching exponents of $n=12$ and $m=24$, with all distance calculations accelerated via a neighbor list ($N_{\mathrm{LIST}}$) featuring a cutoff of \SI{0.6}{\nano\meter} updated every 25 simulation steps.


\begin{thebibliography}{86}%
\makeatletter
\providecommand \@ifxundefined [1]{%
 \@ifx{#1\undefined}
}%
\providecommand \@ifnum [1]{%
 \ifnum #1\expandafter \@firstoftwo
 \else \expandafter \@secondoftwo
 \fi
}%
\providecommand \@ifx [1]{%
 \ifx #1\expandafter \@firstoftwo
 \else \expandafter \@secondoftwo
 \fi
}%
\providecommand \natexlab [1]{#1}%
\providecommand \enquote  [1]{``#1''}%
\providecommand \bibnamefont  [1]{#1}%
\providecommand \bibfnamefont [1]{#1}%
\providecommand \citenamefont [1]{#1}%
\providecommand \href@noop [0]{\@secondoftwo}%
\providecommand \href [0]{\begingroup \@sanitize@url \@href}%
\providecommand \@href[1]{\@@startlink{#1}\@@href}%
\providecommand \@@href[1]{\endgroup#1\@@endlink}%
\providecommand \@sanitize@url [0]{\catcode `\\12\catcode `\$12\catcode
  `\&12\catcode `\#12\catcode `\^12\catcode `\_12\catcode `\%12\relax}%
\providecommand \@@startlink[1]{}%
\providecommand \@@endlink[0]{}%
\providecommand \url  [0]{\begingroup\@sanitize@url \@url }%
\providecommand \@url [1]{\endgroup\@href {#1}{\urlprefix }}%
\providecommand \urlprefix  [0]{URL }%
\providecommand \Eprint [0]{\href }%
\providecommand \doibase [0]{https://doi.org/}%
\providecommand \selectlanguage [0]{\@gobble}%
\providecommand \bibinfo  [0]{\@secondoftwo}%
\providecommand \bibfield  [0]{\@secondoftwo}%
\providecommand \translation [1]{[#1]}%
\providecommand \BibitemOpen [0]{}%
\providecommand \bibitemStop [0]{}%
\providecommand \bibitemNoStop [0]{.\EOS\space}%
\providecommand \EOS [0]{\spacefactor3000\relax}%
\providecommand \BibitemShut  [1]{\csname bibitem#1\endcsname}%
\let\auto@bib@innerbib\@empty
\bibitem [{\citenamefont {Misra}\ and\ \citenamefont
  {Draper}(1998)}]{Misra1998}%
  \BibitemOpen
  \bibfield  {author} {\bibinfo {author} {\bibfnamefont {V.~K.}\ \bibnamefont
  {Misra}}\ and\ \bibinfo {author} {\bibfnamefont {D.~E.}\ \bibnamefont
  {Draper}},\ }\bibfield  {title} {\bibinfo {title} {{On the role of magnesium
  ions in RNA stability}},\ }\href
  {https://doi.org/10.1002/(SICI)1097-0282(1998)48:2<113::AID-BIP3>3.0.CO;2-Y}
  {\bibfield  {journal} {\bibinfo  {journal} {Biopolymers}\ }\textbf {\bibinfo
  {volume} {48}},\ \bibinfo {pages} {113} (\bibinfo {year} {1998})}\BibitemShut
  {NoStop}%
\bibitem [{\citenamefont {Williams}(2000)}]{Williams2000}%
  \BibitemOpen
  \bibfield  {author} {\bibinfo {author} {\bibfnamefont {N.~H.}\ \bibnamefont
  {Williams}},\ }\bibfield  {title} {\bibinfo {title} {{Magnesium Ion Catalyzed
  ATP Hydrolysis}},\ }\href {https://doi.org/10.1021/ja0013374} {\bibfield
  {journal} {\bibinfo  {journal} {Journal of the American Chemical Society}\
  }\textbf {\bibinfo {volume} {122}},\ \bibinfo {pages} {12023} (\bibinfo
  {year} {2000})}\BibitemShut {NoStop}%
\bibitem [{\citenamefont {Cowan}(2002)}]{Cowan2002}%
  \BibitemOpen
  \bibfield  {author} {\bibinfo {author} {\bibfnamefont {J.}~\bibnamefont
  {Cowan}},\ }\bibfield  {title} {\bibinfo {title} {{Structural and catalytic
  chemistry of magnesium-dependent enzymes}},\ }\href
  {https://doi.org/10.1023/A:1016022730880} {\bibfield  {journal} {\bibinfo
  {journal} {Biometals}\ }\textbf {\bibinfo {volume} {15}},\ \bibinfo {pages}
  {225} (\bibinfo {year} {2002})}\BibitemShut {NoStop}%
\bibitem [{\citenamefont {Pyle}(2002)}]{Pyle2002}%
  \BibitemOpen
  \bibfield  {author} {\bibinfo {author} {\bibfnamefont {A.}~\bibnamefont
  {Pyle}},\ }\bibfield  {title} {\bibinfo {title} {{Metal ions in the structure
  and function of RNA}},\ }\href {https://doi.org/10.1007/s00775-002-0387-6}
  {\bibfield  {journal} {\bibinfo  {journal} {JBIC Journal of Biological
  Inorganic Chemistry}\ }\textbf {\bibinfo {volume} {7}},\ \bibinfo {pages}
  {679} (\bibinfo {year} {2002})}\BibitemShut {NoStop}%
\bibitem [{\citenamefont {Born}\ \emph {et~al.}(2009)\citenamefont {Born},
  \citenamefont {Weing\"{a}rtner}, \citenamefont {Br\"{u}ndermann},\ and\
  \citenamefont {Havenith}}]{Born2009}%
  \BibitemOpen
  \bibfield  {author} {\bibinfo {author} {\bibfnamefont {B.}~\bibnamefont
  {Born}}, \bibinfo {author} {\bibfnamefont {H.}~\bibnamefont
  {Weing\"{a}rtner}}, \bibinfo {author} {\bibfnamefont {E.}~\bibnamefont
  {Br\"{u}ndermann}},\ and\ \bibinfo {author} {\bibfnamefont {M.}~\bibnamefont
  {Havenith}},\ }\bibfield  {title} {\bibinfo {title} {{Solvation Dynamics of
  Model Peptides Probed by Terahertz Spectroscopy. Observation of the Onset of
  Collective Network Motions}},\ }\href {https://doi.org/10.1021/ja808997y}
  {\bibfield  {journal} {\bibinfo  {journal} {Journal of the American Chemical
  Society}\ }\textbf {\bibinfo {volume} {131}},\ \bibinfo {pages} {3752}
  (\bibinfo {year} {2009})}\BibitemShut {NoStop}%
\bibitem [{\citenamefont {Stachura}\ \emph {et~al.}(2017)\citenamefont
  {Stachura}, \citenamefont {Chakraborty}, \citenamefont {Gottberg},
  \citenamefont {Ruckthong}, \citenamefont {Pecoraro},\ and\ \citenamefont
  {Hemmingsen}}]{Stachura2017}%
  \BibitemOpen
  \bibfield  {author} {\bibinfo {author} {\bibfnamefont {M.}~\bibnamefont
  {Stachura}}, \bibinfo {author} {\bibfnamefont {S.}~\bibnamefont
  {Chakraborty}}, \bibinfo {author} {\bibfnamefont {A.}~\bibnamefont
  {Gottberg}}, \bibinfo {author} {\bibfnamefont {L.}~\bibnamefont {Ruckthong}},
  \bibinfo {author} {\bibfnamefont {V.~L.}\ \bibnamefont {Pecoraro}},\ and\
  \bibinfo {author} {\bibfnamefont {L.}~\bibnamefont {Hemmingsen}},\ }\bibfield
   {title} {\bibinfo {title} {{Direct Observation of Nanosecond Water Exchange
  Dynamics at a Protein Metal Site}},\ }\href
  {https://doi.org/10.1021/jacs.6b11525} {\bibfield  {journal} {\bibinfo
  {journal} {Journal of the American Chemical Society}\ }\textbf {\bibinfo
  {volume} {139}},\ \bibinfo {pages} {79} (\bibinfo {year} {2017})}\BibitemShut
  {NoStop}%
\bibitem [{\citenamefont {Alln\'{e}r}\ \emph {et~al.}(2012)\citenamefont
  {Alln\'{e}r}, \citenamefont {Nilsson},\ and\ \citenamefont
  {Villa}}]{Allnr2012}%
  \BibitemOpen
  \bibfield  {author} {\bibinfo {author} {\bibfnamefont {O.}~\bibnamefont
  {Alln\'{e}r}}, \bibinfo {author} {\bibfnamefont {L.}~\bibnamefont
  {Nilsson}},\ and\ \bibinfo {author} {\bibfnamefont {A.}~\bibnamefont
  {Villa}},\ }\bibfield  {title} {\bibinfo {title} {{Magnesium Ion–Water
  Coordination and Exchange in Biomolecular Simulations}},\ }\href
  {https://doi.org/10.1021/ct3000734} {\bibfield  {journal} {\bibinfo
  {journal} {Journal of Chemical Theory and Computation}\ }\textbf {\bibinfo
  {volume} {8}},\ \bibinfo {pages} {1493} (\bibinfo {year} {2012})}\BibitemShut
  {NoStop}%
\bibitem [{\citenamefont {Li}\ and\ \citenamefont {Merz}(2014)}]{Li2014}%
  \BibitemOpen
  \bibfield  {author} {\bibinfo {author} {\bibfnamefont {P.}~\bibnamefont
  {Li}}\ and\ \bibinfo {author} {\bibfnamefont {K.~M.}\ \bibnamefont {Merz}},\
  }\bibfield  {title} {\bibinfo {title} {{Taking into Account the Ion-Induced
  Dipole Interaction in the Nonbonded Model of Ions}},\ }\href
  {https://doi.org/10.1021/ct400751u} {\bibfield  {journal} {\bibinfo
  {journal} {Journal of Chemical Theory and Computation}\ }\textbf {\bibinfo
  {volume} {10}},\ \bibinfo {pages} {289} (\bibinfo {year} {2014})}\BibitemShut
  {NoStop}%
\bibitem [{\citenamefont {Dubou\'{e}-Dijon}\ \emph {et~al.}(2018)\citenamefont
  {Dubou\'{e}-Dijon}, \citenamefont {Mason}, \citenamefont {Fischer},\ and\
  \citenamefont {Jungwirth}}]{Dubou-Dijon2018}%
  \BibitemOpen
  \bibfield  {author} {\bibinfo {author} {\bibfnamefont {E.}~\bibnamefont
  {Dubou\'{e}-Dijon}}, \bibinfo {author} {\bibfnamefont {P.~E.}\ \bibnamefont
  {Mason}}, \bibinfo {author} {\bibfnamefont {H.~E.}\ \bibnamefont {Fischer}},\
  and\ \bibinfo {author} {\bibfnamefont {P.}~\bibnamefont {Jungwirth}},\
  }\bibfield  {title} {\bibinfo {title} {{Hydration and Ion Pairing in Aqueous
  Mg 2+ and Zn 2+ Solutions: Force-Field Description Aided by Neutron
  Scattering Experiments and Ab Initio Molecular Dynamics Simulations}},\
  }\href {https://doi.org/10.1021/acs.jpcb.7b09612} {\bibfield  {journal}
  {\bibinfo  {journal} {The Journal of Physical Chemistry B}\ }\textbf
  {\bibinfo {volume} {122}},\ \bibinfo {pages} {3296} (\bibinfo {year}
  {2018})}\BibitemShut {NoStop}%
\bibitem [{\citenamefont {Mamatkulov}\ and\ \citenamefont
  {Schwierz}(2018)}]{Mamatkulov2018}%
  \BibitemOpen
  \bibfield  {author} {\bibinfo {author} {\bibfnamefont {S.}~\bibnamefont
  {Mamatkulov}}\ and\ \bibinfo {author} {\bibfnamefont {N.}~\bibnamefont
  {Schwierz}},\ }\bibfield  {title} {\bibinfo {title} {{Force fields for
  monovalent and divalent metal cations in TIP3P water based on thermodynamic
  and kinetic properties}},\ }\href {https://doi.org/10.1063/1.5017694}
  {\bibfield  {journal} {\bibinfo  {journal} {The Journal of Chemical Physics}\
  }\textbf {\bibinfo {volume} {148}},\ \bibinfo {pages} {074504} (\bibinfo
  {year} {2018})}\BibitemShut {NoStop}%
\bibitem [{\citenamefont {Grotz}\ \emph {et~al.}(2021)\citenamefont {Grotz},
  \citenamefont {Cruz-Le\'{o}n},\ and\ \citenamefont {Schwierz}}]{Grotz2021}%
  \BibitemOpen
  \bibfield  {author} {\bibinfo {author} {\bibfnamefont {K.~K.}\ \bibnamefont
  {Grotz}}, \bibinfo {author} {\bibfnamefont {S.}~\bibnamefont
  {Cruz-Le\'{o}n}},\ and\ \bibinfo {author} {\bibfnamefont {N.}~\bibnamefont
  {Schwierz}},\ }\bibfield  {title} {\bibinfo {title} {{Optimized Magnesium
  Force Field Parameters for Biomolecular Simulations with Accurate Solvation,
  Ion-Binding, and Water-Exchange Properties}},\ }\href
  {https://doi.org/10.1021/acs.jctc.0c01281} {\bibfield  {journal} {\bibinfo
  {journal} {Journal of Chemical Theory and Computation}\ }\textbf {\bibinfo
  {volume} {17}},\ \bibinfo {pages} {2530} (\bibinfo {year}
  {2021})}\BibitemShut {NoStop}%
\bibitem [{\citenamefont {Soniat}\ \emph {et~al.}(2015)\citenamefont {Soniat},
  \citenamefont {Hartman},\ and\ \citenamefont {Rick}}]{Soniat2015}%
  \BibitemOpen
  \bibfield  {author} {\bibinfo {author} {\bibfnamefont {M.}~\bibnamefont
  {Soniat}}, \bibinfo {author} {\bibfnamefont {L.}~\bibnamefont {Hartman}},\
  and\ \bibinfo {author} {\bibfnamefont {S.~W.}\ \bibnamefont {Rick}},\
  }\bibfield  {title} {\bibinfo {title} {{Charge Transfer Models of Zinc and
  Magnesium in Water}},\ }\href {https://doi.org/10.1021/ct501173n} {\bibfield
  {journal} {\bibinfo  {journal} {Journal of Chemical Theory and Computation}\
  }\textbf {\bibinfo {volume} {11}},\ \bibinfo {pages} {1658} (\bibinfo {year}
  {2015})}\BibitemShut {NoStop}%
\bibitem [{\citenamefont {Jing}\ \emph {et~al.}(2018)\citenamefont {Jing},
  \citenamefont {Liu}, \citenamefont {Qi},\ and\ \citenamefont
  {Ren}}]{Jing2018}%
  \BibitemOpen
  \bibfield  {author} {\bibinfo {author} {\bibfnamefont {Z.}~\bibnamefont
  {Jing}}, \bibinfo {author} {\bibfnamefont {C.}~\bibnamefont {Liu}}, \bibinfo
  {author} {\bibfnamefont {R.}~\bibnamefont {Qi}},\ and\ \bibinfo {author}
  {\bibfnamefont {P.}~\bibnamefont {Ren}},\ }\bibfield  {title} {\bibinfo
  {title} {{Many-body effect determines the selectivity for Ca 2+ and Mg 2+ in
  proteins}},\ }\bibfield  {journal} {\bibinfo  {journal} {Proceedings of the
  National Academy of Sciences}\ }\textbf {\bibinfo {volume} {115}},\ \href
  {https://doi.org/10.1073/pnas.1805049115} {10.1073/pnas.1805049115} (\bibinfo
  {year} {2018})\BibitemShut {NoStop}%
\bibitem [{\citenamefont {Mamatkulov}\ \emph {et~al.}(2013)\citenamefont
  {Mamatkulov}, \citenamefont {Fyta},\ and\ \citenamefont
  {Netz}}]{Mamatkulov2013}%
  \BibitemOpen
  \bibfield  {author} {\bibinfo {author} {\bibfnamefont {S.}~\bibnamefont
  {Mamatkulov}}, \bibinfo {author} {\bibfnamefont {M.}~\bibnamefont {Fyta}},\
  and\ \bibinfo {author} {\bibfnamefont {R.~R.}\ \bibnamefont {Netz}},\
  }\bibfield  {title} {\bibinfo {title} {{Force fields for divalent cations
  based on single-ion and ion-pair properties}},\ }\bibfield  {journal}
  {\bibinfo  {journal} {The Journal of Chemical Physics}\ }\textbf {\bibinfo
  {volume} {138}},\ \href {https://doi.org/10.1063/1.4772808}
  {10.1063/1.4772808} (\bibinfo {year} {2013})\BibitemShut {NoStop}%
\bibitem [{\citenamefont {Li}\ \emph {et~al.}(2013)\citenamefont {Li},
  \citenamefont {Roberts}, \citenamefont {Chakravorty},\ and\ \citenamefont
  {Merz}}]{Li2013}%
  \BibitemOpen
  \bibfield  {author} {\bibinfo {author} {\bibfnamefont {P.}~\bibnamefont
  {Li}}, \bibinfo {author} {\bibfnamefont {B.~P.}\ \bibnamefont {Roberts}},
  \bibinfo {author} {\bibfnamefont {D.~K.}\ \bibnamefont {Chakravorty}},\ and\
  \bibinfo {author} {\bibfnamefont {K.~M.}\ \bibnamefont {Merz}},\ }\bibfield
  {title} {\bibinfo {title} {{Rational Design of Particle Mesh Ewald Compatible
  Lennard-Jones Parameters for +2 Metal Cations in Explicit Solvent}},\ }\href
  {https://doi.org/10.1021/ct400146w} {\bibfield  {journal} {\bibinfo
  {journal} {Journal of Chemical Theory and Computation}\ }\textbf {\bibinfo
  {volume} {9}},\ \bibinfo {pages} {2733} (\bibinfo {year} {2013})}\BibitemShut
  {NoStop}%
\bibitem [{\citenamefont {Schwierz}(2020)}]{Schwierz2020}%
  \BibitemOpen
  \bibfield  {author} {\bibinfo {author} {\bibfnamefont {N.}~\bibnamefont
  {Schwierz}},\ }\bibfield  {title} {\bibinfo {title} {{Kinetic pathways of
  water exchange in the first hydration shell of magnesium}},\ }\href
  {https://doi.org/10.1063/1.5144258} {\bibfield  {journal} {\bibinfo
  {journal} {The Journal of Chemical Physics}\ }\textbf {\bibinfo {volume}
  {152}},\ \bibinfo {pages} {224106} (\bibinfo {year} {2020})}\BibitemShut
  {NoStop}%
\bibitem [{\citenamefont {Cruz-Le\'{o}n}\ and\ \citenamefont
  {Schwierz}(2020)}]{Cruz-Len2020}%
  \BibitemOpen
  \bibfield  {author} {\bibinfo {author} {\bibfnamefont {S.}~\bibnamefont
  {Cruz-Le\'{o}n}}\ and\ \bibinfo {author} {\bibfnamefont {N.}~\bibnamefont
  {Schwierz}},\ }\bibfield  {title} {\bibinfo {title} {{Hofmeister Series for
  Metal-Cation–RNA Interactions: The Interplay of Binding Affinity and
  Exchange Kinetics}},\ }\href {https://doi.org/10.1021/acs.langmuir.0c00851}
  {\bibfield  {journal} {\bibinfo  {journal} {Langmuir}\ }\textbf {\bibinfo
  {volume} {36}},\ \bibinfo {pages} {5979} (\bibinfo {year}
  {2020})}\BibitemShut {NoStop}%
\bibitem [{\citenamefont {Fyta}\ and\ \citenamefont {Netz}(2012)}]{Fyta2012}%
  \BibitemOpen
  \bibfield  {author} {\bibinfo {author} {\bibfnamefont {M.}~\bibnamefont
  {Fyta}}\ and\ \bibinfo {author} {\bibfnamefont {R.~R.}\ \bibnamefont
  {Netz}},\ }\bibfield  {title} {\bibinfo {title} {{Ionic force field
  optimization based on single-ion and ion-pair solvation properties: Going
  beyond standard mixing rules}},\ }\bibfield  {journal} {\bibinfo  {journal}
  {The Journal of Chemical Physics}\ }\textbf {\bibinfo {volume} {136}},\ \href
  {https://doi.org/10.1063/1.3693330} {10.1063/1.3693330} (\bibinfo {year}
  {2012})\BibitemShut {NoStop}%
\bibitem [{\citenamefont {Leontyev}\ and\ \citenamefont
  {Stuchebrukhov}(2011)}]{Leontyev2011}%
  \BibitemOpen
  \bibfield  {author} {\bibinfo {author} {\bibfnamefont {I.}~\bibnamefont
  {Leontyev}}\ and\ \bibinfo {author} {\bibfnamefont {A.}~\bibnamefont
  {Stuchebrukhov}},\ }\bibfield  {title} {\bibinfo {title} {{Accounting for
  electronic polarization in non-polarizable force fields}},\ }\href
  {https://doi.org/10.1039/c0cp01971b} {\bibfield  {journal} {\bibinfo
  {journal} {Physical Chemistry Chemical Physics}\ }\textbf {\bibinfo {volume}
  {13}},\ \bibinfo {pages} {2613} (\bibinfo {year} {2011})}\BibitemShut
  {NoStop}%
\bibitem [{\citenamefont {Kohagen}\ \emph {et~al.}(2014)\citenamefont
  {Kohagen}, \citenamefont {Mason},\ and\ \citenamefont
  {Jungwirth}}]{Kohagen2014}%
  \BibitemOpen
  \bibfield  {author} {\bibinfo {author} {\bibfnamefont {M.}~\bibnamefont
  {Kohagen}}, \bibinfo {author} {\bibfnamefont {P.~E.}\ \bibnamefont {Mason}},\
  and\ \bibinfo {author} {\bibfnamefont {P.}~\bibnamefont {Jungwirth}},\
  }\bibfield  {title} {\bibinfo {title} {{Accurate Description of Calcium
  Solvation in Concentrated Aqueous Solutions}},\ }\href
  {https://doi.org/10.1021/jp5005693} {\bibfield  {journal} {\bibinfo
  {journal} {The Journal of Physical Chemistry B}\ }\textbf {\bibinfo {volume}
  {118}},\ \bibinfo {pages} {7902} (\bibinfo {year} {2014})}\BibitemShut
  {NoStop}%
\bibitem [{\citenamefont {Zeron}\ \emph {et~al.}(2019)\citenamefont {Zeron},
  \citenamefont {Abascal},\ and\ \citenamefont {Vega}}]{Zeron2019}%
  \BibitemOpen
  \bibfield  {author} {\bibinfo {author} {\bibfnamefont {I.~M.}\ \bibnamefont
  {Zeron}}, \bibinfo {author} {\bibfnamefont {J.~L.~F.}\ \bibnamefont
  {Abascal}},\ and\ \bibinfo {author} {\bibfnamefont {C.}~\bibnamefont
  {Vega}},\ }\bibfield  {title} {\bibinfo {title} {{A force field of Li+, Na+,
  K+, Mg2+, Ca2+, Cl-, and SO42- in aqueous solution based on the TIP4P/2005
  water model and scaled charges for the ions}},\ }\bibfield  {journal}
  {\bibinfo  {journal} {The Journal of Chemical Physics}\ }\textbf {\bibinfo
  {volume} {151}},\ \href {https://doi.org/10.1063/1.5121392}
  {10.1063/1.5121392} (\bibinfo {year} {2019})\BibitemShut {NoStop}%
\bibitem [{\citenamefont {Li}\ \emph {et~al.}(2015)\citenamefont {Li},
  \citenamefont {Song},\ and\ \citenamefont {Merz}}]{Li2015}%
  \BibitemOpen
  \bibfield  {author} {\bibinfo {author} {\bibfnamefont {P.}~\bibnamefont
  {Li}}, \bibinfo {author} {\bibfnamefont {L.~F.}\ \bibnamefont {Song}},\ and\
  \bibinfo {author} {\bibfnamefont {K.~M.}\ \bibnamefont {Merz}},\ }\bibfield
  {title} {\bibinfo {title} {{Systematic Parameterization of Monovalent Ions
  Employing the Nonbonded Model}},\ }\href {https://doi.org/10.1021/ct500918t}
  {\bibfield  {journal} {\bibinfo  {journal} {Journal of Chemical Theory and
  Computation}\ }\textbf {\bibinfo {volume} {11}},\ \bibinfo {pages} {1645}
  (\bibinfo {year} {2015})}\BibitemShut {NoStop}%
\bibitem [{\citenamefont {Schran}\ \emph {et~al.}(2021)\citenamefont {Schran},
  \citenamefont {Thiemann}, \citenamefont {Rowe}, \citenamefont {M\"{u}ller},
  \citenamefont {Marsalek},\ and\ \citenamefont {Michaelides}}]{Schran2021}%
  \BibitemOpen
  \bibfield  {author} {\bibinfo {author} {\bibfnamefont {C.}~\bibnamefont
  {Schran}}, \bibinfo {author} {\bibfnamefont {F.~L.}\ \bibnamefont
  {Thiemann}}, \bibinfo {author} {\bibfnamefont {P.}~\bibnamefont {Rowe}},
  \bibinfo {author} {\bibfnamefont {E.~A.}\ \bibnamefont {M\"{u}ller}},
  \bibinfo {author} {\bibfnamefont {O.}~\bibnamefont {Marsalek}},\ and\
  \bibinfo {author} {\bibfnamefont {A.}~\bibnamefont {Michaelides}},\
  }\bibfield  {title} {\bibinfo {title} {{Machine learning potentials for
  complex aqueous systems made simple}},\ }\bibfield  {journal} {\bibinfo
  {journal} {Proceedings of the National Academy of Sciences}\ }\textbf
  {\bibinfo {volume} {118}},\ \href {https://doi.org/10.1073/pnas.2110077118}
  {10.1073/pnas.2110077118} (\bibinfo {year} {2021})\BibitemShut {NoStop}%
\bibitem [{\citenamefont {Kocer}\ \emph {et~al.}(2022)\citenamefont {Kocer},
  \citenamefont {Ko},\ and\ \citenamefont {Behler}}]{Kocer2022}%
  \BibitemOpen
  \bibfield  {author} {\bibinfo {author} {\bibfnamefont {E.}~\bibnamefont
  {Kocer}}, \bibinfo {author} {\bibfnamefont {T.~W.}\ \bibnamefont {Ko}},\ and\
  \bibinfo {author} {\bibfnamefont {J.}~\bibnamefont {Behler}},\ }\bibfield
  {title} {\bibinfo {title} {{Neural Network Potentials: A Concise Overview of
  Methods}},\ }\href {https://doi.org/10.1146/annurev-physchem-082720-034254}
  {\bibfield  {journal} {\bibinfo  {journal} {Annual Review of Physical
  Chemistry}\ }\textbf {\bibinfo {volume} {73}},\ \bibinfo {pages} {163}
  (\bibinfo {year} {2022})}\BibitemShut {NoStop}%
\bibitem [{\citenamefont {Omranpour}\ \emph {et~al.}(2024)\citenamefont
  {Omranpour}, \citenamefont {{Montero de Hijes}}, \citenamefont {Behler},\
  and\ \citenamefont {Dellago}}]{Omranpour2024}%
  \BibitemOpen
  \bibfield  {author} {\bibinfo {author} {\bibfnamefont {A.}~\bibnamefont
  {Omranpour}}, \bibinfo {author} {\bibfnamefont {P.}~\bibnamefont {{Montero de
  Hijes}}}, \bibinfo {author} {\bibfnamefont {J.}~\bibnamefont {Behler}},\ and\
  \bibinfo {author} {\bibfnamefont {C.}~\bibnamefont {Dellago}},\ }\bibfield
  {title} {\bibinfo {title} {{Perspective: Atomistic simulations of water and
  aqueous systems with machine learning potentials}},\ }\bibfield  {journal}
  {\bibinfo  {journal} {The Journal of Chemical Physics}\ }\textbf {\bibinfo
  {volume} {160}},\ \href {https://doi.org/10.1063/5.0201241}
  {10.1063/5.0201241} (\bibinfo {year} {2024})\BibitemShut {NoStop}%
\bibitem [{\citenamefont {Daru}\ \emph {et~al.}(2022)\citenamefont {Daru},
  \citenamefont {Forbert}, \citenamefont {Behler},\ and\ \citenamefont
  {Marx}}]{Daru2022}%
  \BibitemOpen
  \bibfield  {author} {\bibinfo {author} {\bibfnamefont {J.}~\bibnamefont
  {Daru}}, \bibinfo {author} {\bibfnamefont {H.}~\bibnamefont {Forbert}},
  \bibinfo {author} {\bibfnamefont {J.}~\bibnamefont {Behler}},\ and\ \bibinfo
  {author} {\bibfnamefont {D.}~\bibnamefont {Marx}},\ }\bibfield  {title}
  {\bibinfo {title} {{Coupled Cluster Molecular Dynamics of Condensed Phase
  Systems Enabled by Machine Learning Potentials: Liquid Water Benchmark}},\
  }\href {https://doi.org/10.1103/PhysRevLett.129.226001} {\bibfield  {journal}
  {\bibinfo  {journal} {Physical Review Letters}\ }\textbf {\bibinfo {volume}
  {129}},\ \bibinfo {pages} {226001} (\bibinfo {year} {2022})}\BibitemShut
  {NoStop}%
\bibitem [{\citenamefont {Liu}\ \emph {et~al.}(2022)\citenamefont {Liu},
  \citenamefont {Lan},\ and\ \citenamefont {He}}]{Liu2022}%
  \BibitemOpen
  \bibfield  {author} {\bibinfo {author} {\bibfnamefont {J.}~\bibnamefont
  {Liu}}, \bibinfo {author} {\bibfnamefont {J.}~\bibnamefont {Lan}},\ and\
  \bibinfo {author} {\bibfnamefont {X.}~\bibnamefont {He}},\ }\bibfield
  {title} {\bibinfo {title} {{Toward High-level Machine Learning Potential for
  Water Based on Quantum Fragmentation and Neural Networks}},\ }\href
  {https://doi.org/10.1021/acs.jpca.2c00601} {\bibfield  {journal} {\bibinfo
  {journal} {The Journal of Physical Chemistry A}\ }\textbf {\bibinfo {volume}
  {126}},\ \bibinfo {pages} {3926} (\bibinfo {year} {2022})}\BibitemShut
  {NoStop}%
\bibitem [{\citenamefont {Batatia}\ \emph {et~al.}(2022)\citenamefont
  {Batatia}, \citenamefont {Kov\'{a}cs}, \citenamefont {Simm}, \citenamefont
  {Ortner},\ and\ \citenamefont {Cs\'{a}nyi}}]{Batatia2022}%
  \BibitemOpen
  \bibfield  {author} {\bibinfo {author} {\bibfnamefont {I.}~\bibnamefont
  {Batatia}}, \bibinfo {author} {\bibfnamefont {D.~P.}\ \bibnamefont
  {Kov\'{a}cs}}, \bibinfo {author} {\bibfnamefont {G.~N.}\ \bibnamefont
  {Simm}}, \bibinfo {author} {\bibfnamefont {C.}~\bibnamefont {Ortner}},\ and\
  \bibinfo {author} {\bibfnamefont {G.}~\bibnamefont {Cs\'{a}nyi}},\ }\bibfield
   {title} {\bibinfo {title} {{MACE: Higher Order Equivariant Message Passing
  Neural Networks for Fast and Accurate Force Fields}},\ }in\ \href@noop {}
  {\emph {\bibinfo {booktitle} {Advances in Neural Information Processing
  Systems}}},\ Vol.~\bibinfo {volume} {35}\ (\bibinfo {year}
  {2022})\BibitemShut {NoStop}%
\bibitem [{\citenamefont {Morawietz}\ \emph {et~al.}(2016)\citenamefont
  {Morawietz}, \citenamefont {Singraber}, \citenamefont {Dellago},\ and\
  \citenamefont {Behler}}]{Morawietz2016}%
  \BibitemOpen
  \bibfield  {author} {\bibinfo {author} {\bibfnamefont {T.}~\bibnamefont
  {Morawietz}}, \bibinfo {author} {\bibfnamefont {A.}~\bibnamefont
  {Singraber}}, \bibinfo {author} {\bibfnamefont {C.}~\bibnamefont {Dellago}},\
  and\ \bibinfo {author} {\bibfnamefont {J.}~\bibnamefont {Behler}},\
  }\bibfield  {title} {\bibinfo {title} {{How van der Waals interactions
  determine the unique properties of water}},\ }\href
  {https://doi.org/10.1073/pnas.1602375113} {\bibfield  {journal} {\bibinfo
  {journal} {Proceedings of the National Academy of Sciences}\ }\textbf
  {\bibinfo {volume} {113}},\ \bibinfo {pages} {8368} (\bibinfo {year}
  {2016})}\BibitemShut {NoStop}%
\bibitem [{\citenamefont {{Montero de Hijes}}\ \emph
  {et~al.}(2024{\natexlab{a}})\citenamefont {{Montero de Hijes}}, \citenamefont
  {Dellago}, \citenamefont {Jinnouchi}, \citenamefont {Schmiedmayer},\ and\
  \citenamefont {Kresse}}]{MonterodeHijes2024b}%
  \BibitemOpen
  \bibfield  {author} {\bibinfo {author} {\bibfnamefont {P.}~\bibnamefont
  {{Montero de Hijes}}}, \bibinfo {author} {\bibfnamefont {C.}~\bibnamefont
  {Dellago}}, \bibinfo {author} {\bibfnamefont {R.}~\bibnamefont {Jinnouchi}},
  \bibinfo {author} {\bibfnamefont {B.}~\bibnamefont {Schmiedmayer}},\ and\
  \bibinfo {author} {\bibfnamefont {G.}~\bibnamefont {Kresse}},\ }\bibfield
  {title} {\bibinfo {title} {{Comparing machine learning potentials for water:
  Kernel-based regression and Behler–Parrinello neural networks}},\
  }\bibfield  {journal} {\bibinfo  {journal} {The Journal of Chemical Physics}\
  }\textbf {\bibinfo {volume} {160}},\ \href
  {https://doi.org/10.1063/5.0197105} {10.1063/5.0197105} (\bibinfo {year}
  {2024}{\natexlab{a}})\BibitemShut {NoStop}%
\bibitem [{\citenamefont {Lightstone}\ \emph {et~al.}(2001)\citenamefont
  {Lightstone}, \citenamefont {Schwegler}, \citenamefont {Hood}, \citenamefont
  {Gygi},\ and\ \citenamefont {Galli}}]{Lightstone2001}%
  \BibitemOpen
  \bibfield  {author} {\bibinfo {author} {\bibfnamefont {F.~C.}\ \bibnamefont
  {Lightstone}}, \bibinfo {author} {\bibfnamefont {E.}~\bibnamefont
  {Schwegler}}, \bibinfo {author} {\bibfnamefont {R.~Q.}\ \bibnamefont {Hood}},
  \bibinfo {author} {\bibfnamefont {F.}~\bibnamefont {Gygi}},\ and\ \bibinfo
  {author} {\bibfnamefont {G.}~\bibnamefont {Galli}},\ }\bibfield  {title}
  {\bibinfo {title} {{A first principles molecular dynamics simulation of the
  hydrated magnesium ion}},\ }\href
  {https://doi.org/10.1016/S0009-2614(01)00735-7} {\bibfield  {journal}
  {\bibinfo  {journal} {Chemical Physics Letters}\ }\textbf {\bibinfo {volume}
  {343}},\ \bibinfo {pages} {549} (\bibinfo {year} {2001})}\BibitemShut
  {NoStop}%
\bibitem [{\citenamefont {Bhattacharjee}\ \emph {et~al.}(2012)\citenamefont
  {Bhattacharjee}, \citenamefont {Pribil}, \citenamefont {Randolf},
  \citenamefont {Rode},\ and\ \citenamefont {Hofer}}]{Bhattacharjee2012}%
  \BibitemOpen
  \bibfield  {author} {\bibinfo {author} {\bibfnamefont {A.}~\bibnamefont
  {Bhattacharjee}}, \bibinfo {author} {\bibfnamefont {A.~B.}\ \bibnamefont
  {Pribil}}, \bibinfo {author} {\bibfnamefont {B.~R.}\ \bibnamefont {Randolf}},
  \bibinfo {author} {\bibfnamefont {B.~M.}\ \bibnamefont {Rode}},\ and\
  \bibinfo {author} {\bibfnamefont {T.~S.}\ \bibnamefont {Hofer}},\ }\bibfield
  {title} {\bibinfo {title} {{Hydration of Mg2+ and its influence on the water
  hydrogen bonding network via ab initio QMCF MD}},\ }\href
  {https://doi.org/10.1016/j.cplett.2012.03.049} {\bibfield  {journal}
  {\bibinfo  {journal} {Chemical Physics Letters}\ }\textbf {\bibinfo {volume}
  {536}},\ \bibinfo {pages} {39} (\bibinfo {year} {2012})}\BibitemShut
  {NoStop}%
\bibitem [{\citenamefont {Wang}\ \emph {et~al.}(2020)\citenamefont {Wang},
  \citenamefont {Toroz}, \citenamefont {Kim}, \citenamefont {Clegg},
  \citenamefont {Park},\ and\ \citenamefont {Tommaso}}]{Wang2020}%
  \BibitemOpen
  \bibfield  {author} {\bibinfo {author} {\bibfnamefont {X.}~\bibnamefont
  {Wang}}, \bibinfo {author} {\bibfnamefont {D.}~\bibnamefont {Toroz}},
  \bibinfo {author} {\bibfnamefont {S.}~\bibnamefont {Kim}}, \bibinfo {author}
  {\bibfnamefont {S.~L.}\ \bibnamefont {Clegg}}, \bibinfo {author}
  {\bibfnamefont {G.-S.}\ \bibnamefont {Park}},\ and\ \bibinfo {author}
  {\bibfnamefont {D.~D.}\ \bibnamefont {Tommaso}},\ }\bibfield  {title}
  {\bibinfo {title} {{Density functional theory based molecular dynamics study
  of solution composition effects on the solvation shell of metal ions}},\
  }\href {https://doi.org/10.1039/D0CP01957G} {\bibfield  {journal} {\bibinfo
  {journal} {Physical Chemistry Chemical Physics}\ }\textbf {\bibinfo {volume}
  {22}},\ \bibinfo {pages} {16301} (\bibinfo {year} {2020})}\BibitemShut
  {NoStop}%
\bibitem [{\citenamefont {Juraskova}\ \emph {et~al.}(2025)\citenamefont
  {Juraskova}, \citenamefont {Tusha}, \citenamefont {Zhang}, \citenamefont
  {Sch\"{a}fer},\ and\ \citenamefont {Duarte}}]{Juraskova2025}%
  \BibitemOpen
  \bibfield  {author} {\bibinfo {author} {\bibfnamefont {V.}~\bibnamefont
  {Juraskova}}, \bibinfo {author} {\bibfnamefont {G.}~\bibnamefont {Tusha}},
  \bibinfo {author} {\bibfnamefont {H.}~\bibnamefont {Zhang}}, \bibinfo
  {author} {\bibfnamefont {L.~V.}\ \bibnamefont {Sch\"{a}fer}},\ and\ \bibinfo
  {author} {\bibfnamefont {F.}~\bibnamefont {Duarte}},\ }\bibfield  {title}
  {\bibinfo {title} {{Modelling ligand exchange in metal complexes with machine
  learning potentials}},\ }\href {https://doi.org/10.1039/D4FD00140K}
  {\bibfield  {journal} {\bibinfo  {journal} {Faraday Discussions}\ }\textbf
  {\bibinfo {volume} {256}},\ \bibinfo {pages} {156} (\bibinfo {year}
  {2025})}\BibitemShut {NoStop}%
\bibitem [{\citenamefont {Ferretti}\ \emph {et~al.}(2025)\citenamefont
  {Ferretti}, \citenamefont {Melani}, \citenamefont {Benedetti}, \citenamefont
  {Sorodoc}, \citenamefont {Fortunelli},\ and\ \citenamefont
  {Brancato}}]{Ferretti2025}%
  \BibitemOpen
  \bibfield  {author} {\bibinfo {author} {\bibfnamefont {A.}~\bibnamefont
  {Ferretti}}, \bibinfo {author} {\bibfnamefont {G.}~\bibnamefont {Melani}},
  \bibinfo {author} {\bibfnamefont {L.}~\bibnamefont {Benedetti}}, \bibinfo
  {author} {\bibfnamefont {R.~A.}\ \bibnamefont {Sorodoc}}, \bibinfo {author}
  {\bibfnamefont {A.}~\bibnamefont {Fortunelli}},\ and\ \bibinfo {author}
  {\bibfnamefont {G.}~\bibnamefont {Brancato}},\ }\bibfield  {title} {\bibinfo
  {title} {{Accurate Simulations of Water and Aqueous Solutions through
  Fine-Tuned Dispersion-Corrected Density Functional Theory and
  Machine-Learning Interatomic Potentials}},\ }\href
  {https://doi.org/10.1021/acs.jcim.5c02079} {\bibfield  {journal} {\bibinfo
  {journal} {Journal of Chemical Information and Modeling}\ }\textbf {\bibinfo
  {volume} {65}},\ \bibinfo {pages} {12437} (\bibinfo {year}
  {2025})}\BibitemShut {NoStop}%
\bibitem [{\citenamefont {O'Neill}\ \emph {et~al.}(2024)\citenamefont
  {O'Neill}, \citenamefont {Shi}, \citenamefont {Fong}, \citenamefont
  {Michaelides},\ and\ \citenamefont {Schran}}]{ONeill2024}%
  \BibitemOpen
  \bibfield  {author} {\bibinfo {author} {\bibfnamefont {N.}~\bibnamefont
  {O'Neill}}, \bibinfo {author} {\bibfnamefont {B.~X.}\ \bibnamefont {Shi}},
  \bibinfo {author} {\bibfnamefont {K.}~\bibnamefont {Fong}}, \bibinfo {author}
  {\bibfnamefont {A.}~\bibnamefont {Michaelides}},\ and\ \bibinfo {author}
  {\bibfnamefont {C.}~\bibnamefont {Schran}},\ }\bibfield  {title} {\bibinfo
  {title} {{To Pair or not to Pair? Machine-Learned Explicitly-Correlated
  Electronic Structure for NaCl in Water}},\ }\href
  {https://doi.org/10.1021/acs.jpclett.4c01030} {\bibfield  {journal} {\bibinfo
   {journal} {The Journal of Physical Chemistry Letters}\ }\textbf {\bibinfo
  {volume} {15}},\ \bibinfo {pages} {6081} (\bibinfo {year}
  {2024})}\BibitemShut {NoStop}%
\bibitem [{\citenamefont {Soyemi}\ and\ \citenamefont
  {Szilv\'{a}si}(2025)}]{Soyemi2025}%
  \BibitemOpen
  \bibfield  {author} {\bibinfo {author} {\bibfnamefont {A.}~\bibnamefont
  {Soyemi}}\ and\ \bibinfo {author} {\bibfnamefont {T.}~\bibnamefont
  {Szilv\'{a}si}},\ }\bibfield  {title} {\bibinfo {title} {{Modeling the
  Behavior of Complex Aqueous Electrolytes Using Machine Learning Interatomic
  Potentials: The Case of Sodium Sulfate}},\ }\href
  {https://doi.org/10.1021/acs.jpcb.5c02306} {\bibfield  {journal} {\bibinfo
  {journal} {The Journal of Physical Chemistry B}\ }\textbf {\bibinfo {volume}
  {129}},\ \bibinfo {pages} {9405} (\bibinfo {year} {2025})}\BibitemShut
  {NoStop}%
\bibitem [{\citenamefont {Cao}\ \emph {et~al.}(2025)\citenamefont {Cao},
  \citenamefont {Kingan}, \citenamefont {Hill}, \citenamefont {Kuang},
  \citenamefont {Wang}, \citenamefont {Zhang}, \citenamefont {Carbone},
  \citenamefont {van Dam}, \citenamefont {Yoo}, \citenamefont {Yan},
  \citenamefont {Takeuchi}, \citenamefont {Takeuchi}, \citenamefont {Wu},
  \citenamefont {Abeykoon}, \citenamefont {Marschilok},\ and\ \citenamefont
  {Lu}}]{Cao2025}%
  \BibitemOpen
  \bibfield  {author} {\bibinfo {author} {\bibfnamefont {C.}~\bibnamefont
  {Cao}}, \bibinfo {author} {\bibfnamefont {A.}~\bibnamefont {Kingan}},
  \bibinfo {author} {\bibfnamefont {R.~C.}\ \bibnamefont {Hill}}, \bibinfo
  {author} {\bibfnamefont {J.}~\bibnamefont {Kuang}}, \bibinfo {author}
  {\bibfnamefont {L.}~\bibnamefont {Wang}}, \bibinfo {author} {\bibfnamefont
  {C.}~\bibnamefont {Zhang}}, \bibinfo {author} {\bibfnamefont {M.~R.}\
  \bibnamefont {Carbone}}, \bibinfo {author} {\bibfnamefont {H.}~\bibnamefont
  {van Dam}}, \bibinfo {author} {\bibfnamefont {S.}~\bibnamefont {Yoo}},
  \bibinfo {author} {\bibfnamefont {S.}~\bibnamefont {Yan}}, \bibinfo {author}
  {\bibfnamefont {E.~S.}\ \bibnamefont {Takeuchi}}, \bibinfo {author}
  {\bibfnamefont {K.~J.}\ \bibnamefont {Takeuchi}}, \bibinfo {author}
  {\bibfnamefont {X.}~\bibnamefont {Wu}}, \bibinfo {author} {\bibfnamefont
  {A.~M.}\ \bibnamefont {Abeykoon}}, \bibinfo {author} {\bibfnamefont {A.~C.}\
  \bibnamefont {Marschilok}},\ and\ \bibinfo {author} {\bibfnamefont
  {D.}~\bibnamefont {Lu}},\ }\bibfield  {title} {\bibinfo {title} {{Resolving
  the Solvation Structure and Transport Properties of Aqueous Zinc Electrolytes
  from Salt-in-Water to Water-in-Salt Using Neural Network Potential}},\ }\href
  {https://doi.org/10.1103/PRXEnergy.4.023004} {\bibfield  {journal} {\bibinfo
  {journal} {PRX Energy}\ }\textbf {\bibinfo {volume} {4}},\ \bibinfo {pages}
  {023004} (\bibinfo {year} {2025})}\BibitemShut {NoStop}%
\bibitem [{\citenamefont {Gillan}\ \emph {et~al.}(2016)\citenamefont {Gillan},
  \citenamefont {Alf\`{e}},\ and\ \citenamefont {Michaelides}}]{Gillan2016}%
  \BibitemOpen
  \bibfield  {author} {\bibinfo {author} {\bibfnamefont {M.~J.}\ \bibnamefont
  {Gillan}}, \bibinfo {author} {\bibfnamefont {D.}~\bibnamefont {Alf\`{e}}},\
  and\ \bibinfo {author} {\bibfnamefont {A.}~\bibnamefont {Michaelides}},\
  }\bibfield  {title} {\bibinfo {title} {{Perspective: How good is DFT for
  water?}},\ }\bibfield  {journal} {\bibinfo  {journal} {The Journal of
  Chemical Physics}\ }\textbf {\bibinfo {volume} {144}},\ \href
  {https://doi.org/10.1063/1.4944633} {10.1063/1.4944633} (\bibinfo {year}
  {2016})\BibitemShut {NoStop}%
\bibitem [{\citenamefont {Palos}\ \emph {et~al.}(2022)\citenamefont {Palos},
  \citenamefont {Lambros}, \citenamefont {Swee}, \citenamefont {Hu},
  \citenamefont {Dasgupta},\ and\ \citenamefont {Paesani}}]{Palos2022}%
  \BibitemOpen
  \bibfield  {author} {\bibinfo {author} {\bibfnamefont {E.}~\bibnamefont
  {Palos}}, \bibinfo {author} {\bibfnamefont {E.}~\bibnamefont {Lambros}},
  \bibinfo {author} {\bibfnamefont {S.}~\bibnamefont {Swee}}, \bibinfo {author}
  {\bibfnamefont {J.}~\bibnamefont {Hu}}, \bibinfo {author} {\bibfnamefont
  {S.}~\bibnamefont {Dasgupta}},\ and\ \bibinfo {author} {\bibfnamefont
  {F.}~\bibnamefont {Paesani}},\ }\bibfield  {title} {\bibinfo {title}
  {{Assessing the Interplay between Functional-Driven and Density-Driven Errors
  in DFT Models of Water}},\ }\href {https://doi.org/10.1021/acs.jctc.2c00050}
  {\bibfield  {journal} {\bibinfo  {journal} {Journal of Chemical Theory and
  Computation}\ }\textbf {\bibinfo {volume} {18}},\ \bibinfo {pages} {3410}
  (\bibinfo {year} {2022})}\BibitemShut {NoStop}%
\bibitem [{\citenamefont {Dasgupta}\ \emph {et~al.}(2021)\citenamefont
  {Dasgupta}, \citenamefont {Lambros}, \citenamefont {Perdew},\ and\
  \citenamefont {Paesani}}]{Dasgupta2021}%
  \BibitemOpen
  \bibfield  {author} {\bibinfo {author} {\bibfnamefont {S.}~\bibnamefont
  {Dasgupta}}, \bibinfo {author} {\bibfnamefont {E.}~\bibnamefont {Lambros}},
  \bibinfo {author} {\bibfnamefont {J.~P.}\ \bibnamefont {Perdew}},\ and\
  \bibinfo {author} {\bibfnamefont {F.}~\bibnamefont {Paesani}},\ }\bibfield
  {title} {\bibinfo {title} {{Elevating density functional theory to chemical
  accuracy for water simulations through a density-corrected many-body
  formalism}},\ }\href {https://doi.org/10.1038/s41467-021-26618-9} {\bibfield
  {journal} {\bibinfo  {journal} {Nature Communications}\ }\textbf {\bibinfo
  {volume} {12}},\ \bibinfo {pages} {6359} (\bibinfo {year}
  {2021})}\BibitemShut {NoStop}%
\bibitem [{\citenamefont {Perdew}\ \emph {et~al.}(1996)\citenamefont {Perdew},
  \citenamefont {Burke},\ and\ \citenamefont {Ernzerhof}}]{Perdew1996}%
  \BibitemOpen
  \bibfield  {author} {\bibinfo {author} {\bibfnamefont {J.~P.}\ \bibnamefont
  {Perdew}}, \bibinfo {author} {\bibfnamefont {K.}~\bibnamefont {Burke}},\ and\
  \bibinfo {author} {\bibfnamefont {M.}~\bibnamefont {Ernzerhof}},\ }\bibfield
  {title} {\bibinfo {title} {{Generalized Gradient Approximation Made
  Simple}},\ }\href {https://doi.org/10.1103/PhysRevLett.77.3865} {\bibfield
  {journal} {\bibinfo  {journal} {Physical Review Letters}\ }\textbf {\bibinfo
  {volume} {77}},\ \bibinfo {pages} {3865} (\bibinfo {year}
  {1996})}\BibitemShut {NoStop}%
\bibitem [{\citenamefont {Zhang}\ and\ \citenamefont {Yang}(1998)}]{Zhang1998}%
  \BibitemOpen
  \bibfield  {author} {\bibinfo {author} {\bibfnamefont {Y.}~\bibnamefont
  {Zhang}}\ and\ \bibinfo {author} {\bibfnamefont {W.}~\bibnamefont {Yang}},\
  }\bibfield  {title} {\bibinfo {title} {{Comment on ``Generalized Gradient
  Approximation Made Simple''}},\ }\href
  {https://doi.org/10.1103/PhysRevLett.80.890} {\bibfield  {journal} {\bibinfo
  {journal} {Physical Review Letters}\ }\textbf {\bibinfo {volume} {80}},\
  \bibinfo {pages} {890} (\bibinfo {year} {1998})}\BibitemShut {NoStop}%
\bibitem [{\citenamefont {Adamo}\ and\ \citenamefont
  {Barone}(1999)}]{Adamo1999}%
  \BibitemOpen
  \bibfield  {author} {\bibinfo {author} {\bibfnamefont {C.}~\bibnamefont
  {Adamo}}\ and\ \bibinfo {author} {\bibfnamefont {V.}~\bibnamefont {Barone}},\
  }\bibfield  {title} {\bibinfo {title} {{Toward reliable density functional
  methods without adjustable parameters: The PBE0 model}},\ }\href
  {https://doi.org/10.1063/1.478522} {\bibfield  {journal} {\bibinfo  {journal}
  {The Journal of Chemical Physics}\ }\textbf {\bibinfo {volume} {110}},\
  \bibinfo {pages} {6158} (\bibinfo {year} {1999})}\BibitemShut {NoStop}%
\bibitem [{\citenamefont {Grimme}\ \emph {et~al.}(2010)\citenamefont {Grimme},
  \citenamefont {Antony}, \citenamefont {Ehrlich},\ and\ \citenamefont
  {Krieg}}]{Grimme2010}%
  \BibitemOpen
  \bibfield  {author} {\bibinfo {author} {\bibfnamefont {S.}~\bibnamefont
  {Grimme}}, \bibinfo {author} {\bibfnamefont {J.}~\bibnamefont {Antony}},
  \bibinfo {author} {\bibfnamefont {S.}~\bibnamefont {Ehrlich}},\ and\ \bibinfo
  {author} {\bibfnamefont {H.}~\bibnamefont {Krieg}},\ }\bibfield  {title}
  {\bibinfo {title} {{A consistent and accurate ab initio parametrization of
  density functional dispersion correction (DFT-D) for the 94 elements H-Pu}},\
  }\bibfield  {journal} {\bibinfo  {journal} {The Journal of Chemical Physics}\
  }\textbf {\bibinfo {volume} {132}},\ \href
  {https://doi.org/10.1063/1.3382344} {10.1063/1.3382344} (\bibinfo {year}
  {2010})\BibitemShut {NoStop}%
\bibitem [{\citenamefont {Grimme}\ \emph {et~al.}(2011)\citenamefont {Grimme},
  \citenamefont {Ehrlich},\ and\ \citenamefont {Goerigk}}]{Grimme2011}%
  \BibitemOpen
  \bibfield  {author} {\bibinfo {author} {\bibfnamefont {S.}~\bibnamefont
  {Grimme}}, \bibinfo {author} {\bibfnamefont {S.}~\bibnamefont {Ehrlich}},\
  and\ \bibinfo {author} {\bibfnamefont {L.}~\bibnamefont {Goerigk}},\
  }\bibfield  {title} {\bibinfo {title} {{Effect of the damping function in
  dispersion corrected density functional theory}},\ }\href
  {https://doi.org/10.1002/jcc.21759} {\bibfield  {journal} {\bibinfo
  {journal} {Journal of Computational Chemistry}\ }\textbf {\bibinfo {volume}
  {32}},\ \bibinfo {pages} {1456} (\bibinfo {year} {2011})}\BibitemShut
  {NoStop}%
\bibitem [{\citenamefont {Lausch}\ \emph {et~al.}(2025)\citenamefont {Lausch},
  \citenamefont {Haouari}, \citenamefont {Trzewik},\ and\ \citenamefont
  {Behler}}]{Lausch2025}%
  \BibitemOpen
  \bibfield  {author} {\bibinfo {author} {\bibfnamefont {K.~N.}\ \bibnamefont
  {Lausch}}, \bibinfo {author} {\bibfnamefont {R.~E.}\ \bibnamefont {Haouari}},
  \bibinfo {author} {\bibfnamefont {D.}~\bibnamefont {Trzewik}},\ and\ \bibinfo
  {author} {\bibfnamefont {J.}~\bibnamefont {Behler}},\ }\bibfield  {title}
  {\bibinfo {title} {{Impact of the damping function in dispersion-corrected
  density functional theory on the properties of liquid water}},\ }\bibfield
  {journal} {\bibinfo  {journal} {The Journal of Chemical Physics}\ }\textbf
  {\bibinfo {volume} {163}},\ \href {https://doi.org/10.1063/5.0275244}
  {10.1063/5.0275244} (\bibinfo {year} {2025})\BibitemShut {NoStop}%
\bibitem [{\citenamefont {Abraham}\ \emph {et~al.}(2015)\citenamefont
  {Abraham}, \citenamefont {Murtola}, \citenamefont {Schulz}, \citenamefont
  {P\'{a}ll}, \citenamefont {Smith}, \citenamefont {Hess},\ and\ \citenamefont
  {Lindahl}}]{Abraham2015}%
  \BibitemOpen
  \bibfield  {author} {\bibinfo {author} {\bibfnamefont {M.~J.}\ \bibnamefont
  {Abraham}}, \bibinfo {author} {\bibfnamefont {T.}~\bibnamefont {Murtola}},
  \bibinfo {author} {\bibfnamefont {R.}~\bibnamefont {Schulz}}, \bibinfo
  {author} {\bibfnamefont {S.}~\bibnamefont {P\'{a}ll}}, \bibinfo {author}
  {\bibfnamefont {J.~C.}\ \bibnamefont {Smith}}, \bibinfo {author}
  {\bibfnamefont {B.}~\bibnamefont {Hess}},\ and\ \bibinfo {author}
  {\bibfnamefont {E.}~\bibnamefont {Lindahl}},\ }\bibfield  {title} {\bibinfo
  {title} {{GROMACS: High performance molecular simulations through multi-level
  parallelism from laptops to supercomputers}},\ }\href
  {https://doi.org/10.1016/j.softx.2015.06.001} {\bibfield  {journal} {\bibinfo
   {journal} {SoftwareX}\ }\textbf {\bibinfo {volume} {1-2}},\ \bibinfo {pages}
  {19} (\bibinfo {year} {2015})}\BibitemShut {NoStop}%
\bibitem [{\citenamefont {Jorgensen}\ \emph {et~al.}(1983)\citenamefont
  {Jorgensen}, \citenamefont {Chandrasekhar}, \citenamefont {Madura},
  \citenamefont {Impey},\ and\ \citenamefont {Klein}}]{Jorgensen1983}%
  \BibitemOpen
  \bibfield  {author} {\bibinfo {author} {\bibfnamefont {W.~L.}\ \bibnamefont
  {Jorgensen}}, \bibinfo {author} {\bibfnamefont {J.}~\bibnamefont
  {Chandrasekhar}}, \bibinfo {author} {\bibfnamefont {J.~D.}\ \bibnamefont
  {Madura}}, \bibinfo {author} {\bibfnamefont {R.~W.}\ \bibnamefont {Impey}},\
  and\ \bibinfo {author} {\bibfnamefont {M.~L.}\ \bibnamefont {Klein}},\
  }\bibfield  {title} {\bibinfo {title} {{Comparison of simple potential
  functions for simulating liquid water}},\ }\href
  {https://doi.org/10.1063/1.445869} {\bibfield  {journal} {\bibinfo  {journal}
  {The Journal of Chemical Physics}\ }\textbf {\bibinfo {volume} {79}},\
  \bibinfo {pages} {926} (\bibinfo {year} {1983})}\BibitemShut {NoStop}%
\bibitem [{\citenamefont {Hutter}\ \emph {et~al.}(2014)\citenamefont {Hutter},
  \citenamefont {Iannuzzi}, \citenamefont {Schiffmann},\ and\ \citenamefont
  {VandeVondele}}]{Hutter2014}%
  \BibitemOpen
  \bibfield  {author} {\bibinfo {author} {\bibfnamefont {J.}~\bibnamefont
  {Hutter}}, \bibinfo {author} {\bibfnamefont {M.}~\bibnamefont {Iannuzzi}},
  \bibinfo {author} {\bibfnamefont {F.}~\bibnamefont {Schiffmann}},\ and\
  \bibinfo {author} {\bibfnamefont {J.}~\bibnamefont {VandeVondele}},\
  }\bibfield  {title} {\bibinfo {title} {{cp2k: atomistic simulations of
  condensed matter systems}},\ }\href {https://doi.org/10.1002/wcms.1159}
  {\bibfield  {journal} {\bibinfo  {journal} {WIREs Computational Molecular
  Science}\ }\textbf {\bibinfo {volume} {4}},\ \bibinfo {pages} {15} (\bibinfo
  {year} {2014})}\BibitemShut {NoStop}%
\bibitem [{\citenamefont {Lippert}\ \emph {et~al.}(1999)\citenamefont
  {Lippert}, \citenamefont {Hutter},\ and\ \citenamefont
  {Parrinello}}]{Lippert1999}%
  \BibitemOpen
  \bibfield  {author} {\bibinfo {author} {\bibfnamefont {G.}~\bibnamefont
  {Lippert}}, \bibinfo {author} {\bibfnamefont {J.}~\bibnamefont {Hutter}},\
  and\ \bibinfo {author} {\bibfnamefont {M.}~\bibnamefont {Parrinello}},\
  }\bibfield  {title} {\bibinfo {title} {{The Gaussian and augmented-plane-wave
  density functional method for ab initio molecular dynamics simulations}},\
  }\href {https://doi.org/10.1007/s002140050523} {\bibfield  {journal}
  {\bibinfo  {journal} {Theoretical Chemistry Accounts: Theory, Computation,
  and Modeling (Theoretica Chimica Acta)}\ }\textbf {\bibinfo {volume} {103}},\
  \bibinfo {pages} {124} (\bibinfo {year} {1999})}\BibitemShut {NoStop}%
\bibitem [{\citenamefont {Goedecker}\ \emph {et~al.}(1996)\citenamefont
  {Goedecker}, \citenamefont {Teter},\ and\ \citenamefont
  {Hutter}}]{Goedecker1996}%
  \BibitemOpen
  \bibfield  {author} {\bibinfo {author} {\bibfnamefont {S.}~\bibnamefont
  {Goedecker}}, \bibinfo {author} {\bibfnamefont {M.}~\bibnamefont {Teter}},\
  and\ \bibinfo {author} {\bibfnamefont {J.}~\bibnamefont {Hutter}},\
  }\bibfield  {title} {\bibinfo {title} {{Separable dual-space Gaussian
  pseudopotentials}},\ }\href {https://doi.org/10.1103/PhysRevB.54.1703}
  {\bibfield  {journal} {\bibinfo  {journal} {Physical Review B}\ }\textbf
  {\bibinfo {volume} {54}},\ \bibinfo {pages} {1703} (\bibinfo {year}
  {1996})}\BibitemShut {NoStop}%
\bibitem [{\citenamefont {Hartwigsen}\ \emph {et~al.}(1998)\citenamefont
  {Hartwigsen}, \citenamefont {Goedecker},\ and\ \citenamefont
  {Hutter}}]{Hartwigsen1998}%
  \BibitemOpen
  \bibfield  {author} {\bibinfo {author} {\bibfnamefont {C.}~\bibnamefont
  {Hartwigsen}}, \bibinfo {author} {\bibfnamefont {S.}~\bibnamefont
  {Goedecker}},\ and\ \bibinfo {author} {\bibfnamefont {J.}~\bibnamefont
  {Hutter}},\ }\bibfield  {title} {\bibinfo {title} {{Relativistic separable
  dual-space Gaussian pseudopotentials from H to Rn}},\ }\href
  {https://doi.org/10.1103/PhysRevB.58.3641} {\bibfield  {journal} {\bibinfo
  {journal} {Physical Review B}\ }\textbf {\bibinfo {volume} {58}},\ \bibinfo
  {pages} {3641} (\bibinfo {year} {1998})}\BibitemShut {NoStop}%
\bibitem [{\citenamefont {Bussi}\ \emph {et~al.}(2007)\citenamefont {Bussi},
  \citenamefont {Donadio},\ and\ \citenamefont {Parrinello}}]{Bussi2007}%
  \BibitemOpen
  \bibfield  {author} {\bibinfo {author} {\bibfnamefont {G.}~\bibnamefont
  {Bussi}}, \bibinfo {author} {\bibfnamefont {D.}~\bibnamefont {Donadio}},\
  and\ \bibinfo {author} {\bibfnamefont {M.}~\bibnamefont {Parrinello}},\
  }\bibfield  {title} {\bibinfo {title} {{Canonical sampling through velocity
  rescaling}},\ }\bibfield  {journal} {\bibinfo  {journal} {Journal of Chemical
  Physics}\ }\textbf {\bibinfo {volume} {126}},\ \href
  {https://doi.org/10.1063/1.2408420} {10.1063/1.2408420} (\bibinfo {year}
  {2007})\BibitemShut {NoStop}%
\bibitem [{\citenamefont {{Montero de Hijes}}\ \emph
  {et~al.}(2024{\natexlab{b}})\citenamefont {{Montero de Hijes}}, \citenamefont
  {Dellago}, \citenamefont {Jinnouchi},\ and\ \citenamefont
  {Kresse}}]{MonterodeHijes2024}%
  \BibitemOpen
  \bibfield  {author} {\bibinfo {author} {\bibfnamefont {P.}~\bibnamefont
  {{Montero de Hijes}}}, \bibinfo {author} {\bibfnamefont {C.}~\bibnamefont
  {Dellago}}, \bibinfo {author} {\bibfnamefont {R.}~\bibnamefont {Jinnouchi}},\
  and\ \bibinfo {author} {\bibfnamefont {G.}~\bibnamefont {Kresse}},\
  }\bibfield  {title} {\bibinfo {title} {{Density isobar of water and melting
  temperature of ice: Assessing common density functionals}},\ }\bibfield
  {journal} {\bibinfo  {journal} {The Journal of Chemical Physics}\ }\textbf
  {\bibinfo {volume} {161}},\ \href {https://doi.org/10.1063/5.0227514}
  {10.1063/5.0227514} (\bibinfo {year} {2024}{\natexlab{b}})\BibitemShut
  {NoStop}%
\bibitem [{\citenamefont {Eastman}\ \emph {et~al.}(2017)\citenamefont
  {Eastman}, \citenamefont {Swails}, \citenamefont {Chodera}, \citenamefont
  {McGibbon}, \citenamefont {Zhao}, \citenamefont {Beauchamp}, \citenamefont
  {Wang}, \citenamefont {Simmonett}, \citenamefont {Harrigan}, \citenamefont
  {Stern}, \citenamefont {Wiewiora}, \citenamefont {Brooks},\ and\
  \citenamefont {Pande}}]{Eastman2017}%
  \BibitemOpen
  \bibfield  {author} {\bibinfo {author} {\bibfnamefont {P.}~\bibnamefont
  {Eastman}}, \bibinfo {author} {\bibfnamefont {J.}~\bibnamefont {Swails}},
  \bibinfo {author} {\bibfnamefont {J.~D.}\ \bibnamefont {Chodera}}, \bibinfo
  {author} {\bibfnamefont {R.~T.}\ \bibnamefont {McGibbon}}, \bibinfo {author}
  {\bibfnamefont {Y.}~\bibnamefont {Zhao}}, \bibinfo {author} {\bibfnamefont
  {K.~A.}\ \bibnamefont {Beauchamp}}, \bibinfo {author} {\bibfnamefont {L.-P.}\
  \bibnamefont {Wang}}, \bibinfo {author} {\bibfnamefont {A.~C.}\ \bibnamefont
  {Simmonett}}, \bibinfo {author} {\bibfnamefont {M.~P.}\ \bibnamefont
  {Harrigan}}, \bibinfo {author} {\bibfnamefont {C.~D.}\ \bibnamefont {Stern}},
  \bibinfo {author} {\bibfnamefont {R.~P.}\ \bibnamefont {Wiewiora}}, \bibinfo
  {author} {\bibfnamefont {B.~R.}\ \bibnamefont {Brooks}},\ and\ \bibinfo
  {author} {\bibfnamefont {V.~S.}\ \bibnamefont {Pande}},\ }\bibfield  {title}
  {\bibinfo {title} {{OpenMM 7: Rapid development of high performance
  algorithms for molecular dynamics}},\ }\href
  {https://doi.org/10.1371/journal.pcbi.1005659} {\bibfield  {journal}
  {\bibinfo  {journal} {PLOS Computational Biology}\ }\textbf {\bibinfo
  {volume} {13}},\ \bibinfo {pages} {e1005659} (\bibinfo {year}
  {2017})}\BibitemShut {NoStop}%
\bibitem [{\citenamefont {Sivak}\ \emph {et~al.}(2014)\citenamefont {Sivak},
  \citenamefont {Chodera},\ and\ \citenamefont {Crooks}}]{Sivak2014}%
  \BibitemOpen
  \bibfield  {author} {\bibinfo {author} {\bibfnamefont {D.~A.}\ \bibnamefont
  {Sivak}}, \bibinfo {author} {\bibfnamefont {J.~D.}\ \bibnamefont {Chodera}},\
  and\ \bibinfo {author} {\bibfnamefont {G.~E.}\ \bibnamefont {Crooks}},\
  }\bibfield  {title} {\bibinfo {title} {{Time Step Rescaling Recovers
  Continuous-Time Dynamical Properties for Discrete-Time Langevin Integration
  of Nonequilibrium Systems}},\ }\href {https://doi.org/10.1021/jp411770f}
  {\bibfield  {journal} {\bibinfo  {journal} {The Journal of Physical Chemistry
  B}\ }\textbf {\bibinfo {volume} {118}},\ \bibinfo {pages} {6466} (\bibinfo
  {year} {2014})}\BibitemShut {NoStop}%
\bibitem [{\citenamefont {Laio}\ and\ \citenamefont
  {Parrinello}(2002)}]{Laio2002}%
  \BibitemOpen
  \bibfield  {author} {\bibinfo {author} {\bibfnamefont {A.}~\bibnamefont
  {Laio}}\ and\ \bibinfo {author} {\bibfnamefont {M.}~\bibnamefont
  {Parrinello}},\ }\bibfield  {title} {\bibinfo {title} {{Escaping free-energy
  minima}},\ }\href {https://doi.org/10.1073/pnas.202427399} {\bibfield
  {journal} {\bibinfo  {journal} {Proceedings of the National Academy of
  Sciences}\ }\textbf {\bibinfo {volume} {99}},\ \bibinfo {pages} {12562}
  (\bibinfo {year} {2002})}\BibitemShut {NoStop}%
\bibitem [{\citenamefont {Barducci}\ \emph {et~al.}(2008)\citenamefont
  {Barducci}, \citenamefont {Bussi},\ and\ \citenamefont
  {Parrinello}}]{Barducci2008}%
  \BibitemOpen
  \bibfield  {author} {\bibinfo {author} {\bibfnamefont {A.}~\bibnamefont
  {Barducci}}, \bibinfo {author} {\bibfnamefont {G.}~\bibnamefont {Bussi}},\
  and\ \bibinfo {author} {\bibfnamefont {M.}~\bibnamefont {Parrinello}},\
  }\bibfield  {title} {\bibinfo {title} {{Well-Tempered Metadynamics: A
  Smoothly Converging and Tunable Free-Energy Method}},\ }\href
  {https://doi.org/10.1103/PhysRevLett.100.020603} {\bibfield  {journal}
  {\bibinfo  {journal} {Physical Review Letters}\ }\textbf {\bibinfo {volume}
  {100}},\ \bibinfo {pages} {020603} (\bibinfo {year} {2008})}\BibitemShut
  {NoStop}%
\bibitem [{\citenamefont {Torrie}\ and\ \citenamefont
  {Valleau}(1977)}]{Torrie1977}%
  \BibitemOpen
  \bibfield  {author} {\bibinfo {author} {\bibfnamefont {G.}~\bibnamefont
  {Torrie}}\ and\ \bibinfo {author} {\bibfnamefont {J.}~\bibnamefont
  {Valleau}},\ }\bibfield  {title} {\bibinfo {title} {{Nonphysical sampling
  distributions in Monte Carlo free-energy estimation: Umbrella sampling}},\
  }\href {https://doi.org/10.1016/0021-9991(77)90121-8} {\bibfield  {journal}
  {\bibinfo  {journal} {Journal of Computational Physics}\ }\textbf {\bibinfo
  {volume} {23}},\ \bibinfo {pages} {187} (\bibinfo {year} {1977})}\BibitemShut
  {NoStop}%
\bibitem [{\citenamefont {Bonomi}\ \emph {et~al.}(2019)\citenamefont {Bonomi},
  \citenamefont {Bussi}, \citenamefont {Camilloni}, \citenamefont {Tribello},
  \citenamefont {Ban\'{a}\v{s}}, \citenamefont {Barducci}, \citenamefont
  {Bernetti}, \citenamefont {Bolhuis}, \citenamefont {Bottaro}, \citenamefont
  {Branduardi}, \citenamefont {Capelli}, \citenamefont {Carloni}, \citenamefont
  {Ceriotti}, \citenamefont {Cesari}, \citenamefont {Chen}, \citenamefont
  {Chen}, \citenamefont {Colizzi}, \citenamefont {De}, \citenamefont {Pierre},
  \citenamefont {Donadio}, \citenamefont {Drobot}, \citenamefont {Ensing},
  \citenamefont {Ferguson}, \citenamefont {Filizola}, \citenamefont {Fraser},
  \citenamefont {Fu}, \citenamefont {Gasparotto}, \citenamefont {Gervasio},
  \citenamefont {Giberti}, \citenamefont {Gil-Ley}, \citenamefont {Giorgino},
  \citenamefont {Heller}, \citenamefont {Hocky}, \citenamefont {Iannuzzi},
  \citenamefont {Invernizzi}, \citenamefont {Jelfs}, \citenamefont {Jussupow},
  \citenamefont {Kirilin}, \citenamefont {Laio}, \citenamefont {Limongelli},
  \citenamefont {Lindorff-Larsen}, \citenamefont {L\"{o}hr}, \citenamefont
  {Marinelli}, \citenamefont {Martin-Samos}, \citenamefont {Masetti},
  \citenamefont {Meyer}, \citenamefont {Michaelides}, \citenamefont {Molteni},
  \citenamefont {Morishita}, \citenamefont {Nava}, \citenamefont {Paissoni},
  \citenamefont {Papaleo}, \citenamefont {Parrinello}, \citenamefont
  {Pfaendtner}, \citenamefont {Piaggi}, \citenamefont {Piccini}, \citenamefont
  {Pietropaolo}, \citenamefont {Pietrucci}, \citenamefont {Pipolo},
  \citenamefont {Provasi}, \citenamefont {Quigley}, \citenamefont {Raiteri},
  \citenamefont {Raniolo}, \citenamefont {Rydzewski}, \citenamefont
  {Salvalaglio}, \citenamefont {Sosso}, \citenamefont {Spiwok}, \citenamefont
  {\v{S}poner}, \citenamefont {Swenson}, \citenamefont {Tiwary}, \citenamefont
  {Valsson}, \citenamefont {Vendruscolo}, \citenamefont {Voth}, \citenamefont
  {White},\ and\ \citenamefont {consortium}}]{Bonomi2019}%
  \BibitemOpen
  \bibfield  {author} {\bibinfo {author} {\bibfnamefont {M.}~\bibnamefont
  {Bonomi}}, \bibinfo {author} {\bibfnamefont {G.}~\bibnamefont {Bussi}},
  \bibinfo {author} {\bibfnamefont {C.}~\bibnamefont {Camilloni}}, \bibinfo
  {author} {\bibfnamefont {G.~A.}\ \bibnamefont {Tribello}}, \bibinfo {author}
  {\bibfnamefont {P.}~\bibnamefont {Ban\'{a}\v{s}}}, \bibinfo {author}
  {\bibfnamefont {A.}~\bibnamefont {Barducci}}, \bibinfo {author}
  {\bibfnamefont {M.}~\bibnamefont {Bernetti}}, \bibinfo {author}
  {\bibfnamefont {P.~G.}\ \bibnamefont {Bolhuis}}, \bibinfo {author}
  {\bibfnamefont {S.}~\bibnamefont {Bottaro}}, \bibinfo {author} {\bibfnamefont
  {D.}~\bibnamefont {Branduardi}}, \bibinfo {author} {\bibfnamefont
  {R.}~\bibnamefont {Capelli}}, \bibinfo {author} {\bibfnamefont
  {P.}~\bibnamefont {Carloni}}, \bibinfo {author} {\bibfnamefont
  {M.}~\bibnamefont {Ceriotti}}, \bibinfo {author} {\bibfnamefont
  {A.}~\bibnamefont {Cesari}}, \bibinfo {author} {\bibfnamefont
  {H.}~\bibnamefont {Chen}}, \bibinfo {author} {\bibfnamefont {W.}~\bibnamefont
  {Chen}}, \bibinfo {author} {\bibfnamefont {F.}~\bibnamefont {Colizzi}},
  \bibinfo {author} {\bibfnamefont {S.}~\bibnamefont {De}}, \bibinfo {author}
  {\bibfnamefont {M.~D.~L.}\ \bibnamefont {Pierre}}, \bibinfo {author}
  {\bibfnamefont {D.}~\bibnamefont {Donadio}}, \bibinfo {author} {\bibfnamefont
  {V.}~\bibnamefont {Drobot}}, \bibinfo {author} {\bibfnamefont
  {B.}~\bibnamefont {Ensing}}, \bibinfo {author} {\bibfnamefont {A.~L.}\
  \bibnamefont {Ferguson}}, \bibinfo {author} {\bibfnamefont {M.}~\bibnamefont
  {Filizola}}, \bibinfo {author} {\bibfnamefont {J.~S.}\ \bibnamefont
  {Fraser}}, \bibinfo {author} {\bibfnamefont {H.}~\bibnamefont {Fu}}, \bibinfo
  {author} {\bibfnamefont {P.}~\bibnamefont {Gasparotto}}, \bibinfo {author}
  {\bibfnamefont {F.~L.}\ \bibnamefont {Gervasio}}, \bibinfo {author}
  {\bibfnamefont {F.}~\bibnamefont {Giberti}}, \bibinfo {author} {\bibfnamefont
  {A.}~\bibnamefont {Gil-Ley}}, \bibinfo {author} {\bibfnamefont
  {T.}~\bibnamefont {Giorgino}}, \bibinfo {author} {\bibfnamefont {G.~T.}\
  \bibnamefont {Heller}}, \bibinfo {author} {\bibfnamefont {G.~M.}\
  \bibnamefont {Hocky}}, \bibinfo {author} {\bibfnamefont {M.}~\bibnamefont
  {Iannuzzi}}, \bibinfo {author} {\bibfnamefont {M.}~\bibnamefont
  {Invernizzi}}, \bibinfo {author} {\bibfnamefont {K.~E.}\ \bibnamefont
  {Jelfs}}, \bibinfo {author} {\bibfnamefont {A.}~\bibnamefont {Jussupow}},
  \bibinfo {author} {\bibfnamefont {E.}~\bibnamefont {Kirilin}}, \bibinfo
  {author} {\bibfnamefont {A.}~\bibnamefont {Laio}}, \bibinfo {author}
  {\bibfnamefont {V.}~\bibnamefont {Limongelli}}, \bibinfo {author}
  {\bibfnamefont {K.}~\bibnamefont {Lindorff-Larsen}}, \bibinfo {author}
  {\bibfnamefont {T.}~\bibnamefont {L\"{o}hr}}, \bibinfo {author}
  {\bibfnamefont {F.}~\bibnamefont {Marinelli}}, \bibinfo {author}
  {\bibfnamefont {L.}~\bibnamefont {Martin-Samos}}, \bibinfo {author}
  {\bibfnamefont {M.}~\bibnamefont {Masetti}}, \bibinfo {author} {\bibfnamefont
  {R.}~\bibnamefont {Meyer}}, \bibinfo {author} {\bibfnamefont
  {A.}~\bibnamefont {Michaelides}}, \bibinfo {author} {\bibfnamefont
  {C.}~\bibnamefont {Molteni}}, \bibinfo {author} {\bibfnamefont
  {T.}~\bibnamefont {Morishita}}, \bibinfo {author} {\bibfnamefont
  {M.}~\bibnamefont {Nava}}, \bibinfo {author} {\bibfnamefont {C.}~\bibnamefont
  {Paissoni}}, \bibinfo {author} {\bibfnamefont {E.}~\bibnamefont {Papaleo}},
  \bibinfo {author} {\bibfnamefont {M.}~\bibnamefont {Parrinello}}, \bibinfo
  {author} {\bibfnamefont {J.}~\bibnamefont {Pfaendtner}}, \bibinfo {author}
  {\bibfnamefont {P.}~\bibnamefont {Piaggi}}, \bibinfo {author} {\bibfnamefont
  {G.}~\bibnamefont {Piccini}}, \bibinfo {author} {\bibfnamefont
  {A.}~\bibnamefont {Pietropaolo}}, \bibinfo {author} {\bibfnamefont
  {F.}~\bibnamefont {Pietrucci}}, \bibinfo {author} {\bibfnamefont
  {S.}~\bibnamefont {Pipolo}}, \bibinfo {author} {\bibfnamefont
  {D.}~\bibnamefont {Provasi}}, \bibinfo {author} {\bibfnamefont
  {D.}~\bibnamefont {Quigley}}, \bibinfo {author} {\bibfnamefont
  {P.}~\bibnamefont {Raiteri}}, \bibinfo {author} {\bibfnamefont
  {S.}~\bibnamefont {Raniolo}}, \bibinfo {author} {\bibfnamefont
  {J.}~\bibnamefont {Rydzewski}}, \bibinfo {author} {\bibfnamefont
  {M.}~\bibnamefont {Salvalaglio}}, \bibinfo {author} {\bibfnamefont {G.~C.}\
  \bibnamefont {Sosso}}, \bibinfo {author} {\bibfnamefont {V.}~\bibnamefont
  {Spiwok}}, \bibinfo {author} {\bibfnamefont {J.}~\bibnamefont {\v{S}poner}},
  \bibinfo {author} {\bibfnamefont {D.~W.~H.}\ \bibnamefont {Swenson}},
  \bibinfo {author} {\bibfnamefont {P.}~\bibnamefont {Tiwary}}, \bibinfo
  {author} {\bibfnamefont {O.}~\bibnamefont {Valsson}}, \bibinfo {author}
  {\bibfnamefont {M.}~\bibnamefont {Vendruscolo}}, \bibinfo {author}
  {\bibfnamefont {G.~A.}\ \bibnamefont {Voth}}, \bibinfo {author}
  {\bibfnamefont {A.}~\bibnamefont {White}},\ and\ \bibinfo {author}
  {\bibfnamefont {T.~P.}\ \bibnamefont {consortium}},\ }\bibfield  {title}
  {\bibinfo {title} {{Promoting transparency and reproducibility in enhanced
  molecular simulations}},\ }\href {https://doi.org/10.1038/s41592-019-0506-8}
  {\bibfield  {journal} {\bibinfo  {journal} {Nature Methods}\ }\textbf
  {\bibinfo {volume} {16}},\ \bibinfo {pages} {670} (\bibinfo {year}
  {2019})}\BibitemShut {NoStop}%
\bibitem [{\citenamefont {Tribello}\ \emph {et~al.}(2014)\citenamefont
  {Tribello}, \citenamefont {Bonomi}, \citenamefont {Branduardi}, \citenamefont
  {Camilloni},\ and\ \citenamefont {Bussi}}]{Tribello2014}%
  \BibitemOpen
  \bibfield  {author} {\bibinfo {author} {\bibfnamefont {G.~A.}\ \bibnamefont
  {Tribello}}, \bibinfo {author} {\bibfnamefont {M.}~\bibnamefont {Bonomi}},
  \bibinfo {author} {\bibfnamefont {D.}~\bibnamefont {Branduardi}}, \bibinfo
  {author} {\bibfnamefont {C.}~\bibnamefont {Camilloni}},\ and\ \bibinfo
  {author} {\bibfnamefont {G.}~\bibnamefont {Bussi}},\ }\bibfield  {title}
  {\bibinfo {title} {{PLUMED 2: New feathers for an old bird}},\ }\href
  {https://doi.org/10.1016/j.cpc.2013.09.018} {\bibfield  {journal} {\bibinfo
  {journal} {Computer Physics Communications}\ }\textbf {\bibinfo {volume}
  {185}},\ \bibinfo {pages} {604} (\bibinfo {year} {2014})}\BibitemShut
  {NoStop}%
\bibitem [{\citenamefont {Kirkwood}\ and\ \citenamefont
  {Buff}(1951)}]{Kirkwood1951}%
  \BibitemOpen
  \bibfield  {author} {\bibinfo {author} {\bibfnamefont {J.~G.}\ \bibnamefont
  {Kirkwood}}\ and\ \bibinfo {author} {\bibfnamefont {F.~P.}\ \bibnamefont
  {Buff}},\ }\bibfield  {title} {\bibinfo {title} {{The Statistical Mechanical
  Theory of Solutions. I}},\ }\href {https://doi.org/10.1063/1.1748352}
  {\bibfield  {journal} {\bibinfo  {journal} {The Journal of Chemical Physics}\
  }\textbf {\bibinfo {volume} {19}},\ \bibinfo {pages} {774} (\bibinfo {year}
  {1951})}\BibitemShut {NoStop}%
\bibitem [{\citenamefont {Karwounopoulos}\ \emph {et~al.}(2024)\citenamefont
  {Karwounopoulos}, \citenamefont {Wu}, \citenamefont {Tkaczyk}, \citenamefont
  {Wang}, \citenamefont {Baskerville}, \citenamefont {Ranasinghe},
  \citenamefont {Langer}, \citenamefont {Wood}, \citenamefont {Wieder},\ and\
  \citenamefont {Boresch}}]{Karwounopoulos2024}%
  \BibitemOpen
  \bibfield  {author} {\bibinfo {author} {\bibfnamefont {J.}~\bibnamefont
  {Karwounopoulos}}, \bibinfo {author} {\bibfnamefont {Z.}~\bibnamefont {Wu}},
  \bibinfo {author} {\bibfnamefont {S.}~\bibnamefont {Tkaczyk}}, \bibinfo
  {author} {\bibfnamefont {S.}~\bibnamefont {Wang}}, \bibinfo {author}
  {\bibfnamefont {A.}~\bibnamefont {Baskerville}}, \bibinfo {author}
  {\bibfnamefont {K.}~\bibnamefont {Ranasinghe}}, \bibinfo {author}
  {\bibfnamefont {T.}~\bibnamefont {Langer}}, \bibinfo {author} {\bibfnamefont
  {G.~P.~F.}\ \bibnamefont {Wood}}, \bibinfo {author} {\bibfnamefont
  {M.}~\bibnamefont {Wieder}},\ and\ \bibinfo {author} {\bibfnamefont
  {S.}~\bibnamefont {Boresch}},\ }\bibfield  {title} {\bibinfo {title}
  {{Insights and Challenges in Correcting Force Field Based Solvation Free
  Energies Using a Neural Network Potential}},\ }\href
  {https://doi.org/10.1021/acs.jpcb.4c01417} {\bibfield  {journal} {\bibinfo
  {journal} {The Journal of Physical Chemistry B}\ }\textbf {\bibinfo {volume}
  {128}},\ \bibinfo {pages} {6693} (\bibinfo {year} {2024})}\BibitemShut
  {NoStop}%
\bibitem [{\citenamefont {Picha}\ \emph {et~al.}(2025)\citenamefont {Picha},
  \citenamefont {Tkaczyk}, \citenamefont {Langer}, \citenamefont {Wieder},\
  and\ \citenamefont {Boresch}}]{Picha2025}%
  \BibitemOpen
  \bibfield  {author} {\bibinfo {author} {\bibfnamefont {A.~K.}\ \bibnamefont
  {Picha}}, \bibinfo {author} {\bibfnamefont {S.}~\bibnamefont {Tkaczyk}},
  \bibinfo {author} {\bibfnamefont {T.}~\bibnamefont {Langer}}, \bibinfo
  {author} {\bibfnamefont {M.}~\bibnamefont {Wieder}},\ and\ \bibinfo {author}
  {\bibfnamefont {S.}~\bibnamefont {Boresch}},\ }\bibfield  {title} {\bibinfo
  {title} {{Architecture-Independent Absolute Solvation Free Energy
  Calculations with Neural Network Potentials}},\ }\href
  {https://doi.org/10.1021/acs.jpclett.5c02980} {\bibfield  {journal} {\bibinfo
   {journal} {The Journal of Physical Chemistry Letters}\ }\textbf {\bibinfo
  {volume} {16}},\ \bibinfo {pages} {12080} (\bibinfo {year}
  {2025})}\BibitemShut {NoStop}%
\bibitem [{\citenamefont {Bolhuis}\ \emph {et~al.}(2002)\citenamefont
  {Bolhuis}, \citenamefont {Chandler}, \citenamefont {Dellago},\ and\
  \citenamefont {Geissler}}]{Bolhuis2002}%
  \BibitemOpen
  \bibfield  {author} {\bibinfo {author} {\bibfnamefont {P.~G.}\ \bibnamefont
  {Bolhuis}}, \bibinfo {author} {\bibfnamefont {D.}~\bibnamefont {Chandler}},
  \bibinfo {author} {\bibfnamefont {C.}~\bibnamefont {Dellago}},\ and\ \bibinfo
  {author} {\bibfnamefont {P.~L.}\ \bibnamefont {Geissler}},\ }\bibfield
  {title} {\bibinfo {title} {{Transition Path Sampling: Throwing Ropes Over
  Rough Mountain Passes, in the Dark}},\ }\href
  {https://doi.org/10.1146/annurev.physchem.53.082301.113146} {\bibfield
  {journal} {\bibinfo  {journal} {Annual Review of Physical Chemistry}\
  }\textbf {\bibinfo {volume} {53}},\ \bibinfo {pages} {291} (\bibinfo {year}
  {2002})}\BibitemShut {NoStop}%
\bibitem [{\citenamefont {van Erp}\ \emph {et~al.}(2003)\citenamefont {van
  Erp}, \citenamefont {Moroni},\ and\ \citenamefont {Bolhuis}}]{vanErp2003}%
  \BibitemOpen
  \bibfield  {author} {\bibinfo {author} {\bibfnamefont {T.~S.}\ \bibnamefont
  {van Erp}}, \bibinfo {author} {\bibfnamefont {D.}~\bibnamefont {Moroni}},\
  and\ \bibinfo {author} {\bibfnamefont {P.~G.}\ \bibnamefont {Bolhuis}},\
  }\bibfield  {title} {\bibinfo {title} {{A novel path sampling method for the
  calculation of rate constants}},\ }\href {https://doi.org/10.1063/1.1562614}
  {\bibfield  {journal} {\bibinfo  {journal} {The Journal of Chemical Physics}\
  }\textbf {\bibinfo {volume} {118}},\ \bibinfo {pages} {7762} (\bibinfo {year}
  {2003})}\BibitemShut {NoStop}%
\bibitem [{\citenamefont {van Erp}(2007)}]{vanErp2007}%
  \BibitemOpen
  \bibfield  {author} {\bibinfo {author} {\bibfnamefont {T.~S.}\ \bibnamefont
  {van Erp}},\ }\bibfield  {title} {\bibinfo {title} {{Reaction Rate
  Calculation by Parallel Path Swapping}},\ }\href
  {https://doi.org/10.1103/PhysRevLett.98.268301} {\bibfield  {journal}
  {\bibinfo  {journal} {Physical Review Letters}\ }\textbf {\bibinfo {volume}
  {98}},\ \bibinfo {pages} {268301} (\bibinfo {year} {2007})}\BibitemShut
  {NoStop}%
\bibitem [{\citenamefont {Kumar}\ \emph {et~al.}(1992)\citenamefont {Kumar},
  \citenamefont {Rosenberg}, \citenamefont {Bouzida}, \citenamefont
  {Swendsen},\ and\ \citenamefont {Kollman}}]{Kumar1992}%
  \BibitemOpen
  \bibfield  {author} {\bibinfo {author} {\bibfnamefont {S.}~\bibnamefont
  {Kumar}}, \bibinfo {author} {\bibfnamefont {J.~M.}\ \bibnamefont
  {Rosenberg}}, \bibinfo {author} {\bibfnamefont {D.}~\bibnamefont {Bouzida}},
  \bibinfo {author} {\bibfnamefont {R.~H.}\ \bibnamefont {Swendsen}},\ and\
  \bibinfo {author} {\bibfnamefont {P.~A.}\ \bibnamefont {Kollman}},\
  }\bibfield  {title} {\bibinfo {title} {{THE weighted histogram analysis
  method for free-energy calculations on biomolecules. I. The method}},\ }\href
  {https://doi.org/10.1002/jcc.540130812} {\bibfield  {journal} {\bibinfo
  {journal} {Journal of Computational Chemistry}\ }\textbf {\bibinfo {volume}
  {13}},\ \bibinfo {pages} {1011} (\bibinfo {year} {1992})}\BibitemShut
  {NoStop}%
\bibitem [{\citenamefont {Marcus}(1988)}]{Marcus1988}%
  \BibitemOpen
  \bibfield  {author} {\bibinfo {author} {\bibfnamefont {Y.}~\bibnamefont
  {Marcus}},\ }\bibfield  {title} {\bibinfo {title} {{Ionic radii in aqueous
  solutions}},\ }\href {https://doi.org/10.1021/cr00090a003} {\bibfield
  {journal} {\bibinfo  {journal} {Chemical Reviews}\ }\textbf {\bibinfo
  {volume} {88}},\ \bibinfo {pages} {1475} (\bibinfo {year}
  {1988})}\BibitemShut {NoStop}%
\bibitem [{\citenamefont {Marcus}(1997)}]{Marcus1997}%
  \BibitemOpen
  \bibfield  {author} {\bibinfo {author} {\bibfnamefont {Y.}~\bibnamefont
  {Marcus}},\ }\href@noop {} {\emph {\bibinfo {title} {{Ion Properties}}}}\
  (\bibinfo  {publisher} {Marcel Dekker},\ \bibinfo {year} {1997})\BibitemShut
  {NoStop}%
\bibitem [{\citenamefont {Robinson}\ and\ \citenamefont
  {Stokes}(2002)}]{Robinson2002}%
  \BibitemOpen
  \bibfield  {author} {\bibinfo {author} {\bibfnamefont {R.~A.}\ \bibnamefont
  {Robinson}}\ and\ \bibinfo {author} {\bibfnamefont {R.~H.}\ \bibnamefont
  {Stokes}},\ }\href@noop {} {\emph {\bibinfo {title} {{Electrolyte
  Solutions}}}},\ \bibinfo {edition} {2nd}\ ed.\ (\bibinfo  {publisher} {Dover
  Publications},\ \bibinfo {year} {2002})\BibitemShut {NoStop}%
\bibitem [{\citenamefont {Bleuzen}\ \emph {et~al.}(1997)\citenamefont
  {Bleuzen}, \citenamefont {Pittet}, \citenamefont {Helm},\ and\ \citenamefont
  {Merbach}}]{Bleuzen1997}%
  \BibitemOpen
  \bibfield  {author} {\bibinfo {author} {\bibfnamefont {A.}~\bibnamefont
  {Bleuzen}}, \bibinfo {author} {\bibfnamefont {P.-A.}\ \bibnamefont {Pittet}},
  \bibinfo {author} {\bibfnamefont {L.}~\bibnamefont {Helm}},\ and\ \bibinfo
  {author} {\bibfnamefont {A.~E.}\ \bibnamefont {Merbach}},\ }\bibfield
  {title} {\bibinfo {title} {{Water exchange on magnesium(II) in aqueous
  solution: a variable temperature and pressure17O NMR study}},\ }\href
  {https://doi.org/10.1002/(SICI)1097-458X(199711)35:11<765::AID-OMR169>3.0.CO;2-F}
  {\bibfield  {journal} {\bibinfo  {journal} {Magnetic Resonance in Chemistry}\
  }\textbf {\bibinfo {volume} {35}},\ \bibinfo {pages} {765} (\bibinfo {year}
  {1997})}\BibitemShut {NoStop}%
\bibitem [{\citenamefont {Neely}\ and\ \citenamefont
  {Connick}(1970)}]{Neely1970}%
  \BibitemOpen
  \bibfield  {author} {\bibinfo {author} {\bibfnamefont {J.}~\bibnamefont
  {Neely}}\ and\ \bibinfo {author} {\bibfnamefont {R.}~\bibnamefont
  {Connick}},\ }\bibfield  {title} {\bibinfo {title} {{Rate of water exchange
  from hydrated magnesium ion}},\ }\href {https://doi.org/10.1021/ja00714a048}
  {\bibfield  {journal} {\bibinfo  {journal} {Journal of the American Chemical
  Society}\ }\textbf {\bibinfo {volume} {92}},\ \bibinfo {pages} {3476}
  (\bibinfo {year} {1970})}\BibitemShut {NoStop}%
\bibitem [{\citenamefont {Helm}\ and\ \citenamefont
  {Merbach}(1999)}]{Helm1999}%
  \BibitemOpen
  \bibfield  {author} {\bibinfo {author} {\bibfnamefont {L.}~\bibnamefont
  {Helm}}\ and\ \bibinfo {author} {\bibfnamefont {A.}~\bibnamefont {Merbach}},\
  }\bibfield  {title} {\bibinfo {title} {{Water exchange on metal ions:
  experiments and simulations}},\ }\href
  {https://doi.org/10.1016/S0010-8545(99)90232-1} {\bibfield  {journal}
  {\bibinfo  {journal} {Coordination Chemistry Reviews}\ }\textbf {\bibinfo
  {volume} {187}},\ \bibinfo {pages} {151} (\bibinfo {year}
  {1999})}\BibitemShut {NoStop}%
\bibitem [{\citenamefont {Mondal}\ \emph {et~al.}(2023)\citenamefont {Mondal},
  \citenamefont {Kussainova}, \citenamefont {Yue},\ and\ \citenamefont
  {Panagiotopoulos}}]{Mondal2023}%
  \BibitemOpen
  \bibfield  {author} {\bibinfo {author} {\bibfnamefont {A.}~\bibnamefont
  {Mondal}}, \bibinfo {author} {\bibfnamefont {D.}~\bibnamefont {Kussainova}},
  \bibinfo {author} {\bibfnamefont {S.}~\bibnamefont {Yue}},\ and\ \bibinfo
  {author} {\bibfnamefont {A.~Z.}\ \bibnamefont {Panagiotopoulos}},\ }\bibfield
   {title} {\bibinfo {title} {{Modeling Chemical Reactions in Alkali
  Carbonate–Hydroxide Electrolytes with Deep Learning Potentials}},\ }\href
  {https://doi.org/10.1021/acs.jctc.2c00816} {\bibfield  {journal} {\bibinfo
  {journal} {Journal of Chemical Theory and Computation}\ }\textbf {\bibinfo
  {volume} {19}},\ \bibinfo {pages} {4584} (\bibinfo {year}
  {2023})}\BibitemShut {NoStop}%
\bibitem [{\citenamefont {Park}\ \emph {et~al.}(2025)\citenamefont {Park},
  \citenamefont {Ryu},\ and\ \citenamefont {Lee}}]{Park2025}%
  \BibitemOpen
  \bibfield  {author} {\bibinfo {author} {\bibfnamefont {A.}~\bibnamefont
  {Park}}, \bibinfo {author} {\bibfnamefont {J.}~\bibnamefont {Ryu}},\ and\
  \bibinfo {author} {\bibfnamefont {W.~B.}\ \bibnamefont {Lee}},\ }\bibfield
  {title} {\bibinfo {title} {{Ionic Liquid Molecular Dynamics Simulation with
  Machine Learning Force Fields: DPMD and MACE}},\ }\href@noop {} {\bibfield
  {journal} {\bibinfo  {journal} {arXiv:2503.18249}\ } (\bibinfo {year}
  {2025})}\BibitemShut {NoStop}%
\bibitem [{\citenamefont {Jiao}\ \emph {et~al.}(2006)\citenamefont {Jiao},
  \citenamefont {King}, \citenamefont {Grossfield}, \citenamefont {Darden},\
  and\ \citenamefont {Ren}}]{Jiao2006}%
  \BibitemOpen
  \bibfield  {author} {\bibinfo {author} {\bibfnamefont {D.}~\bibnamefont
  {Jiao}}, \bibinfo {author} {\bibfnamefont {C.}~\bibnamefont {King}}, \bibinfo
  {author} {\bibfnamefont {A.}~\bibnamefont {Grossfield}}, \bibinfo {author}
  {\bibfnamefont {T.~A.}\ \bibnamefont {Darden}},\ and\ \bibinfo {author}
  {\bibfnamefont {P.}~\bibnamefont {Ren}},\ }\bibfield  {title} {\bibinfo
  {title} {{Simulation of Ca 2+ and Mg 2+ Solvation Using Polarizable Atomic
  Multipole Potential}},\ }\href {https://doi.org/10.1021/jp062230r} {\bibfield
   {journal} {\bibinfo  {journal} {The Journal of Physical Chemistry B}\
  }\textbf {\bibinfo {volume} {110}},\ \bibinfo {pages} {18553} (\bibinfo
  {year} {2006})}\BibitemShut {NoStop}%
\bibitem [{\citenamefont {Yeh}\ and\ \citenamefont {Hummer}(2004)}]{Yeh2004}%
  \BibitemOpen
  \bibfield  {author} {\bibinfo {author} {\bibfnamefont {I.-C.}\ \bibnamefont
  {Yeh}}\ and\ \bibinfo {author} {\bibfnamefont {G.}~\bibnamefont {Hummer}},\
  }\bibfield  {title} {\bibinfo {title} {{System-Size Dependence of Diffusion
  Coefficients and Viscosities from Molecular Dynamics Simulations with
  Periodic Boundary Conditions}},\ }\href {https://doi.org/10.1021/jp0477147}
  {\bibfield  {journal} {\bibinfo  {journal} {The Journal of Physical Chemistry
  B}\ }\textbf {\bibinfo {volume} {108}},\ \bibinfo {pages} {15873} (\bibinfo
  {year} {2004})}\BibitemShut {NoStop}%
\bibitem [{\citenamefont {Falkner}\ and\ \citenamefont
  {Schwierz}(2021)}]{Falkner2021}%
  \BibitemOpen
  \bibfield  {author} {\bibinfo {author} {\bibfnamefont {S.}~\bibnamefont
  {Falkner}}\ and\ \bibinfo {author} {\bibfnamefont {N.}~\bibnamefont
  {Schwierz}},\ }\bibfield  {title} {\bibinfo {title} {{Kinetic pathways of
  water exchange in the first hydration shell of magnesium: Influence of water
  model and ionic force field}},\ }\href {https://doi.org/10.1063/5.0060896}
  {\bibfield  {journal} {\bibinfo  {journal} {The Journal of Chemical Physics}\
  }\textbf {\bibinfo {volume} {155}},\ \bibinfo {pages} {084503} (\bibinfo
  {year} {2021})}\BibitemShut {NoStop}%
\bibitem [{\citenamefont {Langford}(1965)}]{CHLangford1965}%
  \BibitemOpen
  \bibfield  {author} {\bibinfo {author} {\bibfnamefont {H.~B. G. C.~H.}\
  \bibnamefont {Langford}},\ }\href@noop {} {\emph {\bibinfo {title} {{Ligand
  Substitution Processes}}}}\ (\bibinfo  {publisher} {W. A. Benjamin, Inc.},\
  \bibinfo {year} {1965})\BibitemShut {NoStop}%
\bibitem [{\citenamefont {Schwaab}\ \emph {et~al.}(2019)\citenamefont
  {Schwaab}, \citenamefont {Sebastiani},\ and\ \citenamefont
  {Havenith}}]{Schwaab2019}%
  \BibitemOpen
  \bibfield  {author} {\bibinfo {author} {\bibfnamefont {G.}~\bibnamefont
  {Schwaab}}, \bibinfo {author} {\bibfnamefont {F.}~\bibnamefont
  {Sebastiani}},\ and\ \bibinfo {author} {\bibfnamefont {M.}~\bibnamefont
  {Havenith}},\ }\bibfield  {title} {\bibinfo {title} {{Ion Hydration and Ion
  Pairing as Probed by THz Spectroscopy}},\ }\href
  {https://doi.org/10.1002/anie.201805261} {\bibfield  {journal} {\bibinfo
  {journal} {Angewandte Chemie International Edition}\ }\textbf {\bibinfo
  {volume} {58}},\ \bibinfo {pages} {3000} (\bibinfo {year}
  {2019})}\BibitemShut {NoStop}%
\bibitem [{\citenamefont {Anstine}\ and\ \citenamefont
  {Isayev}(2023)}]{Anstine2023}%
  \BibitemOpen
  \bibfield  {author} {\bibinfo {author} {\bibfnamefont {D.~M.}\ \bibnamefont
  {Anstine}}\ and\ \bibinfo {author} {\bibfnamefont {O.}~\bibnamefont
  {Isayev}},\ }\bibfield  {title} {\bibinfo {title} {{Machine Learning
  Interatomic Potentials and Long-Range Physics}},\ }\href
  {https://doi.org/10.1021/acs.jpca.2c06778} {\bibfield  {journal} {\bibinfo
  {journal} {The Journal of Physical Chemistry A}\ }\textbf {\bibinfo {volume}
  {127}},\ \bibinfo {pages} {2417} (\bibinfo {year} {2023})}\BibitemShut
  {NoStop}%
\bibitem [{\citenamefont {Maruf}\ \emph {et~al.}(2025)\citenamefont {Maruf},
  \citenamefont {Kim},\ and\ \citenamefont {Ahmad}}]{Maruf2025}%
  \BibitemOpen
  \bibfield  {author} {\bibinfo {author} {\bibfnamefont {M.~U.}\ \bibnamefont
  {Maruf}}, \bibinfo {author} {\bibfnamefont {S.}~\bibnamefont {Kim}},\ and\
  \bibinfo {author} {\bibfnamefont {Z.}~\bibnamefont {Ahmad}},\ }\bibfield
  {title} {\bibinfo {title} {{Learning Long-Range Interactions in Equivariant
  Machine Learning Interatomic Potentials via Electronic Degrees of Freedom}},\
  }\href {https://doi.org/10.1021/acs.jpclett.5c02352} {\bibfield  {journal}
  {\bibinfo  {journal} {The Journal of Physical Chemistry Letters}\ }\textbf
  {\bibinfo {volume} {16}},\ \bibinfo {pages} {9078} (\bibinfo {year}
  {2025})}\BibitemShut {NoStop}%
\bibitem [{\citenamefont {Ko}\ \emph {et~al.}(2021)\citenamefont {Ko},
  \citenamefont {Finkler}, \citenamefont {Goedecker},\ and\ \citenamefont
  {Behler}}]{Ko2021}%
  \BibitemOpen
  \bibfield  {author} {\bibinfo {author} {\bibfnamefont {T.~W.}\ \bibnamefont
  {Ko}}, \bibinfo {author} {\bibfnamefont {J.~A.}\ \bibnamefont {Finkler}},
  \bibinfo {author} {\bibfnamefont {S.}~\bibnamefont {Goedecker}},\ and\
  \bibinfo {author} {\bibfnamefont {J.}~\bibnamefont {Behler}},\ }\bibfield
  {title} {\bibinfo {title} {{A fourth-generation high-dimensional neural
  network potential with accurate electrostatics including non-local charge
  transfer}},\ }\href {https://doi.org/10.1038/s41467-020-20427-2} {\bibfield
  {journal} {\bibinfo  {journal} {Nature Communications}\ }\textbf {\bibinfo
  {volume} {12}},\ \bibinfo {pages} {398} (\bibinfo {year} {2021})}\BibitemShut
  {NoStop}%
\bibitem [{\citenamefont {Vondr\'{a}k}\ \emph {et~al.}(2026)\citenamefont
  {Vondr\'{a}k}, \citenamefont {Baldwin}, \citenamefont {Cs\'{a}nyi},
  \citenamefont {Reuter},\ and\ \citenamefont {Margraf}}]{Vondrk2026}%
  \BibitemOpen
  \bibfield  {author} {\bibinfo {author} {\bibfnamefont {M.}~\bibnamefont
  {Vondr\'{a}k}}, \bibinfo {author} {\bibfnamefont {W.~J.}\ \bibnamefont
  {Baldwin}}, \bibinfo {author} {\bibfnamefont {G.}~\bibnamefont {Cs\'{a}nyi}},
  \bibinfo {author} {\bibfnamefont {K.}~\bibnamefont {Reuter}},\ and\ \bibinfo
  {author} {\bibfnamefont {J.~T.}\ \bibnamefont {Margraf}},\ }\href
  {https://doi.org/10.26434/chemrxiv.15000377/v1} {\bibinfo {title}
  {{Integrating Charge Equilibration with Equivariant Machine-Learning
  Interatomic Potentials}}} (\bibinfo {year} {2026})\BibitemShut {NoStop}%
\end{thebibliography}

\begin{thebibliography}{26}%
\makeatletter
\providecommand \@ifxundefined [1]{%
 \@ifx{#1\undefined}
}%
\providecommand \@ifnum [1]{%
 \ifnum #1\expandafter \@firstoftwo
 \else \expandafter \@secondoftwo
 \fi
}%
\providecommand \@ifx [1]{%
 \ifx #1\expandafter \@firstoftwo
 \else \expandafter \@secondoftwo
 \fi
}%
\providecommand \natexlab [1]{#1}%
\providecommand \enquote  [1]{``#1''}%
\providecommand \bibnamefont  [1]{#1}%
\providecommand \bibfnamefont [1]{#1}%
\providecommand \citenamefont [1]{#1}%
\providecommand \href@noop [0]{\@secondoftwo}%
\providecommand \href [0]{\begingroup \@sanitize@url \@href}%
\providecommand \@href[1]{\@@startlink{#1}\@@href}%
\providecommand \@@href[1]{\endgroup#1\@@endlink}%
\providecommand \@sanitize@url [0]{\catcode `\\12\catcode `\$12\catcode
  `\&12\catcode `\#12\catcode `\^12\catcode `\_12\catcode `\%12\relax}%
\providecommand \@@startlink[1]{}%
\providecommand \@@endlink[0]{}%
\providecommand \url  [0]{\begingroup\@sanitize@url \@url }%
\providecommand \@url [1]{\endgroup\@href {#1}{\urlprefix }}%
\providecommand \urlprefix  [0]{URL }%
\providecommand \Eprint [0]{\href }%
\providecommand \doibase [0]{https://doi.org/}%
\providecommand \selectlanguage [0]{\@gobble}%
\providecommand \bibinfo  [0]{\@secondoftwo}%
\providecommand \bibfield  [0]{\@secondoftwo}%
\providecommand \translation [1]{[#1]}%
\providecommand \BibitemOpen [0]{}%
\providecommand \bibitemStop [0]{}%
\providecommand \bibitemNoStop [0]{.\EOS\space}%
\providecommand \EOS [0]{\spacefactor3000\relax}%
\providecommand \BibitemShut  [1]{\csname bibitem#1\endcsname}%
\let\auto@bib@innerbib\@empty
\bibitem [{\citenamefont {Abraham}\ \emph {et~al.}(2015)\citenamefont
  {Abraham}, \citenamefont {Murtola}, \citenamefont {Schulz}, \citenamefont
  {P\'{a}ll}, \citenamefont {Smith}, \citenamefont {Hess},\ and\ \citenamefont
  {Lindahl}}]{SAbraham2015}%
  \BibitemOpen
  \bibfield  {author} {\bibinfo {author} {\bibfnamefont {M.~J.}\ \bibnamefont
  {Abraham}}, \bibinfo {author} {\bibfnamefont {T.}~\bibnamefont {Murtola}},
  \bibinfo {author} {\bibfnamefont {R.}~\bibnamefont {Schulz}}, \bibinfo
  {author} {\bibfnamefont {S.}~\bibnamefont {P\'{a}ll}}, \bibinfo {author}
  {\bibfnamefont {J.~C.}\ \bibnamefont {Smith}}, \bibinfo {author}
  {\bibfnamefont {B.}~\bibnamefont {Hess}},\ and\ \bibinfo {author}
  {\bibfnamefont {E.}~\bibnamefont {Lindahl}},\ }\bibfield  {title} {\bibinfo
  {title} {{GROMACS: High performance molecular simulations through multi-level
  parallelism from laptops to supercomputers}},\ }\href
  {https://doi.org/10.1016/j.softx.2015.06.001} {\bibfield  {journal} {\bibinfo
   {journal} {SoftwareX}\ }\textbf {\bibinfo {volume} {1-2}},\ \bibinfo {pages}
  {19} (\bibinfo {year} {2015})}\BibitemShut {NoStop}%
\bibitem [{\citenamefont {Mamatkulov}\ and\ \citenamefont
  {Schwierz}(2018)}]{SMamatkulov2018}%
  \BibitemOpen
  \bibfield  {author} {\bibinfo {author} {\bibfnamefont {S.}~\bibnamefont
  {Mamatkulov}}\ and\ \bibinfo {author} {\bibfnamefont {N.}~\bibnamefont
  {Schwierz}},\ }\bibfield  {title} {\bibinfo {title} {{Force fields for
  monovalent and divalent metal cations in TIP3P water based on thermodynamic
  and kinetic properties}},\ }\href {https://doi.org/10.1063/1.5017694}
  {\bibfield  {journal} {\bibinfo  {journal} {The Journal of Chemical Physics}\
  }\textbf {\bibinfo {volume} {148}},\ \bibinfo {pages} {074504} (\bibinfo
  {year} {2018})}\BibitemShut {NoStop}%
\bibitem [{\citenamefont {Jorgensen}\ \emph {et~al.}(1983)\citenamefont
  {Jorgensen}, \citenamefont {Chandrasekhar}, \citenamefont {Madura},
  \citenamefont {Impey},\ and\ \citenamefont {Klein}}]{SJorgensen1983}%
  \BibitemOpen
  \bibfield  {author} {\bibinfo {author} {\bibfnamefont {W.~L.}\ \bibnamefont
  {Jorgensen}}, \bibinfo {author} {\bibfnamefont {J.}~\bibnamefont
  {Chandrasekhar}}, \bibinfo {author} {\bibfnamefont {J.~D.}\ \bibnamefont
  {Madura}}, \bibinfo {author} {\bibfnamefont {R.~W.}\ \bibnamefont {Impey}},\
  and\ \bibinfo {author} {\bibfnamefont {M.~L.}\ \bibnamefont {Klein}},\
  }\bibfield  {title} {\bibinfo {title} {{Comparison of simple potential
  functions for simulating liquid water}},\ }\href
  {https://doi.org/10.1063/1.445869} {\bibfield  {journal} {\bibinfo  {journal}
  {The Journal of Chemical Physics}\ }\textbf {\bibinfo {volume} {79}},\
  \bibinfo {pages} {926} (\bibinfo {year} {1983})}\BibitemShut {NoStop}%
\bibitem [{\citenamefont {Darden}\ \emph {et~al.}(1993)\citenamefont {Darden},
  \citenamefont {York},\ and\ \citenamefont {Pedersen}}]{SDarden1993}%
  \BibitemOpen
  \bibfield  {author} {\bibinfo {author} {\bibfnamefont {T.}~\bibnamefont
  {Darden}}, \bibinfo {author} {\bibfnamefont {D.}~\bibnamefont {York}},\ and\
  \bibinfo {author} {\bibfnamefont {L.}~\bibnamefont {Pedersen}},\ }\bibfield
  {title} {\bibinfo {title} {{Particle mesh Ewald: An N$\cdot$log(N) method for
  Ewald sums in large systems}},\ }\href {https://doi.org/10.1063/1.464397}
  {\bibfield  {journal} {\bibinfo  {journal} {The Journal of Chemical Physics}\
  }\textbf {\bibinfo {volume} {98}},\ \bibinfo {pages} {10089} (\bibinfo {year}
  {1993})}\BibitemShut {NoStop}%
\bibitem [{\citenamefont {Bussi}\ \emph {et~al.}(2007)\citenamefont {Bussi},
  \citenamefont {Donadio},\ and\ \citenamefont {Parrinello}}]{SBussi2007}%
  \BibitemOpen
  \bibfield  {author} {\bibinfo {author} {\bibfnamefont {G.}~\bibnamefont
  {Bussi}}, \bibinfo {author} {\bibfnamefont {D.}~\bibnamefont {Donadio}},\
  and\ \bibinfo {author} {\bibfnamefont {M.}~\bibnamefont {Parrinello}},\
  }\bibfield  {title} {\bibinfo {title} {{Canonical sampling through velocity
  rescaling}},\ }\bibfield  {journal} {\bibinfo  {journal} {Journal of Chemical
  Physics}\ }\textbf {\bibinfo {volume} {126}},\ \href
  {https://doi.org/10.1063/1.2408420} {10.1063/1.2408420} (\bibinfo {year}
  {2007})\BibitemShut {NoStop}%
\bibitem [{\citenamefont {Berendsen}\ \emph {et~al.}(1984)\citenamefont
  {Berendsen}, \citenamefont {Postma}, \citenamefont {Gunsteren}, \citenamefont
  {Dinola},\ and\ \citenamefont {Haak}}]{SBerendsen1984}%
  \BibitemOpen
  \bibfield  {author} {\bibinfo {author} {\bibfnamefont {H.~J.}\ \bibnamefont
  {Berendsen}}, \bibinfo {author} {\bibfnamefont {J.~P.}\ \bibnamefont
  {Postma}}, \bibinfo {author} {\bibfnamefont {W.~F.~V.}\ \bibnamefont
  {Gunsteren}}, \bibinfo {author} {\bibfnamefont {A.}~\bibnamefont {Dinola}},\
  and\ \bibinfo {author} {\bibfnamefont {J.~R.}\ \bibnamefont {Haak}},\
  }\bibfield  {title} {\bibinfo {title} {{Molecular dynamics with coupling to
  an external bath}},\ }\href {https://doi.org/10.1063/1.448118} {\bibfield
  {journal} {\bibinfo  {journal} {The Journal of Chemical Physics}\ }\textbf
  {\bibinfo {volume} {81}},\ \bibinfo {pages} {3684} (\bibinfo {year}
  {1984})}\BibitemShut {NoStop}%
\bibitem [{\citenamefont {Bernetti}\ and\ \citenamefont
  {Bussi}(2020)}]{SBernetti2020}%
  \BibitemOpen
  \bibfield  {author} {\bibinfo {author} {\bibfnamefont {M.}~\bibnamefont
  {Bernetti}}\ and\ \bibinfo {author} {\bibfnamefont {G.}~\bibnamefont
  {Bussi}},\ }\bibfield  {title} {\bibinfo {title} {{Pressure control using
  stochastic cell rescaling}},\ }\bibfield  {journal} {\bibinfo  {journal} {The
  Journal of Chemical Physics}\ }\textbf {\bibinfo {volume} {153}},\ \href
  {https://doi.org/10.1063/5.0020514} {10.1063/5.0020514} (\bibinfo {year}
  {2020})\BibitemShut {NoStop}%
\bibitem [{\citenamefont {Hutter}\ \emph {et~al.}(2014)\citenamefont {Hutter},
  \citenamefont {Iannuzzi}, \citenamefont {Schiffmann},\ and\ \citenamefont
  {VandeVondele}}]{SHutter2014}%
  \BibitemOpen
  \bibfield  {author} {\bibinfo {author} {\bibfnamefont {J.}~\bibnamefont
  {Hutter}}, \bibinfo {author} {\bibfnamefont {M.}~\bibnamefont {Iannuzzi}},
  \bibinfo {author} {\bibfnamefont {F.}~\bibnamefont {Schiffmann}},\ and\
  \bibinfo {author} {\bibfnamefont {J.}~\bibnamefont {VandeVondele}},\
  }\bibfield  {title} {\bibinfo {title} {{cp2k: atomistic simulations of
  condensed matter systems}},\ }\href {https://doi.org/10.1002/wcms.1159}
  {\bibfield  {journal} {\bibinfo  {journal} {WIREs Computational Molecular
  Science}\ }\textbf {\bibinfo {volume} {4}},\ \bibinfo {pages} {15} (\bibinfo
  {year} {2014})}\BibitemShut {NoStop}%
\bibitem [{\citenamefont {Lippert}\ \emph {et~al.}(1999)\citenamefont
  {Lippert}, \citenamefont {Hutter},\ and\ \citenamefont
  {Parrinello}}]{SLippert1999}%
  \BibitemOpen
  \bibfield  {author} {\bibinfo {author} {\bibfnamefont {G.}~\bibnamefont
  {Lippert}}, \bibinfo {author} {\bibfnamefont {J.}~\bibnamefont {Hutter}},\
  and\ \bibinfo {author} {\bibfnamefont {M.}~\bibnamefont {Parrinello}},\
  }\bibfield  {title} {\bibinfo {title} {{The Gaussian and augmented-plane-wave
  density functional method for ab initio molecular dynamics simulations}},\
  }\href {https://doi.org/10.1007/s002140050523} {\bibfield  {journal}
  {\bibinfo  {journal} {Theoretical Chemistry Accounts: Theory, Computation,
  and Modeling (Theoretica Chimica Acta)}\ }\textbf {\bibinfo {volume} {103}},\
  \bibinfo {pages} {124} (\bibinfo {year} {1999})}\BibitemShut {NoStop}%
\bibitem [{\citenamefont {Perdew}\ \emph {et~al.}(1996)\citenamefont {Perdew},
  \citenamefont {Burke},\ and\ \citenamefont {Ernzerhof}}]{SPerdew1996}%
  \BibitemOpen
  \bibfield  {author} {\bibinfo {author} {\bibfnamefont {J.~P.}\ \bibnamefont
  {Perdew}}, \bibinfo {author} {\bibfnamefont {K.}~\bibnamefont {Burke}},\ and\
  \bibinfo {author} {\bibfnamefont {M.}~\bibnamefont {Ernzerhof}},\ }\bibfield
  {title} {\bibinfo {title} {{Generalized Gradient Approximation Made
  Simple}},\ }\href {https://doi.org/10.1103/PhysRevLett.77.3865} {\bibfield
  {journal} {\bibinfo  {journal} {Physical Review Letters}\ }\textbf {\bibinfo
  {volume} {77}},\ \bibinfo {pages} {3865} (\bibinfo {year}
  {1996})}\BibitemShut {NoStop}%
\bibitem [{\citenamefont {Zhang}\ and\ \citenamefont {Yang}(1998)}]{SZhang1998}%
  \BibitemOpen
  \bibfield  {author} {\bibinfo {author} {\bibfnamefont {Y.}~\bibnamefont
  {Zhang}}\ and\ \bibinfo {author} {\bibfnamefont {W.}~\bibnamefont {Yang}},\
  }\bibfield  {title} {\bibinfo {title} {{Comment on ``Generalized Gradient
  Approximation Made Simple''}},\ }\href
  {https://doi.org/10.1103/PhysRevLett.80.890} {\bibfield  {journal} {\bibinfo
  {journal} {Physical Review Letters}\ }\textbf {\bibinfo {volume} {80}},\
  \bibinfo {pages} {890} (\bibinfo {year} {1998})}\BibitemShut {NoStop}%
\bibitem [{\citenamefont {Grimme}\ \emph {et~al.}(2010)\citenamefont {Grimme},
  \citenamefont {Antony}, \citenamefont {Ehrlich},\ and\ \citenamefont
  {Krieg}}]{SGrimme2010}%
  \BibitemOpen
  \bibfield  {author} {\bibinfo {author} {\bibfnamefont {S.}~\bibnamefont
  {Grimme}}, \bibinfo {author} {\bibfnamefont {J.}~\bibnamefont {Antony}},
  \bibinfo {author} {\bibfnamefont {S.}~\bibnamefont {Ehrlich}},\ and\ \bibinfo
  {author} {\bibfnamefont {H.}~\bibnamefont {Krieg}},\ }\bibfield  {title}
  {\bibinfo {title} {{A consistent and accurate ab initio parametrization of
  density functional dispersion correction (DFT-D) for the 94 elements H-Pu}},\
  }\bibfield  {journal} {\bibinfo  {journal} {The Journal of Chemical Physics}\
  }\textbf {\bibinfo {volume} {132}},\ \href
  {https://doi.org/10.1063/1.3382344} {10.1063/1.3382344} (\bibinfo {year}
  {2010})\BibitemShut {NoStop}%
\bibitem [{\citenamefont {Grimme}\ \emph {et~al.}(2011)\citenamefont {Grimme},
  \citenamefont {Ehrlich},\ and\ \citenamefont {Goerigk}}]{SGrimme2011}%
  \BibitemOpen
  \bibfield  {author} {\bibinfo {author} {\bibfnamefont {S.}~\bibnamefont
  {Grimme}}, \bibinfo {author} {\bibfnamefont {S.}~\bibnamefont {Ehrlich}},\
  and\ \bibinfo {author} {\bibfnamefont {L.}~\bibnamefont {Goerigk}},\
  }\bibfield  {title} {\bibinfo {title} {{Effect of the damping function in
  dispersion corrected density functional theory}},\ }\href
  {https://doi.org/10.1002/jcc.21759} {\bibfield  {journal} {\bibinfo
  {journal} {Journal of Computational Chemistry}\ }\textbf {\bibinfo {volume}
  {32}},\ \bibinfo {pages} {1456} (\bibinfo {year} {2011})}\BibitemShut
  {NoStop}%
\bibitem [{\citenamefont {Goedecker}\ \emph {et~al.}(1996)\citenamefont
  {Goedecker}, \citenamefont {Teter},\ and\ \citenamefont
  {Hutter}}]{SGoedecker1996}%
  \BibitemOpen
  \bibfield  {author} {\bibinfo {author} {\bibfnamefont {S.}~\bibnamefont
  {Goedecker}}, \bibinfo {author} {\bibfnamefont {M.}~\bibnamefont {Teter}},\
  and\ \bibinfo {author} {\bibfnamefont {J.}~\bibnamefont {Hutter}},\
  }\bibfield  {title} {\bibinfo {title} {{Separable dual-space Gaussian
  pseudopotentials}},\ }\href {https://doi.org/10.1103/PhysRevB.54.1703}
  {\bibfield  {journal} {\bibinfo  {journal} {Physical Review B}\ }\textbf
  {\bibinfo {volume} {54}},\ \bibinfo {pages} {1703} (\bibinfo {year}
  {1996})}\BibitemShut {NoStop}%
\bibitem [{\citenamefont {Hartwigsen}\ \emph {et~al.}(1998)\citenamefont
  {Hartwigsen}, \citenamefont {Goedecker},\ and\ \citenamefont
  {Hutter}}]{SHartwigsen1998}%
  \BibitemOpen
  \bibfield  {author} {\bibinfo {author} {\bibfnamefont {C.}~\bibnamefont
  {Hartwigsen}}, \bibinfo {author} {\bibfnamefont {S.}~\bibnamefont
  {Goedecker}},\ and\ \bibinfo {author} {\bibfnamefont {J.}~\bibnamefont
  {Hutter}},\ }\bibfield  {title} {\bibinfo {title} {{Relativistic separable
  dual-space Gaussian pseudopotentials from H to Rn}},\ }\href
  {https://doi.org/10.1103/PhysRevB.58.3641} {\bibfield  {journal} {\bibinfo
  {journal} {Physical Review B}\ }\textbf {\bibinfo {volume} {58}},\ \bibinfo
  {pages} {3641} (\bibinfo {year} {1998})}\BibitemShut {NoStop}%
\bibitem [{\citenamefont {Karwounopoulos}\ \emph {et~al.}(2024)\citenamefont
  {Karwounopoulos}, \citenamefont {Wu}, \citenamefont {Tkaczyk}, \citenamefont
  {Wang}, \citenamefont {Baskerville}, \citenamefont {Ranasinghe},
  \citenamefont {Langer}, \citenamefont {Wood}, \citenamefont {Wieder},\ and\
  \citenamefont {Boresch}}]{SKarwounopoulos2024}%
  \BibitemOpen
  \bibfield  {author} {\bibinfo {author} {\bibfnamefont {J.}~\bibnamefont
  {Karwounopoulos}}, \bibinfo {author} {\bibfnamefont {Z.}~\bibnamefont {Wu}},
  \bibinfo {author} {\bibfnamefont {S.}~\bibnamefont {Tkaczyk}}, \bibinfo
  {author} {\bibfnamefont {S.}~\bibnamefont {Wang}}, \bibinfo {author}
  {\bibfnamefont {A.}~\bibnamefont {Baskerville}}, \bibinfo {author}
  {\bibfnamefont {K.}~\bibnamefont {Ranasinghe}}, \bibinfo {author}
  {\bibfnamefont {T.}~\bibnamefont {Langer}}, \bibinfo {author} {\bibfnamefont
  {G.~P.~F.}\ \bibnamefont {Wood}}, \bibinfo {author} {\bibfnamefont
  {M.}~\bibnamefont {Wieder}},\ and\ \bibinfo {author} {\bibfnamefont
  {S.}~\bibnamefont {Boresch}},\ }\bibfield  {title} {\bibinfo {title}
  {{Insights and Challenges in Correcting Force Field Based Solvation Free
  Energies Using a Neural Network Potential}},\ }\href
  {https://doi.org/10.1021/acs.jpcb.4c01417} {\bibfield  {journal} {\bibinfo
  {journal} {The Journal of Physical Chemistry B}\ }\textbf {\bibinfo {volume}
  {128}},\ \bibinfo {pages} {6693} (\bibinfo {year} {2024})}\BibitemShut
  {NoStop}%
\bibitem [{\citenamefont {Grotz}\ \emph {et~al.}(2021)\citenamefont {Grotz},
  \citenamefont {Cruz-Le\'{o}n},\ and\ \citenamefont {Schwierz}}]{SGrotz2021}%
  \BibitemOpen
  \bibfield  {author} {\bibinfo {author} {\bibfnamefont {K.~K.}\ \bibnamefont
  {Grotz}}, \bibinfo {author} {\bibfnamefont {S.}~\bibnamefont
  {Cruz-Le\'{o}n}},\ and\ \bibinfo {author} {\bibfnamefont {N.}~\bibnamefont
  {Schwierz}},\ }\bibfield  {title} {\bibinfo {title} {{Optimized Magnesium
  Force Field Parameters for Biomolecular Simulations with Accurate Solvation,
  Ion-Binding, and Water-Exchange Properties}},\ }\href
  {https://doi.org/10.1021/acs.jctc.0c01281} {\bibfield  {journal} {\bibinfo
  {journal} {Journal of Chemical Theory and Computation}\ }\textbf {\bibinfo
  {volume} {17}},\ \bibinfo {pages} {2530} (\bibinfo {year}
  {2021})}\BibitemShut {NoStop}%
\bibitem [{\citenamefont {Picha}\ \emph {et~al.}(2025)\citenamefont {Picha},
  \citenamefont {Tkaczyk}, \citenamefont {Langer}, \citenamefont {Wieder},\
  and\ \citenamefont {Boresch}}]{SPicha2025}%
  \BibitemOpen
  \bibfield  {author} {\bibinfo {author} {\bibfnamefont {A.~K.}\ \bibnamefont
  {Picha}}, \bibinfo {author} {\bibfnamefont {S.}~\bibnamefont {Tkaczyk}},
  \bibinfo {author} {\bibfnamefont {T.}~\bibnamefont {Langer}}, \bibinfo
  {author} {\bibfnamefont {M.}~\bibnamefont {Wieder}},\ and\ \bibinfo {author}
  {\bibfnamefont {S.}~\bibnamefont {Boresch}},\ }\bibfield  {title} {\bibinfo
  {title} {{Architecture-Independent Absolute Solvation Free Energy
  Calculations with Neural Network Potentials}},\ }\href
  {https://doi.org/10.1021/acs.jpclett.5c02980} {\bibfield  {journal} {\bibinfo
   {journal} {The Journal of Physical Chemistry Letters}\ }\textbf {\bibinfo
  {volume} {16}},\ \bibinfo {pages} {12080} (\bibinfo {year}
  {2025})}\BibitemShut {NoStop}%
\bibitem [{\citenamefont {Bennett}(1976)}]{SBennett1976}%
  \BibitemOpen
  \bibfield  {author} {\bibinfo {author} {\bibfnamefont {C.~H.}\ \bibnamefont
  {Bennett}},\ }\bibfield  {title} {\bibinfo {title} {{Efficient estimation of
  free energy differences from Monte Carlo data}},\ }\href
  {https://doi.org/10.1016/0021-9991(76)90078-4} {\bibfield  {journal}
  {\bibinfo  {journal} {Journal of Computational Physics}\ }\textbf {\bibinfo
  {volume} {22}},\ \bibinfo {pages} {245} (\bibinfo {year} {1976})}\BibitemShut
  {NoStop}%
\bibitem [{\citenamefont {Jung}\ \emph {et~al.}(2017)\citenamefont {Jung},
  \citenamefont {ichi Okazaki},\ and\ \citenamefont {Hummer}}]{SJung2017}%
  \BibitemOpen
  \bibfield  {author} {\bibinfo {author} {\bibfnamefont {H.}~\bibnamefont
  {Jung}}, \bibinfo {author} {\bibfnamefont {K.}~\bibnamefont {ichi Okazaki}},\
  and\ \bibinfo {author} {\bibfnamefont {G.}~\bibnamefont {Hummer}},\
  }\bibfield  {title} {\bibinfo {title} {{Transition path sampling of rare
  events by shooting from the top}},\ }\href
  {https://doi.org/10.1063/1.4997378} {\bibfield  {journal} {\bibinfo
  {journal} {The Journal of Chemical Physics}\ }\textbf {\bibinfo {volume}
  {147}},\ \bibinfo {pages} {152716} (\bibinfo {year} {2017})}\BibitemShut
  {NoStop}%
\bibitem [{\citenamefont {van Erp}(2007)}]{SvanErp2007}%
  \BibitemOpen
  \bibfield  {author} {\bibinfo {author} {\bibfnamefont {T.~S.}\ \bibnamefont
  {van Erp}},\ }\bibfield  {title} {\bibinfo {title} {{Reaction Rate
  Calculation by Parallel Path Swapping}},\ }\href
  {https://doi.org/10.1103/PhysRevLett.98.268301} {\bibfield  {journal}
  {\bibinfo  {journal} {Physical Review Letters}\ }\textbf {\bibinfo {volume}
  {98}},\ \bibinfo {pages} {268301} (\bibinfo {year} {2007})}\BibitemShut
  {NoStop}%
\bibitem [{\citenamefont {Kirkwood}\ and\ \citenamefont
  {Buff}(1951)}]{SKirkwood1951}%
  \BibitemOpen
  \bibfield  {author} {\bibinfo {author} {\bibfnamefont {J.~G.}\ \bibnamefont
  {Kirkwood}}\ and\ \bibinfo {author} {\bibfnamefont {F.~P.}\ \bibnamefont
  {Buff}},\ }\bibfield  {title} {\bibinfo {title} {{The Statistical Mechanical
  Theory of Solutions. I}},\ }\href {https://doi.org/10.1063/1.1748352}
  {\bibfield  {journal} {\bibinfo  {journal} {The Journal of Chemical Physics}\
  }\textbf {\bibinfo {volume} {19}},\ \bibinfo {pages} {774} (\bibinfo {year}
  {1951})}\BibitemShut {NoStop}%
\bibitem [{\citenamefont {Torrie}\ and\ \citenamefont
  {Valleau}(1977)}]{STorrie1977}%
  \BibitemOpen
  \bibfield  {author} {\bibinfo {author} {\bibfnamefont {G.}~\bibnamefont
  {Torrie}}\ and\ \bibinfo {author} {\bibfnamefont {J.}~\bibnamefont
  {Valleau}},\ }\bibfield  {title} {\bibinfo {title} {{Nonphysical sampling
  distributions in Monte Carlo free-energy estimation: Umbrella sampling}},\
  }\href {https://doi.org/10.1016/0021-9991(77)90121-8} {\bibfield  {journal}
  {\bibinfo  {journal} {Journal of Computational Physics}\ }\textbf {\bibinfo
  {volume} {23}},\ \bibinfo {pages} {187} (\bibinfo {year} {1977})}\BibitemShut
  {NoStop}%
\bibitem [{\citenamefont {Kumar}\ \emph {et~al.}(1992)\citenamefont {Kumar},
  \citenamefont {Rosenberg}, \citenamefont {Bouzida}, \citenamefont
  {Swendsen},\ and\ \citenamefont {Kollman}}]{SKumar1992}%
  \BibitemOpen
  \bibfield  {author} {\bibinfo {author} {\bibfnamefont {S.}~\bibnamefont
  {Kumar}}, \bibinfo {author} {\bibfnamefont {J.~M.}\ \bibnamefont
  {Rosenberg}}, \bibinfo {author} {\bibfnamefont {D.}~\bibnamefont {Bouzida}},
  \bibinfo {author} {\bibfnamefont {R.~H.}\ \bibnamefont {Swendsen}},\ and\
  \bibinfo {author} {\bibfnamefont {P.~A.}\ \bibnamefont {Kollman}},\
  }\bibfield  {title} {\bibinfo {title} {{THE weighted histogram analysis
  method for free-energy calculations on biomolecules. I. The method}},\ }\href
  {https://doi.org/10.1002/jcc.540130812} {\bibfield  {journal} {\bibinfo
  {journal} {Journal of Computational Chemistry}\ }\textbf {\bibinfo {volume}
  {13}},\ \bibinfo {pages} {1011} (\bibinfo {year} {1992})}\BibitemShut
  {NoStop}%
\bibitem [{\citenamefont {Laio}\ and\ \citenamefont
  {Parrinello}(2002)}]{SLaio2002}%
  \BibitemOpen
  \bibfield  {author} {\bibinfo {author} {\bibfnamefont {A.}~\bibnamefont
  {Laio}}\ and\ \bibinfo {author} {\bibfnamefont {M.}~\bibnamefont
  {Parrinello}},\ }\bibfield  {title} {\bibinfo {title} {{Escaping free-energy
  minima}},\ }\href {https://doi.org/10.1073/pnas.202427399} {\bibfield
  {journal} {\bibinfo  {journal} {Proceedings of the National Academy of
  Sciences}\ }\textbf {\bibinfo {volume} {99}},\ \bibinfo {pages} {12562}
  (\bibinfo {year} {2002})}\BibitemShut {NoStop}%
\bibitem [{\citenamefont {Barducci}\ \emph {et~al.}(2008)\citenamefont
  {Barducci}, \citenamefont {Bussi},\ and\ \citenamefont
  {Parrinello}}]{SBarducci2008}%
  \BibitemOpen
  \bibfield  {author} {\bibinfo {author} {\bibfnamefont {A.}~\bibnamefont
  {Barducci}}, \bibinfo {author} {\bibfnamefont {G.}~\bibnamefont {Bussi}},\
  and\ \bibinfo {author} {\bibfnamefont {M.}~\bibnamefont {Parrinello}},\
  }\bibfield  {title} {\bibinfo {title} {{Well-Tempered Metadynamics: A
  Smoothly Converging and Tunable Free-Energy Method}},\ }\href
  {https://doi.org/10.1103/PhysRevLett.100.020603} {\bibfield  {journal}
  {\bibinfo  {journal} {Physical Review Letters}\ }\textbf {\bibinfo {volume}
  {100}},\ \bibinfo {pages} {020603} (\bibinfo {year} {2008})}\BibitemShut
  {NoStop}%

\end{thebibliography}
\end{document}